\documentclass[12pt]{article}
\usepackage{epsfig}
\usepackage{psfrag}
\usepackage{latexsym}
\usepackage{indentfirst}
\usepackage{fancyhdr}
\usepackage{amssymb}
\usepackage{amsfonts}
\usepackage{cite}
\usepackage{bbold}
\usepackage[footnotesize]{caption2}
\usepackage{graphicx}
\usepackage[center,footnotesize,hang]{subfigure}
\usepackage{url}
\usepackage{color}
\newcommand{\be}{\beta}

\newcommand{\La}{\Lambda}
\newcommand{\om}{\omega}

\newcommand{\mean}[1]{\langle#1\rangle}

\newcommand{\beq}{\begin{equation}}
\newcommand{\eeq}{\end{equation}}
\newcommand{\bac}{\beq\begin{array}}
\newcommand{\eac}{\end{array}\eeq}
\newcommand{\ba}{\begin{array}}
\newcommand{\ea}{\end{array}}
\newcommand{\bea}{\begin{eqnarray}}
\newcommand{\eea}{\end{eqnarray}}
\newcommand{\beaa}{\begin{eqnarray*}}
\newcommand{\eeaa}{\end{eqnarray*}}

\newcommand{\nn}{\nonumber}

\def\cC{{\cal C}}

\def\beq{\begin{equation}}
\def\eeq{\end{equation}}
\def\bea{\begin{eqnarray}}
\def\eea{\end{eqnarray}}
\def\bet{\begin{tabular}}
\def\eet{\end{tabular}}
\def\bes{\begin{subequations}\bea}
\def\ees{\eea\end{subequations}}

\def\gappeq{\mathrel{\rlap {\raise.5ex\hbox{$>$}} {\lower.5ex\hbox{$\sim$}}}}
\def\lappeq{\mathrel{\rlap{\raise.5ex\hbox{$<$}} {\lower.5ex\hbox{$\sim$}}}}
\def\lsim{\ \rlap{\raise 3pt \hbox{$<$}}{\lower 3pt \hbox{$\sim$}}\ }
\def\gsim{\ \rlap{\raise 3pt \hbox{$>$}}{\lower 3pt \hbox{$\sim$}}\ }
\def\cY{{\cal Y}}
\def\hcY{\hat{\cal Y}}

\def\be{\begin{equation}}
\def\ee{\end{equation}}
\def\bc{\begin{center}}
\def\ec{\end{center}}
\def\bea{\begin{eqnarray}}
\def\eea{\end{eqnarray}}
\def\dd{\displaystyle}
\def\nn{\nonumber}

\def\ov{\overline}

\catcode`@=11
\def\marginnote#1{}
\newcount\hour
\newcount\minute
\newtoks\amorpm
\hour=\time\divide\hour by60
\minute=\time{\multiply\hour by60 \global\advance\minute by-\hour}
\edef\standardtime{{\ifnum\hour<12 \global\amorpm={am}%
        \else\global\amorpm={pm}\advance\hour by-12 \fi
        \ifnum\hour=0 \hour=12 \fi
        \number\hour:\ifnum\minute<10 0\fi\number\minute\the\amorpm}}
\edef\militarytime{\number\hour:\ifnum\minute<10 0\fi\number\minute}
\def\draftlabel#1{{\@bsphack\if@filesw {\let\thepage\relax
   \xdef\@gtempa{\write\@auxout{\string
      \newlabel{#1}{{\@currentlabel}{\thepage}}}}}\@gtempa
   \if@nobreak \ifvmode\nobreak\fi\fi\fi\@esphack}
        \gdef\@eqnlabel{#1}}
\def\@eqnlabel{}
\def\@vacuum{}
\def\draftmarginnote#1{\marginpar{\raggedright\scriptsize\tt#1}}
\def\draft{\oddsidemargin 0.0truein
        \def\@oddfoot{\sl preliminary draft \hfil
        \rm\thepage\hfil\sl\today\quad\militarytime}
        \let\@evenfoot\@oddfoot \overfullrule 3pt
        \let\label=\draftlabel
        \let\marginnote=\draftmarginnote
   \def\@eqnnum{(\theequation)\rlap{\kern\marginparsep\tt\@eqnlabel}%
\global\let\@eqnlabel\@vacuum}  }
\catcode`@=12
\begin{document}
\begin{titlepage}
\vspace*{-1cm}
\phantom{hep-ph/***}

\hfill{RM3-TH/10-01}
\hfill{CERN-PH-TH/2010-016}
\hfill{DFPD-10/TH/02}

\vskip 2.5cm
\begin{center}
{\Large\bf Discrete Flavor Symmetries}

\vskip 0.2 cm
{\Large\bf and Models of Neutrino Mixing }
\end{center}
\vskip 0.2  cm
\vskip 0.5  cm
\begin{center}
{\large Guido Altarelli}~\footnote{e-mail address: guido.altarelli@cern.ch}
\\
\vskip .1cm
Dipartimento di Fisica `E.~Amaldi', Universit\`a di Roma Tre
\\
INFN, Sezione di Roma Tre, I-00146 Rome, Italy
\\
\vskip .1cm
and
\\
CERN, Department of Physics, Theory Division
\\
CH-1211 Geneva 23, Switzerland
\\

\vskip .2cm
{\large Ferruccio Feruglio}~\footnote{e-mail address: feruglio@pd.infn.it} 
\\
\vskip .1cm
Dipartimento di Fisica `G.~Galilei', Universit\`a di Padova
\\
INFN, Sezione di Padova, Via Marzolo~8, I-35131 Padua, Italy
\\
\end{center}
\vskip 0.7cm
\begin{abstract}
\noindent
We review the application of non abelian discrete groups to the theory of neutrino masses and mixing, which is strongly suggested by the agreement of the Tri-Bimaximal (TB) mixing pattern with experiment. After summarizing the motivation and the formalism, we discuss specific models, based on $A_4$, $S_4$ and other finite groups, and their phenomenological implications, including lepton flavor violating processes, leptogenesis and the extension to quarks. In alternative to TB mixing the application of discrete flavor symmetries to quark-lepton complementarity and Bimaximal Mixing (BM) is also considered.
\end{abstract}
\end{titlepage}
\setcounter{footnote}{0}
\vskip2truecm
%
%
\section{Introduction}

Experiments on neutrino oscillations, which measure differences of squared masses and mixing angles \cite{Altarelli:2004, Altarelli:2009, Mohapatra:2006, Mohapatra:2007, Grimus:2006, Gonzalez-Garcia:2008} 
have established that neutrinos have a mass. We refer in particular to ref. \cite{Altarelli:2004} for an introduction to the subject, the main results, the basic formalism and all definitions and notations. Two distinct oscillation frequencies have been first measured in solar \cite{Super-Kamiokande:2006, Super-Kamiokande:2008, SNO:2009} and atmospheric  \cite{Super-Kamiokande:2006a, Super-Kamiokande:2006b} neutrino oscillations and later confirmed by experiments on earth, like KamLAND \cite{KamLAND:2008}, K2K \cite{K2K:2006}, MINOS \cite{Kafka:2010zz} and OPERA \cite{Agafonova:2010dc}. A signal corresponding to a third mass difference was claimed by the LSND experiment \cite{LSND:1996, LSND:1998, LSND:1998a} but not confirmed by KARMEN \cite{KARMEN:2002} and recently by MiniBooNE \cite{MiniBooNE:2009, MiniBooNE:2009a}. Two well separated differences need at least three different neutrino mass eigenstates involved in oscillations. Actually the three known neutrino species can be sufficient. At least two $\nu$'s must be massive while, in principle, the third one could still be massless. In the following we will assume the simplest picture with three active neutrinos, no sterile neutrinos and CPT invariance. The mass eigenstates involved in solar oscillations are $m_1$ and $m_2$ and, by definition, $|m_2|> |m_1|$, so that $\Delta m^2_{sun}=\Delta m^2_{21}=|m_2|^2-|m_1|^2>0$. The atmospheric neutrino oscillations involve $m_3$:  $\Delta m^2_{atm}=|\Delta m^2_{31}|$ with $\Delta m^2_{31}=|m_3|^2-|m_1|^2$ either positive (normal hierarchy) or negative (inverse hierarchy). The present data \cite{Strumia:2006, Gonzalez-Garcia:2008a, Bandyopadhyay:2008, Fogli:2008, Fogli:2008a, Schwetz:2008, Maltoni:2008}  are compatible with both cases. The degenerate spectrum occurs when the average absolute value of the masses is much larger than all mass squared differences: $|m_i|^2 >> |\Delta m^2_{hk}|$. With the standard set of notations and definitions \cite{Altarelli:2004} the present data are summarised in Table 1.

\begin{table}[h]
\begin{center}
\begin{tabular}{|c|c|c|}
  \hline
  Quantity & ref. \cite{Fogli:2008, Fogli:2008a} & ref. \cite{Schwetz:2008, Maltoni:2008} \\
  \hline
  $\Delta m^2_{sun}~(10^{-5}~{\rm eV}^2)$ &$7.67^{+0.16}_{-0.19}$ & $7.65^{+0.23}_{-0.20}$  \\
  $\Delta m^2_{atm}~(10^{-3}~{\rm eV}^2)$ &$2.39^{+0.11}_{-0.08}$ & $2.40^{+0.12}_{-0.11}$  \\
  $\sin^2\theta_{12}$ &$0.312^{+0.019}_{-0.018}$ & $0.304^{+0.022}_{-0.016}$ \\
  $\sin^2\theta_{23}$ &$0.466^{+0.073}_{-0.058}$ &  $0.50^{+0.07}_{-0.06}$ \\
  $\sin^2\theta_{13}$ &$0.016\pm0.010$ &$0.010^{+0.016}_{-0.011}$  \\
  \hline
  \end{tabular}
\end{center}
\caption{\label{tab:data} Fits to neutrino oscillation data.}
\end{table}

Oscillation experiments do not provide information about either the absolute neutrino mass scale or the Dirac/Majorana nature of neutrinos.
Limits on the mass scale are obtained \cite{Altarelli:2004} from the endpoint of the tritium beta decay spectrum, from cosmology (see, for example \cite{Lesgourgues:2006}) and from neutrinoless double beta decay ($0\nu \beta \beta$) (for a recent review, see, for example \cite {Avignone:2008}). From tritium we have an absolute upper limit of
2.2 eV (at 95\% C.L.) on the mass of electron  antineutrino \cite{Kraus:2005}, which, combined with the observed oscillation
frequencies under the assumption of three CPT-invariant light neutrinos, represents also an upper bound on the masses of
the other active neutrinos. Complementary information on the sum of neutrino masses is also provided by the galaxy power
spectrum combined with measurements of the cosmic  microwave background anisotropies. According to recent analyses of the most reliable data \cite{Fogli:2008b}
$\sum_i \vert m_i\vert < 0.60\div 0.75$ eV (at 95\% C.L.) depending on the retained data (the numbers for the sum have to be divided by 3 in order to obtain a limit on the mass of each neutrino).
The discovery of $0\nu \beta \beta$ decay would be very important because it would establish lepton number violation and
the Majorana nature of $\nu$'s, and provide direct information on the absolute
scale of neutrino masses.
The present limit from $0\nu \beta \beta$  (with large ambiguities from nuclear matrix elements) is about $\vert m_{ee}\vert < (0.3\div 0.8)$ eV \cite {Avignone:2008, Fogli:2008b} (see eq. (\ref{3nu1gen})). 

After KamLAND \cite{KamLAND:2008}, SNO \cite{SNO:2009} and the upper limits on the absolute value of neutrino masses not too much hierarchy in the spectrum of neutrinos is indicated by experiments: 
\bea
r = \Delta m_{sol}^2/\Delta m_{atm}^2 \sim 1/30~~~.
\label{r}
\eea
Precisely $r=0.032^{+0.006}_{-0.005}$ at $3\sigma$'s  \cite{Fogli:2008, Fogli:2008a, Schwetz:2008, Maltoni:2008}. Thus, for a hierarchical spectrum, $m_2/m_3 \sim \sqrt{r} \sim 0.2$, which is comparable to the Cabibbo angle $\lambda_C \sim 0.22$ or to its leptonic analogue $\sqrt{m_{\mu}/m_{\tau}} \sim 0.24$. This suggests that the same hierarchy parameter (raised to powers with $\mathcal{O}(1)$ exponents) may apply for quark, charged lepton and neutrino mass matrices. This in turn indicates that, in the absence of some special dynamical reason, we do not expect quantities like $\theta_{13}$ or the deviation of  $\theta_{23}$ from its maximal value to be too small. Indeed it would be very important to know how small the mixing angle $\theta_{13}$  is and how close to maximal $\theta_{23}$ is. 

Given that neutrino masses are certainly extremely
small, it is really difficult from the theory point of view to avoid the conclusion that the lepton number $L$ conservation is probably violated and that $\nu$'s are Majorana fermions.
In this case the smallness of neutrino masses can be naturally explained as inversely proportional
to the large scale where $L$ conservation is violated. 
If neutrinos are Majorana particles, their masses arise  from the generic dimension-five non renormalizable operator of the form \cite{Weinberg:1979}: 
\be
O_5=\frac{(H l)^T_i \eta_{ij} (H l)_j}{M}+~h.c.~~~,
\label{O5}
\ee  
with $H$ being the ordinary Higgs doublet, $l_i$ the SU(2) lepton doublets, $\eta$ a matrix in  flavor space, $M$ a large scale of mass and a charge conjugation matrix $C$
between the lepton fields is understood. 
For $\eta_{ij}\approx \mathcal{O}(1)$, neutrino masses generated by $O_5$ are of the order
$m_{\nu}\approx v^2/M$  where $v\sim {\rm O}(100~{\rm GeV})$ is the vacuum
expectation value of the ordinary Higgs. A particular realization of this effective mass operator is given by the see-saw mechanism \cite{Minkowski:1977, Yanagida:1979, Gell-Mann:1979, Glashow:1980, Mohapatra:1980} , where $M$ derives from the exchange of heavy neutral objects of weak isospin 0 or 1. In the simplest case the exchanged particle is the right-handed (RH) neutrino $\nu^c$ (a gauge singlet fermion here described through its charge conjugate field), and the resulting neutrino mass matrix reads (type I see-saw ) \cite{Altarelli:2004}:
\be  
m_{\nu}=m_D^T M^{-1}m_D~~~,
\ee 
where $m_D$ and $M$ denote the Dirac neutrino mass matrix (defined as $ {\nu^c}^T m_D \nu$) and the Majorana mass matrix of $\nu^c$ (defined as $ {\nu^c}^T M \nu^c$), respectively.
As one sees,  the light neutrino masses are quadratic in the Dirac
masses and inversely proportional to the large Majorana mass.  For
$m_{\nu}\approx \sqrt{\Delta m^2_{atm}}\approx 0.05$ eV and 
$m_{\nu}\approx m_D^2/M$ with $m_D\approx v
\approx 200$~GeV we find $M\approx 10^{15}$~GeV which indeed is an impressive indication that the scale for lepton number violation is close to the grand unified scale
$M_{GUT}$. Thus probably neutrino masses are a probe into the physics near $M_{GUT}$. This argument, in our opinion, strongly discourages models where neutrino masses are generated near the weak scale and are suppressed by some special mechanism.

Oscillation experiments cannot distinguish between
Dirac and Majorana neutrinos.
The detection of neutrino-less double beta decay would provide direct evidence of $L$ non conservation, and the Majorana nature of neutrinos. It would also offer a way to possibly disentangle the 3 cases of degenerate, normal or inverse hierachy neutrino spectrum.  The quantity which is bound by experiments on $0\nu \beta \beta$
is the 11 entry of the
$\nu$ mass matrix, which in general, from $m_{\nu}=U^* m_{diag} U^\dagger$, is given by :
\bea 
\vert m_{ee}\vert~=\vert(1-s^2_{13})~(m_1 c^2_{12}~+~m_2 s^2_{12})+m_3 e^{2 i\phi} s^2_{13}\vert~~~,
\label{3nu1gen}
\eea
where $U\equiv U_{PMNS}$ is the mixing matrix, $m_{1,2}$ are complex masses (including Majorana phases) while $m_3$ can be taken as real and positive and $\phi$ is the $U_{PMNS}$ phase measurable from CP violation in oscillation experiments. Starting from this general formula it is simple to
derive the bounds for degenerate, inverse hierarchy or normal hierarchy mass patterns shown in Fig. 1 \cite{Feruglio:2002}. In the next few years a new generation of experiments will reach a larger sensitivity on $0\nu \beta \beta$ by about an order of magnitude. If these experiments will observe a signal this will be compatible with both type of neutrino mass ordering, if not, then the normal hierarchy case remains a possibility. Establishing that $L$ is violated in particle interactions
would also strongly support the possibility that the observed baryon asymmetry is generated via leptogenesis, through the out-of-equilibrium, CP and $L$ violating decays of the heavy RH neutrinos (see Sect. 10).

\begin{figure}
\centering
\includegraphics [width=10.0 cm]{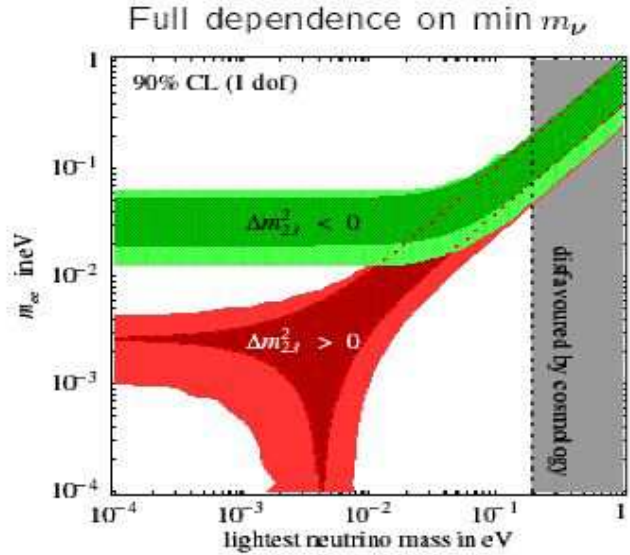}    
\caption{A plot \cite{Feruglio:2002} of $m_{ee}$ in eV, the quantity measured in neutrino-less double beta decay, given in eq. (\ref{3nu1gen}), versus the lightest neutrino mass $m_1$, also in eV. The upper (lower) band is for inverse (normal) hierarchy.}
\end{figure}

Neutrino mixing is important because it could in principle provide new clues for the understanding of the flavor problem. Even more so since neutrino mixing angles show a pattern that is completely different than that of quark mixing: for quarks all mixing angles are small, for neutrinos two angles are large (one is even compatible with the maximal value) and only the third one is small. For building up theoretical models of neutrino mixing one must guess which features of the data are really relevant in order to identify the basic principles for the formulation of the model.  In particular, it is an experimental fact \cite{Strumia:2006, Gonzalez-Garcia:2008a, Bandyopadhyay:2008, Fogli:2008, Fogli:2008a, Schwetz:2008, Maltoni:2008} that within measurement errors
the observed neutrino mixing matrix \cite{Altarelli:2004} is compatible with
the so called Tri-Bimaximal (TB) form in eq. (\ref{2}) \cite{Harrison:2002, Harrison:2002a, Harrison:2003, Harrison:2004}. The best measured neutrino mixing angle $\theta_{12}$ is just about 1$\sigma$ below the TB value $\tan^2{\theta_{12}}=1/2$, while the other two angles are well inside the 1$\sigma$ interval (see table \ref{tab:data}). Thus, one possibility is that one takes this coincidence seriously and only considers models where TB mixing is automatically a good first approximation. Alternatively one can assume that the agreement of the data with TB mixing is accidental. Indeed there are many models that fit the data and yet TB mixing does not play any role in their architecture. For example, in ref. \cite{Albright:2008} there is a list of Grand Unified SO(10) models with parameters that can be fitted to the  neutrino mixing angles leading to a good agreement with the data although most of these models  have no built-in relation with TB mixing (see also \cite{Bertolini:2006}). Another class of examples is found in ref. \cite{Plentinger:2008}. Clearly, for this type of models, in most cases different mixing angles could also be accommodated by simply varying the fitted values of the parameters. If instead we assume that TB mixing has a real physical meaning, then we are led to consider models that naturally produce TB mixing in first approximation and only a very special dynamics can lead to this peculiar mixing matrix. Discrete non abelian groups (for an introduction see, for example, \cite{Frampton:1995,Ishimori:2010au}) naturally emerge as suitable flavor symmetries. In fact the TB mixing matrix immediately suggests rotations by fixed, discrete angles. It has been found that a broken flavor symmetry based on the discrete
group $A_4$ (the group of even permutations of 4 elements, which can be seen as the invariance group of a rigid regular tetrahedron) appears to be particularly suitable to reproduce this specific mixing pattern in leading order (LO). A non exhaustive list of papers that discuss the application of $A_4$ to neutrino mixing is given by \cite{Ma:2001, Ma:2002, Babu:2003, Hirsch:2004, Ma:2004, Ma:2004a, Chen:2005, Altarelli:2005, Ma:2005, Hirsch:2005, Babu:2005, Ma:2005a, Zee:2005, Ma:2006, He:2006, Adhikary:2006, Altarelli:2006, Lavoura:2006, Ma:2007, Hirsch:2007, Altarelli:2007, Yin:2007, Bazzocchi:2008, Bazzocchi:2008a, Honda:2008, Brahmachari:2008, Adhikary:2008, Hirsch:2008, Frampton:2008, Csaki:2008, Altarelli:2008, Morisi:2009, Lin:2009, Lin:2009a, Altarelli:2009a, Ma:2005b, Ma:2006a, Ma:2006b, Morisi:2007, Grimus:2008, Ciafaloni:2009, Bazzocchi:2009, Bazzocchi:2008b, delAguila:2010,Kadosh:2010rm,Antusch:2010es}. The choice of this particular discrete group is not unique and, for example, other solutions based on alternative discrete flavor symmetries (for example, $T$' \cite{Frampton:1995, Aranda:2000, Aranda:2000a, Carr:2000,  Aranda:2007, Frampton:2007, Frampton:2009, Ding:2009,Feruglio:2007, Chen:2007}, $S_4$ \cite{Mohapatra:2004, Hagedorn:2006, Cai:2006, Ma:2007a, Bazzocchi:2008c, Ishimori:2009, Bazzocchi:2009a, Bazzocchi:2009b, Meloni:2009, Dutta:2009, Dutta:2009a,Ding:2010,Morisi:2010,Hagedorn:2010th,Ishimori:2010xk}, $\Delta(27) $ \cite{deMedeirosVarzielas:2007, Ma:2007b, Grimus:2007, Luhn:2007, Bazzocchi:2009c, Ding:2010}  and other groups \cite{Everett:2009, Luhn:2007a, King:2009, King:2009a, Luhn:2007b}) or  continuous flavor symmetries \cite{King:2005, King:2006, deMedeirosVarzielas:2006, deMedeirosVarzielas:2007a, Adulpravitchai:2009c, Berger:2009}  have also been considered  (for other approaches to TB mixing see \cite{Xing:2002, Matias:2005, Luo:2005, Grimus:2005, Koide:2007, Grimus:2009,Babu:2010bx}), but the $A_4$ models have a particularly economical and attractive structure, e.g. in terms of group representations and of field content. 
In most of the models $A_4$ is accompanied by additional flavor symmetries, either discrete like $Z_N$ or continuous like U(1), which are necessary to eliminate unwanted couplings, to ensure the needed vacuum alignment and to reproduce the observed mass hierarchies. Given the set of flavor symmetries and having specified the field content, the non leading corrections to TB mixing arising from higher order effects can be evaluated in a well defined expansion. In the absence of specific dynamical tricks, in a generic model, all the three mixing angles receive corrections of the same order of magnitude. Since the experimentally allowed departures of $\theta_{12}$ from the TB value $\sin^2{\theta_{12}}=1/3$ are small, at most of $\mathcal{O}(\lambda_C^2)$ , with $\lambda_C$ the Cabibbo angle, it follows that both $\theta_{13}$ and the deviation of $\theta_{23}$ from the maximal value are typically expected in these models to also be at most of $\mathcal{O}(\lambda_C^2)$ \footnote{By $\mathcal{O}(\lambda_C^2)$ we mean numerically of order $\lambda_C^2$. As $\lambda_C \sim 0.22$ a linear term in $\lambda_C$ with a smallish coefficient can easily be $\mathcal{O}(\lambda_C^2)$}. A value of $\theta_{13} \sim \mathcal{O}(\lambda_C^2)$ is within the sensitivity of the experiments which are now in preparation and will take data in the near future. 

\begin{figure}
\centering
\includegraphics [width=10.0 cm]{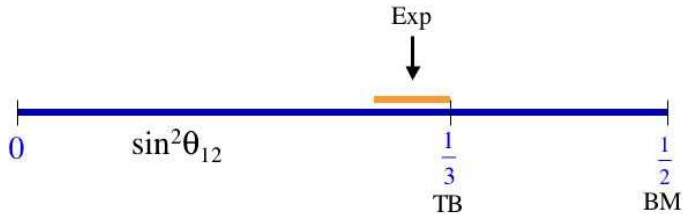}    
\caption{The values of $\sin^2{\theta_{12}}$ for TB o BM mixing are compared with the data}
\label{TBBM}
\end{figure}

Going back to the possibility that the agreement of the data with TB mixing is accidental, we observe that the present data do not exclude a value for $\theta_{13}$, i.e. $\theta_{13} \sim \mathcal{O}(\lambda_C)$, larger than generally implied by models with approximate TB mixing. In fact, recent analysis of the available data lead to
$\sin^2{\theta_{13}}=0.016\pm0.010$ at 1$\sigma$ \cite{Fogli:2008, Fogli:2008a}, $\sin^2{\theta_{13}}=0.010^{+0.016}_{-0.011}$ at 1$\sigma$ \cite{Schwetz:2008, Maltoni:2008},
$\sin^2{\theta_{13}}=0.014^{+0.013}_{-0.011}$  at 1$\sigma$
\cite{GonzalezGarcia:2010} and $\sin^2{\theta_{13}}=0.010^{+0.013}_{-0.009}$ at 1$\sigma$
\cite{GonzalezGarcia:2010}, which are compatible with both options.
If experimentally it is found that $\theta_{13}$ is near its present upper bound, this could be interpreted as an indication that the agreement with the TB mixing is accidental.  In fact a different empirical observation is that $\theta_{12}+\lambda_C\sim \pi/4$, a relation known as quark-lepton complementarity \cite{Raidal:2004, Minakata:2004}, or similarly $\theta_{12}+\sqrt{m_\mu/m_\tau} \sim \pi/4$. No compelling model leading, without parameter fixing, to the exact complementarity relation has been produced so far. Probably the exact complementarity relation is to be replaced with something like $\theta_{12}+\mathcal{O}(\lambda_C)\sim \pi/4$ or $\theta_{12}+\mathcal{O}(\sqrt{m_\mu/m_\tau})\sim \pi/4$ (which we could call "weak" complementarity \cite{Altarelli:2004a, Frampton:2005, Ferrandis:2005, Kang:2005, Minakata:2005, Li:2005, Cheung:2005, Xing:2005, Datta:2005, Ohlsson:2005, Antusch:2005, Lindner:2005, King:2005a, Dighe:2006, Schmidt:2006, Chauhan:2007, Hochmuth:2007, Plentinger:2007, Plentinger:2008a, Altarelli:2009b}. If we take any of these complementarity relations as a serious hint then a scheme would be relevant where  Bimaximal (BM) mixing, instead of TB mixing, is the correct first approximation, modified by terms of $\mathcal{O}(\lambda_C)$. A comparison of the TB or BM mixing values with the data on $\sin^2{\theta_{12}}$ is shown in Fig. (\ref{TBBM}). 

A very special dynamics is also needed for BM mixing and again discrete symmetry groups offer possible solutions. For example, a model \cite{Altarelli:2009b} based on $S_4$, the permutation group of 4 elements, naturally leads to BM mixing in LO. This model is built in such a way that the dominant corrections to the BM mixing only arise from the charged lepton sector at Next-to-the-Leading-Order (NLO)  and naturally inherit $\lambda_C$ as the relevant expansion parameter. As a result the mixing angles deviate from the BM values by terms of  $\mathcal{O}(\lambda_C)$ (at most), and weak complementarity holds. A crucial feature of this particular model is that only $\theta_{12}$ and $\theta_{13}$ are corrected by terms of $\mathcal{O}(\lambda_C)$ while $\theta_{23}$ is unchanged at this order (which is essential to make the model agree with the present data). 

Other types of LO approximations for the lepton mixing pattern have been suggested. For instance a viable first approximation
of the solar mixing angle is also $\theta_{12}=\tan^{-1}(1/\varphi)$ where $\varphi=(1+\sqrt{5})/2$ is the golden ratio \cite{Kajiyama:2007}. This leads to
$\sin^2\theta_{12}=1/(1+\varphi^2)\approx 0.276$, not far from the allowed range. 
Another possible connection with the golden ratio has been proposed in ref. \cite{Rodejohann:2009}. In this case $\cos\theta_{12} = \varphi/2$, or $\sin^2\theta_{12}= 1/4 (3 - \varphi) \approx  0.345$. 
There have been attempts to reproduce these values 
by exploiting flavor symmetries of icosahedral type \cite{Everett:2009}, for the first possibility, or of dihedral type \cite{Adulpravitchai:2009} for the second case.

Thus discrete flavor symmetries may play an important role in models of neutrino mixing. In particular this is the case if some special patterns indicated by the data as possible first approximations, like TB or BM mixing or others, are indeed physically relevant. A list of the simplest discrete groups that have been considered for neutrino mixing, with some of their properties, is shown in Table 2.  In the present review we will discuss the formalism and the physics of a non exhaustive list of models of neutrino mixing based on discrete symmetries.

\begin{table}[t]

\begin{center}
\begin{tabular}{|l|c|c|c|c|}
\hline \textbf{Group} & d & Irr. Repr.'s&Presentation&Ref.'s\\
\hline $D_3\sim S_3$ & 6 & 1, $1'$, 2& $A^3=B^2=(AB)^2=1$&[i]\\
\hline $D_4$ & 8&$1_1,...1_4, 2$& $A^4=B^2=(AB)^2=1$&[ii]\\
\hline $D_7$ & 14&1, $1'$, $2$, $2'$, $2''$& $A^7=B^2=(AB)^2=1$&[iii]\\
\hline $A_4$ & 12& 1, $1'$, $1''$, 3&$A^3=B^2=(AB)^3=1$&[iv]\\
\hline $A_5 \sim PSL_2(5)$ & 60& 1, 3, $3'$, 4, 5&$A^3=B^2=(BA)^5=1$&[v]\\
\hline $T'$ & 24& 1, $1'$, $1''$, $2$, $2'$, $2''$, 3&$A^3=(AB)^3=R^2=1,~ B^2=R$&[vi]\\
\hline $S_4$ & 24 &  1, $1'$, 2, 3, $3'$&$BM: A^4=B^2=(AB)^3=1$&\\
$$ &  &&$TB: A^3=B^4=(BA^2)^2=1$&[vii]\\ 
\hline $\Delta(27) \sim Z_3 ~\rtimes~ Z_3$ &27& $1_1,...1_9, 3, \overline{3}$&&[viii]\\
\hline $PSL_2(7)$ &168&$ 1, 3,\overline{3}, 6, 7, 8$&$A^3=B^2=(BA)^7=(B^{-1}A^{-1}BA)^4=1$&[ix]\\
\hline $T_7 \sim Z_7~\rtimes~Z_3$ & 21&$ 1, 1', \overline{1'}, 3, \overline{3}$&$A^7=B^3=1,~AB=BA^4$&[x]\\
\hline
\end{tabular}
\caption{Some small discrete groups used for model building. [i]\cite{Kubo:2003, Kubo:2004, Kubo:2006, Chen:2004, Lavoura:2005, Dermisek:2005, Caravaglios:2005, Caravaglios:2005a, Grimus:2006a, Koide:2006, Teshima:2006, Haba:2006, Tanimoto:2006, Koide:2006a, Morisi:2006, Picariello:2006, Mohapatra:2006a, Mohapatra:2006b, Kaneko:2007, Koide:2007a, Chen:2008, Feruglio:2007a}; [ii]\cite{Grimus:2004, Adulpravitchai:2009a}; [iii]\cite{Blum:2004, Blum:2008};[iv]\cite{Ma:2001, Ma:2002, Babu:2003, Hirsch:2004, Ma:2004, Ma:2004a, Chen:2005, Altarelli:2005, Ma:2005, Hirsch:2005, Babu:2005, Ma:2005a, Zee:2005, Ma:2006, He:2006, Adhikary:2006, Altarelli:2006, Lavoura:2006, Ma:2007, Hirsch:2007, Altarelli:2007, Yin:2007, Bazzocchi:2008, Bazzocchi:2008a, Honda:2008, Brahmachari:2008, Adhikary:2008, Hirsch:2008, Frampton:2008, Csaki:2008, Altarelli:2008, Morisi:2009, Lin:2009, Lin:2009a, Altarelli:2009a, Ma:2005b, Ma:2006a, Ma:2006b, Morisi:2007, Grimus:2008, Ciafaloni:2009, Bazzocchi:2009, Bazzocchi:2008b, delAguila:2010,Kadosh:2010rm,Antusch:2010es}; [v]\cite{Everett:2009}; [vi]\cite{Frampton:1995, Aranda:2000, Aranda:2000a, Carr:2000,  Aranda:2007, Frampton:2007, Frampton:2009, Ding:2009,Feruglio:2007, Chen:2007}; [vii]\cite{Mohapatra:2004, Hagedorn:2006, Cai:2006, Zhang:2007,Ma:2007a, Bazzocchi:2008c, Ishimori:2009, Bazzocchi:2009a, Bazzocchi:2009b, Meloni:2009, Dutta:2009, Dutta:2009a, Ding:2010, Morisi:2010, Hagedorn:2010th,Ishimori:2010xk}; [viii]\cite{deMedeirosVarzielas:2007, Ma:2007b, Grimus:2007, Luhn:2007, Bazzocchi:2009c}; [ix]\cite{Luhn:2007a, King:2009, King:2009a}; [x]\cite{Luhn:2007b}.}
\label{groups}
\end{center}
\end{table}

%
%
\section{Special patterns of neutrino mixing}

Given the PNMS mixing matrix $U$ (we refer the reader to ref. \cite{Altarelli:2004} for its general definition and parametrisation), the general form of the neutrino mass matrix, in terms of the (complex
\footnote{We absorb the Majorana phases in the mass eigenvalues $m_i$, rather than in the mixing matrix $U$.
The dependence on these phases drops in neutrino oscillations.}) mass eigenvalues $m_1, m_2, m_3$, in the basis where charged leptons are diagonal, is given  by:
\beq
m_{\nu}=U^* {\rm diag}(m_1,m_2,m_3) U^\dagger~~~.
\label{numass}
\eeq
We will present here a number of particularly relevant forms of $U$ and $m_\nu$ that will be important in the following. We start by the most general mass matrix that corresponds to $\theta_{13}=0$ and $\theta_{23}$ maximal, that is to $U$ given by  (in a particular phase convention):
\begin{equation}
U= \left(\matrix{
c_{12}&s_{12}&0\cr
-\dd\frac{s_{12}}{\sqrt 2}&\dd\frac{c_{12}}{\sqrt 2}&-\dd\frac{1}{\sqrt 2}\cr
-\dd\frac{s_{12}}{\sqrt 2}&\dd\frac{c_{12}}{\sqrt 2}&\dd\frac{1}{\sqrt 2}}\right)~~~,
\label{2.1}
\end{equation}
with $c_{12}\equiv \cos{\theta_{12}}$ and $s_{12}\equiv \sin{\theta_{12}}$. 
By applying eq. (\ref{numass}) we obtain a matrix of the form \cite{Fukuyama:1997, Mohapatra:1999, Ma:2001a, Lam:2001, Kitabayashi:2003, Grimus:2003, Koide:2004, Ghosal:2003, Grimus:2005a, deGouvea:2004, Mohapatra:2005, Kitabayashi:2005, Mohapatra:2005a, Mohapatra:2005b, Mohapatra:2005c, Ahn:2006}:
\begin{equation}
m=\left(\matrix{
x&y&y\cr
y&z&w\cr
y&w&z}\right),
\label{gl}
\end{equation}
with complex coefficients $x$, $y$, $z$ and $w$.
This matrix is the most general one that is symmetric under 2-3 (or $\mu - \tau$) exchange or:
\bea
m_\nu=A_{23}m_\nu A_{23}~~\label{inv.1}~~~,
\label{mmutau}
\eea
where $A_{23}$ is given by:
\be
A_{23}=\left(
\begin{array}{ccc}
1&0&0\\
0&0&1\\
0&1&0
\end{array}
\right)~~~.
\label{Amutau}
\ee
The solar mixing angle $\theta_{12}$ is given by
\begin{eqnarray}
\sin^2 2 \theta_{12}&=&\dd\frac{8\vert x^* y+y^*(w+z)\vert^2}{8\vert x^* y+y^*(w+z)\vert^2+(\vert w+z\vert^2-\vert x\vert^2)^2}\nn\\
&=&\dd\frac{8 y^2}{(x-w-z)^2+8 y^2}~~~
\label{teta12}
\end{eqnarray}
where the second equality applies to real parameters. 
Since $\theta_{13}=0$ there is no CP violation in neutrino oscillations, and the only physical phases
are the Majorana ones, accounted for by the general case of complex parameters.
We restrict here our consideration to real parameters. There are four of them in eq. (\ref{gl}) which correspond to the three mass eigenvalues and one remaining mixing angle, $\theta_{12}$. 
Models with $\mu$-$\tau$ symmetry have been extensively studied \cite{Fukuyama:1997, Mohapatra:1999, Ma:2001a, Lam:2001, Kitabayashi:2003, Grimus:2003, Koide:2004, Ghosal:2003, Grimus:2005a, deGouvea:2004, Mohapatra:2005, Kitabayashi:2005, Mohapatra:2005a, Mohapatra:2005b, Mohapatra:2005c, Ahn:2006, Ge:2010}.

The particularly important case of TB mixing is obtained when $\sin^2{2\theta_{12}}=8/9$ or $x+y=w+z$
\footnote{The other solution $x-y=w+z$ gives rise to TB mixing in another phase convention and is physically equivalent to  $x+y=w+z$.}
. In this case the matrix $m_\nu$ takes the form:
\begin{equation}
m_\nu=\left(\matrix{
x&y&y\cr
y&x+v&y-v\cr
y&y-v&x+v}\right)~~~,
\label{gl21}
\end{equation}
In fact, in this case, $U=U_{TB}$ is given by \cite{Harrison:2002, Harrison:2002a, Harrison:2003, Harrison:2004}:
\begin{equation}
U_{TB}= \left(\matrix{
\dd\sqrt{\frac{2}{3}}&\dd\frac{1}{\sqrt 3}&0\cr
-\dd\frac{1}{\sqrt 6}&\dd\frac{1}{\sqrt 3}&-\dd\frac{1}{\sqrt 2}\cr
-\dd\frac{1}{\sqrt 6}&\dd\frac{1}{\sqrt 3}&\dd\frac{1}{\sqrt 2}}\right)~~~, 
\label{2}
\end{equation}
and, from eq. (\ref{numass}) one obtains:
\begin{equation}
m_{\nu}=  m_1\Phi_1 \Phi_1^T + m_2\Phi_2 \Phi_2^T + m_3\Phi_3 \Phi_3^T~~~, 
\label{1k1}
\end{equation}
where
\be
\Phi_1^T=\frac{1}{\sqrt{6}}(2,-1,-1)~~~,~~~~~
\Phi_2^T=\frac{1}{\sqrt{3}}(1,1,1)~~~,~~~~~\Phi_3^T=\frac{1}{\sqrt{2}}(0,-1,1)
\label{4k1}
\ee
are the respective columns of $U_{TB}$ and $m_i$ are the neutrino mass eigenvalues ($m_1=x-y$, $m_2=x+2y$ and $m_3=x-y+2v$).
It is easy to see that the TB mass matrix in eqs. (\ref{1k1},\ref{4k1}) is indeed of the form in eq. (\ref{gl21}).
All patterns for the neutrino spectrum are in principle possible. For a hierarchical spectrum $m_3>>m_2>>m_1$, $m_3^2 \sim \Delta m^2_{atm}$, $m_2^2/m_3^2 \sim \Delta m^2_{sol}/\Delta m^2_{atm}$ and $m_1$ could be negligible. But also degenerate masses and inverse hierarchy can be reproduced: for example, by taking $m_3= - m_2=m_1$  we have a degenerate model, while for $m_1= - m_2$ and $m_3=0$ an inverse hierarchy case is realized (stability under renormalization group running (for a review see for example \cite{Chankowski:2001mx}) strongly prefers opposite signs for the first and the second eigenvalue which are related to solar oscillations and have the smallest mass squared splitting).

Note that the mass matrix for TB mixing, in the basis where charged leptons are diagonal, as given in eq. (\ref{gl21}), can be specified as the most general matrix which is invariant under $\mu-\tau$ (or 2-3) symmetry (see eqs. (\ref{mmutau}),(\ref{Amutau})) and, in addition, under the action of a unitary symmetric matrix $S_{TB}$ (actually $S_{TB}^2=1$ and $[S_{TB},A_{23}]=0$):
\bea
m_\nu=S_{TB}m_\nu S_{TB}~~~,~~~~~m_\nu=A_{23}m_\nu A_{23}~~~,
\label{inv}
\eea
where $S_{TB}$ is given by:
\bea
\label{trep}
S_{TB}&=\dd\frac{1}{3} \left(\matrix{
-1&2&2\cr
2&-1&2\cr
2&2&-1}\right)~~~.
\eea

As a last example consider the case of BM where, in addition to $\theta_{13}=0$ and $\theta_{23}$ maximal, one also has $\sin^2{2\theta_{12}}=1$. 
The BM mixing matrix is given by:
\begin{equation}
U_{BM}= \left(\matrix{
\dd\frac{1}{\sqrt 2}&\dd-\frac{1}{\sqrt 2}&0\cr
\dd\frac{1}{2}&\dd\frac{1}{2}&-\dd\frac{1}{\sqrt 2}\cr
\dd\frac{1}{2}&\dd\frac{1}{2}&\dd\frac{1}{\sqrt 2}}\right)~~~.
\label{21}
\end{equation}
In the basis where charged lepton masses are
diagonal, from eq. (\ref{numass}), we derive the effective neutrino mass matrix in the BM case:
\begin{equation}
m_{\nu}=  m_1\Phi_1 \Phi_1^T + m_2\Phi_2 \Phi_2^T + m_3\Phi_3 \Phi_3^T~~~, 
\label{1k}
\end{equation}
where
\be
\Phi_1^T=\frac{1}{2}(\sqrt{2},1,1)~~~,~~~~~
\Phi_2^T=\frac{1}{2}(-\sqrt{2},1,1)~~~,~~~~~\Phi_3^T=\frac{1}{\sqrt{2}}(0,-1,1)
\label{4k}
\ee

As we see the most general mass matrix leading to BM mixing is of the form:
\begin{equation}
m_{\nu}=\left(\matrix{
x&y&y\cr
y&z&x-z\cr
y&x-z&z}\right)\;,
\label{gl2}
\end{equation}
The resulting matrix can be completely characterized by the requirement of being invariant under the action of $A_{23}$ and also of the unitary, real, symmetric matrix $S_{BM}$ (satisfying $S_{BM}^2=1$ and $[S_{BM},A_{23}]=0$):
\beq
m_{\nu}= S_{BM} m_{\nu}S_{BM}~~~, ~~~~~~m_{\nu}=A_{23}m_{\nu}A_{23}~~~,
\label{invS}
\eeq
with $S_{BM}$ given by:
\beq
S_{BM}=\left(
    \begin{array}{ccc}
      0 & -\dd\frac{1}{\sqrt{2}} & -\dd\frac{1}{\sqrt{2}}  \\
      -\dd\frac{1}{\sqrt{2}} & \dd\frac{1}{2} & -\dd\frac{1}{2} \\
      -\dd\frac{1}{\sqrt{2}} & -\dd\frac{1}{2} & \dd\frac{1}{2} \\
    \end{array}
  \right)~~~.
  \label{matS}
\eeq

The $m_\nu$ mass matrices of the previous examples were all derived in the basis where charged leptons are diagonal. It is useful to consider the product $m^2=m_e^\dagger m_e$, where $m_e$ is the charged lepton mass matrix (defined as $\overline \psi_R m_e \psi_L$), because this product transforms as $m'^2=U_e^\dagger m^2 U_e$, with $U_e$ the unitary matrix that rotates the left-handed (LH) charged lepton fields. The most general diagonal $m^2$ is invariant under a diagonal phase matrix with 3 different phase factors:
\beq
m_e^\dagger m_e= T^\dagger m_e^\dagger m_e T
\label{Tdiag}
\eeq
and conversely a matrix $m_e^\dagger m_e$ satisfying the above requirement is diagonal. If $T^n=1$
the matrix $T$ generates a cyclic group $Z_n$.
In the simplest case n=3 and we get $Z_3$ but $n>3$ is equally possible. Examples are:
\bea
\label{ta4}
T_{TB}=\left(\matrix{
1&0&0\cr
0&\omega&0\cr
0&0&\omega^2}
\right).
\eea
where $\omega^3=1$, so that $T_{TB}^3=1$, or 
\bea
\label{ts4}
T_{BM}=\left(
    \begin{array}{ccc}
      -1 & 0 & 0 \\
      0 & -i & 0 \\
      0 & 0 & i \\
    \end{array}
  \right).
\eea
with $T_{BM}^4=1$.

We are now in a position to explain the role of finite groups and to formulate the general strategy to obtain one of the previous special mass matrices, for example that of TB mixing. We must find a group $G_f$ which, for simplicity, must be as small as possible but large enough to contain the $S$ and $T$ transformations. A limited number of products of $S$ and $T$ close a finite group $G_f$.  Hence the group $G_f$ contains the subgroups $G_S$ and $G_T$ generated by monomials in $S$ and $T$, respectively. We assume that the theory is invariant under the spontaneously broken symmetry described by $G_f$. Then we must arrange a breaking of $G_f$ such that, in leading order, $G_f$ is broken down to $G_S$ in the neutrino mass sector and down to $G_T$ in the charged lepton mass sector. In a good model this step must be realized in a natural way as a consequence of the stated basic principles, and not put in by hand. The symmetry under $A_{23}$ in some cases is also part of $G_f$ (this the case of $S_4$) and then must be preserved in the neutrino sector along with $S$ by the $G_f$ breaking or it could arise as a consequence of a special feature of the $G_f$ breaking (for example, in $A_4$ it is obtained by allowing only some transformation properties for the flavons with non vanishing VEV's). The explicit example of $A_4$ is discussed in the next section.
Note that, along the same line, a model with $\mu-\tau$ symmetry  can be realized in terms of the group $S_3$ generated by products of $A_{23}$ and $T$ (see, for example, \cite{Feruglio:2007a}).
\section{The $A_4$ group}
$A_4$ is the group of the even permutations of 4 objects. It has 4!/2=12 elements. Geometrically, it can be seen as the invariance group of a tetrahedron (the odd permutations, for example the exchange of two vertices, cannot be obtained by moving a rigid solid). Let us denote a generic permutation $(1,2,3,4)\rightarrow (n_1,n_2,n_3,n_4)$ simply by $(n_1n_2n_3n_4)$. $A_4$ can be generated by two basic permutations $S$ and $T$ given by $S=(4321)$ and $T=(2314)$. One checks immediately that:
\be\label{pres}
S^2=T^3=(ST)^3=1
\ee
This is called a "presentation" of the group. The 12 even permutations belong to 4 equivalence classes ($h$ and $k$ belong to the same class if there is a $g$ in the group such that $ghg^{-1}=k$) and are generated from $S$ and $T$ as follows:
\bea \label{class}
&C1&: I=(1234)\\ \nonumber
&C2&: T=(2314),ST=(4132),TS=(3241),STS=(1423)\\ \nonumber
&C3&: T^2=(3124),ST^2=(4213),T^2S=(2431),TST=(1342)\\ \nonumber
&C4&: S=(4321),T^2ST=(3412),TST^2=(2143)\\ \nonumber
\eea
Note that, except for the identity $I$ which always forms an equivalence class in itself, the other classes are according to the powers of $T$ (in C4 $S$ could as well be seen as $ST^3$).

The characters of a group $\chi_g^R$ are defined, for each element $g$, as the trace of the matrix that maps the element in a given representation $R$. From the invariance of traces under similarity transformations it follows that equivalent representations have the same characters and that characters have the same value for all elements in an equivalence class. Characters satisfy $\sum_g \chi_g^R \chi_g^{S*}= N \delta^{RS}$, where $N$ is the number of transformations in the group ($N=12$ in $A_4$).  Also, for each element $h$, the character of $h$ in a direct product of representations is the product of the characters: $\chi_h^{R\otimes S}=\chi_h^R \chi_h^S$ and also is equal to the sum of the characters in each representation that appears in the decomposition of $R\otimes S$. In a finite group the squared dimensions of the inequivalent irreducible representations add up to $N$. The character table of $A_4$ is given in Table 3. 
From this table one derives that $A_4$ has four inequivalent representations: three of dimension one, $1$, $1'$ and $1"$ and one of dimension $3$. 

It is immediate to see that the one-dimensional unitary representations are obtained by:
\bea \label{uni}
1&S=1&T=1\\ \nonumber
1'&S=1&T=e^{\dd i 2 \pi/3}\equiv\omega\\\nonumber
1''&S=1&T=e^{\dd i 4\pi/3}\equiv\omega^2\\\nonumber
\eea
Note that $\omega=-1/2+i \sqrt{3}/2$ is the cubic root of 1 and satisfies $\omega^2=\omega^*$, $1+\omega+\omega^2=0$.

\begin{table}[t]
\begin{center}
\begin{tabular}{|l|c|c|c|c|}
\hline \textbf{Class} & \textbf{$\chi^1$} & \textbf{$\chi^{1'}$} &\textbf{$\chi^{1"}$}&\textbf{$\chi^3$}\\
\hline $C_1$ & 1 & 1& 1&3\\
\hline $C_2$ & 1 &$\omega$&$\omega^2$&0\\
\hline $C_3$ & 1 & $\omega^2$&$\omega$&0\\
\hline $C_4$ & 1 & 1& 1&-1\\
\hline
\end{tabular}
\caption{Characters of $A_4$}
\label{tcar}
\end{center}
\end{table}

The three-dimensional unitary representation, in a basis
where the element $S=S'$ is diagonal, is built up from:
\begin{equation}\label{tre}
S'=\left(\matrix{
1&0&0\cr
0&-1&0\cr
0&0&-1}\right),~
T'=\left(\matrix{
0&1&0\cr
0&0&1\cr
1&0&0}
\right).
\ee

The multiplication rules are as follows: the product of two 3 gives $3 \times 3 = 1 + 1' + 1'' + 3 + 3$ and $1' \times 1' = 1''$, $1' \times 1'' = 1$, $1'' \times 1'' = 1'$ etc.
If $3\sim (a_1,a_2,a_3)$ is a triplet transforming by the matrices in eq. (\ref{tre}) we have that under $S'$: $S'(a_1,a_2,a_3)^t= (a_1,-a_2,-a_3)^t$ (here the upper index $t$ indicates transposition)  and under $T'$: $T'(a_1,a_2,a_3)^t= (a_2,a_3,a_1)^t$. Then, from two such triplets $3_a\sim (a_1,a_2,a_3)$, $3_b\sim (b_1,b_2,b_3)$ the irreducible representations obtained from their product are:
\begin{equation}
1=a_1b_1+a_2b_2+a_3b_3
\end{equation}
\begin{equation}
1'=a_1b_1+\omega^2 a_2b_2+\omega a_3b_3
\end{equation}
\begin{equation}
1"=a_1b_1+\omega a_2b_2+\omega^2 a_3b_3
\end{equation}
\begin{equation}
3\sim (a_2b_3, a_3b_1, a_1b_2)
\end{equation}
\begin{equation}
3\sim (a_3b_2, a_1b_3, a_2b_1)
\end{equation}
In fact, take for example the expression for $1"=a_1b_1+\omega a_2b_2+\omega^2 a_3b_3$. Under $S'$ it is invariant and under $T'$ it goes into $a_2b_2+\omega a_3b_3+\omega^2 a_1b_1=\omega^2[a_1b_1+\omega a_2b_2+\omega^2 a_3b_3]$ which is exactly the transformation corresponding to $1"$. 

In eq. (\ref{tre}) we have the representation 3 in a basis where $S$ is diagonal. We shall see that for   our purposes it is convenient to go to a basis where instead it is $T$ that is diagonal. This is obtained through the unitary transformation:
\bea
T&=&VT'V^\dagger=\left(\matrix{
1&0&0\cr
0&\omega&0\cr
0&0&\omega^2}
\right),\label{mT}\\
S&=&VS'V^\dagger=\frac{1}{3} \left(\matrix{
-1&2&2\cr
2&-1&2\cr
2&2&-1}\right).
\label{mS}
\eea
where:
\be \label {vu}
V=\frac{1}{\sqrt{3}} \left(\matrix{
1&1&1\cr
1&\omega^2&\omega\cr
1&\omega&\omega^2}\right).
\ee
The matrix $V$ is special in that it is a 3x3 unitary matrix with all entries of unit absolute value. It is interesting that this matrix was proposed long ago as a possible mixing matrix for neutrinos \cite{Cabibbo:1978, Wolfenstein:1978}. We shall see in the following that  in the $T$ diagonal basis the charged lepton mass matrix (to be precise the matrix $m_e^\dagger m_e$) is diagonal. Notice that the matrices $(S,T)$ of eqs. (\ref{mT}-\ref{mS}) coincide
with the matrices $(S_{TB},T_{TB})$ of the previous section.

In this basis the product rules of two triplets, ($\psi_1,\psi_2,\psi_3$) and ($\varphi_1,\varphi_2,\varphi_3$) of $A_4$, according to the multiplication rule $3\times 3=1+1'+1"+3+3$ are different than in the $S$ diagonal basis (because for Majorana mass matrices the relevant scalar product is $(ab)$ and not  $(a^\dagger b$) and are given by:
\bea \label{tensorproda4}
&\psi_1\varphi_1+\psi_2\varphi_3+\psi_3\varphi_2 \sim 1 ~,\nn \\
&\psi_3\varphi_3+\psi_1\varphi_2+\psi_2\varphi_1 \sim 1' ~,\nn \\
&\psi_2\varphi_2+\psi_3\varphi_1+\psi_1\varphi_3 \sim 1'' ~,\nn
\eea
  \be
   \left( 
 \ba{c}
2\psi_1\varphi_1-\psi_2\varphi_3-\psi_3\varphi_2 \\
2\psi_3\varphi_3-\psi_1\varphi_2-\psi_2\varphi_1 \\
2\psi_2\varphi_2-\psi_1\varphi_3-\psi_3\varphi_1 \\
  \ea
  \right) \sim 3_S~, \qquad
  \left( 
 \ba{c}
\psi_2\varphi_3-\psi_3\varphi_2 \\
\psi_1\varphi_2-\psi_2\varphi_1 \\
\psi_3\varphi_1-\psi_1\varphi_3 \\
  \ea
  \right) \sim 3_A~.
  \label{tensorp}
   \ee
   
In the following we will work in the $T$ diagonal basis, unless otherwise stated. In this basis the 12 matrices of the 3-dimensional representation of $A_4$ are given by:
\begin{center}
\begin{tabular}{ll}
  $\cC_1$ : &$1= \left(
                 \begin{array}{ccc}
                   1 & 0 & 0 \\
                   0 & 1 & 0 \\
                   0 & 0 & 1 \\
                 \end{array}
               \right)$, \\
  \\
  $\cC_2$ : &$T=
               \left(
                 \begin{array}{ccc}
                   1 & 0 & 0 \\
                   0 & \om & 0 \\
                   0 & 0 & \om^2 \\
                 \end{array}
               \right),
                 ST=\frac{1}{3}\left(
                 \begin{array}{ccc}
                   -1 & 2\om & 2\om^2 \\
                   2 & -\om & 2\om^2 \\
                   2 & 2\om & -\om^2 \\
                 \end{array}
               \right),
               $,\\[0.6cm]
             & $TS=\frac{1}{3}\left(
                 \begin{array}{ccc}
                   -1 & 2 & 2 \\
                   2\om & -\om & 2\om \\
                   2\om^2 & 2\om^2 & -\om^2 \\
                 \end{array}
               \right), 
               STS=\frac{1}{3}\left(
                 \begin{array}{ccc}
                   -1 & 2\om^2 & 2\om \\
                   2\om^2 & -\om & 2 \\
                   2\om & 2 & -\om^2 \\
                 \end{array}
               \right)$,
\\
\\
   $\cC_3$ : & $T^2=\left(
                 \begin{array}{ccc}
                   1 & 0 & 0 \\
                   0 & \om^2 & 0 \\
                   0 & 0 & \om \\
                 \end{array}
               \right),
              
              ST^2=\frac{1}{3}\left(
                 \begin{array}{ccc}
                   -1 & 2\om^2 & 2\om \\
                   2 & -\om^2 & 2\om \\
                   2 & 2\om^2 & -\om \\
                 \end{array}
               \right)$,\\[0.6cm]
               &$T^2S=\frac{1}{3}\left(
                 \begin{array}{ccc}
                   -1 & 2 & 2 \\
                   2\om^2 & -\om^2 & 2\om^2 \\
                   2\om & 2\om & -\om \\
                 \end{array}
               \right),             
               TST=\frac{1}{3}\left(
                 \begin{array}{ccc}
                   -1 & 2\om & 2\om^2 \\
                   2\om & -\om^2 & 2 \\
                   2\om^2 & 2 & -\om \\
                 \end{array}
               \right)$,
\end{tabular}
\end{center}
\begin{center}
\begin{tabular}{ll}

$\cC_4$ : & $S=\frac{1}{3}\left(
                 \begin{array}{ccc}
                   -1 & 2 & 2 \\
                   2 & -1 & 2 \\
                   2 & 2 & -1 \\
                 \end{array}
               \right),
               T^2ST=\frac{1}{3}\left(
                 \begin{array}{ccc}
                   -1 & 2\om & 2\om^2 \\
                   2\om^2 & -1 & 2\om \\
                   2\om & 2\om^2 & -1 \\
                 \end{array}
               \right)$,\\[0.6cm]
               &$
               TST^2=\frac{1}{3}\left(
                 \begin{array}{ccc}
                   -1 & 2\om^2 & 2\om \\
                   2\om & -1 & 2\om^2 \\
                   2\om^2 & 2\om & -1 \\
                 \end{array}
               \right)$. 
\end{tabular}
\end{center}

We can now see why $A_4$ works for TB mixing. In section 2 we have already mentioned that the most general mass matrix for TB mixing in eq. (\ref{gl21}), in the basis where charged leptons are diagonal, can be specified as one which is invariant under the 2-3 (or $\mu-\tau$) symmetry  and under the $S$ unitary transformation, as stated in eq. (\ref{inv}) (note that $S_{TB}$ in eqs. (\ref{inv}, \ref{trep}) coincides with $S$ in eq. (\ref{mS})). 
This observation plays a key role in leading to $A_4$ as a candidate group for TB mixing, because $S$ is a matrix of $A_4$. Instead the matrix $A_{23}$ is not an element of $A_4$ 
(because the 2-3 exchange is an odd permutation). 
We shall see that in $A_4$ models the 2-3 symmetry is maintained by imposing that there are no flavons transforming as $1'$ or $1''$ that break $A_4$ with two different VEV's (in particular one can assume that there are no flavons in the model transforming as $1'$ or $1''$).
It is also clear that a generic diagonal charged lepton matrix $m_e^\dagger m_e$ is characterized by the invariance under $T$, or $T^\dagger m_e^\dagger m_e T=m_e^\dagger m_e$.
	
The group $A_4$ has two obvious subgroups: $G_S$, which is a reflection subgroup
generated by $S$ and $G_T$, which is the group generated by $T$, which is isomorphic to $Z_3$.
If the flavor symmetry associated to $A_4$ is broken by the VEV of a triplet
$\varphi=(\varphi_1,\varphi_2,\varphi_3)$ of scalar fields,
there are two interesting breaking pattern. The VEV
\be
\langle\varphi\rangle=(v_S,v_S,v_S)
\label{unotre}
\ee
breaks $A_4$ down to $G_S$, while
\be
\langle\varphi\rangle=(v_T,0,0)
\label{unozero}
\ee
breaks $A_4$ down to $G_T$. As we will see, $G_S$ and $G_T$ are the relevant low-energy symmetries
of the neutrino and the charged-lepton sectors, respectively. Indeed we have already seen that the TB mass matrix is invariant under $G_S$
and a diagonal charged lepton mass $m_e^\dagger m_e$ is invariant under $G_T$.

\section{Applying $A_4$ to lepton masses and mixings}
In the lepton sector a typical $A_4$ model works as follows \cite{Altarelli:2006}.  One assigns
leptons to the four inequivalent
representations of $A_4$: LH lepton doublets $l$ transform
as a triplet $3$, while the RH charged leptons $e^c$,
$\mu^c$ and $\tau^c$ transform as $1$, $1''$ and $1'$, respectively. 
Here we consider a  see-saw realization, so we also introduce conjugate neutrino fields $\nu^c$ transforming as a triplet of $A_4$. We adopt a supersymmetric (SUSY) context also to make contact 
with Grand Unification (flavor symmetries are supposed to act near the GUT scale). In fact, as well known, SUSY is important in GUT's for offering a solution to the hierarchy problem, for improving coupling unification and for making the theory compatible with bounds on proton decay. But in models of lepton mixing SUSY also helps for obtaining the vacuum alignment, because the SUSY constraints are very strong and limit the form of the superpotential very much. Thus SUSY is not necessary but it is a plausible and useful ingredient. The flavor symmetry is broken by two triplets
$\varphi_S$ and $\varphi_T$ and by one or more singlets $\xi$. All these fields are invariant under the SM gauge symmetry.
Two Higgs doublets $h_{u,d}$, invariant under $A_4$, are
also introduced. One can obtain  the observed hierarchy among $m_e$, $m_\mu$ and
$m_\tau$ by introducing an additional U(1)$_{FN}$ flavor symmetry \cite{Froggatt:1979} under
which only the  RH  lepton sector is charged (recently some models were proposed with a different VEV alignment such that the charged lepton hierarchies are obtained without introducing a $U(1)$ symmetry \cite{Lin:2009, Lin:2009a, Altarelli:2009a}).
We recall that $U(1)_{FN}$ is a simplest flavor symmetry where particles in different generations are assigned (in general) different values of an Abelian charge.  Also Higgs fields may get a non zero charge. When the symmetry is spontaneously broken the entries of mass matrices are suppressed if there is a charge mismatch and more so if the corresponding mismatch is larger.
We assign FN-charges $0$, $q$ and $2q$ to $\tau^c$, $\mu^c$ and
$e^c$, respectively. There is some freedom in the choice of $q$.
Here we take $q=2$.
By assuming that a flavon $\theta$, carrying
a negative unit of FN charge, acquires a VEV 
$\langle \theta \rangle/\Lambda\equiv\lambda<1$, the Yukawa couplings
become field dependent quantities $y_{e,\mu,\tau}=y_{e,\mu,\tau}(\theta)$
and we have
\be
y_\tau\approx \mathcal{O}(1)~~~,~~~~~~~y_\mu\approx O(\lambda^2)~~~,
~~~~~~~y_e\approx O(\lambda^{4})~~~.
\ee
Had we chosen $q=1$, we would have needed  $\langle \theta \rangle/\Lambda$ of order $\lambda^2$, to reproduce the above result.
The superpotential term for lepton masses, $w_l$ is given by:
\be
w_l=y_e e^c (\varphi_T l)+y_\mu \mu^c (\varphi_T l)'+
y_\tau \tau^c (\varphi_T l)''+ y (\nu^c l)+
(x_A\xi+\tilde{x}_A\tilde{\xi}) (\nu^c\nu^c)+x_B (\varphi_S \nu^c\nu^c)+h.c.+...
\label{wlss}
\ee
with dots denoting higher   
dimensional operators that lead to corrections to the LO   
approximation. In our notation, the product of 2 triplets  $(3 3)$ transforms as $1$, 
$(3 3)'$ transforms as $1'$ and $(3 3)''$ transforms as $1''$. 
To keep our formulae compact, we omit to write the Higgs and flavon  fields
$h_{u,d}$, $\theta$ and the cut-off scale $\Lambda$. For instance 
$y_e e^c (\varphi_T l)$ stands for $y_e e^c (\varphi_T l) h_d \theta^4/\Lambda^5$. The parameters of the superpotential $w_l$ are complex, in particular those responsible for the
heavy neutrino Majorana masses, $x_{A,B}$. Some terms allowed by the $A_4$ symmetry, such as the terms 
obtained by the exchange $\varphi_T\leftrightarrow \varphi_S$, 
(or the term $(\nu^c\nu^c)$) are missing in $w_l$. 
Their absence is crucial and, in each version of $A_4$ models, is
motivated by additional symmetries (In ref. \cite{Altarelli:2005} a natural solution of this problem based on a formulation with extra dimensions was discussed; for a similar approach see also \cite{Csaki:2008,Kadosh:2010rm}). In the present version the additional symmetry is $Z_3$. A $U(1)_R$ symmetry related to R-parity and the presence of driving fields in the flavon superpotential are common features of supersymmetric formulations. Eventually, after the inclusion of N = 1 SUSY breaking effects, the $U(1)_R$ symmetry will be broken at the low energy scale $m_{SUSY}$ down to the discrete R-parity.
Supersymmetry also helps producing and maintaining the hierarchy $\langle h_{u,d}\rangle=v_{u,d}\ll \Lambda$ where $\Lambda$ is the cut-off scale of the theory.
The fields in the model and their classification under the symmetry are summarized in Table \ref{table:TransformationsA}.

\begin{table}[h]
\begin{center}
\begin{tabular}{|c||c|c|c|c|c|c||c|c|c|c||c|c|c|}
  \hline
  &&&&&&&&&&&&&\\[-0,3cm]
  & $l$ & $e^c$ & $\mu^c$ & $\tau^c$ & $\nu^c$ & $h_{u,d}$ & $\theta$ & $\varphi_T$ & $\varphi_S$ &  $\xi$ &$\varphi_0^T$ & $\varphi_0^S$ &  $\xi_0$ \\
 &&&&&&&&&&&&&\\[-0,3cm]
  \hline
  &&&&&&&&&&&&&\\[-0,3cm]
  $A_4$ & 3 & 1 & $1''$ & $1'$ & 3 & 1 & 1 & 3 & $3$ & 1 & 3 & 3 & 1  \\
   &&&&&&&&&&&&&\\[-0,3cm]
  $Z_3$ & $\omega$ &$\omega^2$ & $\omega^2$& $\omega^2$& $\omega^2$ & 1 & 1 &1& $\omega^2$&$\omega^2$& 1 &$\omega^2$ & $\omega^2$  \\
   &&&&&&&&&&&&&\\[-0,3cm]
  $U(1)_{FN}$ & 0 & 4 & 2 & 0 & 0 & 0 & -1 & 0 & 0 & 0 & 0 & 0 & 0  \\
  &&&&&&&&&&&&&\\[-0,3cm]
  $U(1)_R$ & 1 & 1 & 1 & 1 & 1 & 0 & 0 & 0 & 0 & 0 & 2 & 2 & 2  \\
  \hline
  \end{tabular}
\end{center}
\caption{\label{table:TransformationsA}Transformation properties of all the fields.}
\end{table}
In this set up it can be shown that the fields $\varphi_T$,
$\varphi_S$ and $\xi$ develop a VEV along the directions:
\bea
\langle \varphi_T \rangle&=&(v_T,0,0)\nn\\ 
\langle \varphi_S\rangle&=&(v_S,v_S,v_S)\nn\\
\langle \xi \rangle&=&u~~~. 
\label{align}
\eea 
A crucial part of all serious $A_4$ models is the dynamical generation of this alignment in a natural way. We refer to ref.  \cite{Altarelli:2006} for a proof that the above alignment naturally follows from the most general LO superpotential implied by the symmetries of the model. As already mentioned, the group $A_4$ has two obvious subgroups: $G_S$, which is a reflection subgroup
generated by $S$ and $G_T$, which is the group generated by $T$, isomorphic to $Z_3$.
In the basis where $S$ and $T$ are given by eq. (\ref{trep}), the VEV $\langle \varphi_T \rangle=(v_T,0,0)$ breaks $A_4$ down to $G_T$, while
$\langle \varphi_S\rangle=(v_S,v_S,v_S)$
breaks $A_4$ down to $G_S$.  

If the alignment in eq. (\ref{align}) is realized, at the leading order of the $1/\Lambda$ expansion,
the mass matrices $m_l$ and $m_\nu$ for charged leptons and 
neutrinos correspond to TB   
mixing. The charged lepton mass matrix is diagonal:
\be
m_l=v_d\frac{v_T}{\Lambda}\left(
\begin{array}{ccc}
y_e& 0& 0\\
0& y_\mu & 0 \\
0& 0& y_\tau 
\end{array}
\right)~~~,
\label{mch}
\ee
The charged fermion masses are given by:
\be \label{chmasses}
m_e= y_e v_d \frac{v_T}{\Lambda}~~~,~~~~~~~
m_\mu= y_\mu v_d \frac{v_T}{\Lambda}~~~,~~~~~~~
m_\tau=y_\tau v_d \frac{v_T}{\Lambda}~~~,
\ee
where the suppression coming from the breaking of $U(1)_{FN}$ is understood. For example $y_e$ stands for $y_e \theta^4/\Lambda^4$.
In the neutrino sector, after electroweak and $A_4$ symmetry breaking we have Dirac
and Majorana masses:
\be
m_\nu^D=\left(
          \begin{array}{ccc}
            1 & 0 & 0 \\
            0 & 0 & 1 \\
            0 & 1 & 0 \\
          \end{array}
        \right)yv_u\qquad\qquad,~~
M=\left(
\begin{array}{ccc}
A+2 B/3& -B/3& -B/3\\
-B/3& 2B/3& A-B/3\\
-B/3& A-B/3& 2 B/3
\end{array}
\right) u ~~~,
\ee
where 
\be
A\equiv 2 x_A ~~~,~~~~~~~B\equiv 2 x_B \frac{v_S}{u}~~~.
\label{add}
\ee
The eigenvalues of $M$ are 
\be
M_1=(A+B)u~~~,~~~~~~~ M_2=Au~~~,~~~~~~~M_3=(-A+B)u.
\ee
The mass matrix for light neutrinos is $m_\nu=(m^D_\nu)^T M^{-1} m^D_\nu$ with eigenvalues
\be
m_1=\frac{y^2 v_u^2}{M_1}~~~,~~~~~~~
m_2=\frac{y^2 v_u^2}{M_2}~~~,~~~~~~~
m_3=\frac{y^2 v_u^2}{M_3}~~~.
\label{nueig}
\ee
The mixing matrix is $U_{TB}$, eq. (\ref{2}). 
Both normal and inverted hierarchies 
in the neutrino mass spectrum can be realized. It is interesting that $A_4$ models with the see-saw mechanism typically lead to a light neutrino spectrum which satisfies the sum rule (among complex masses):
\be
\frac{1}{m_3}=\frac{1}{m_1}-\frac{2}{m_2}~~~.\\
\label{sumr}
\ee
The phases of the complex parameters $A$ and $B$ do not produce any CP violation in
neutrino oscillations, since $\theta_{13}=0$, but are quite important to make the above
sum rule compatible with the present data on neutrino masses. A detailed discussion of a spectrum of this type can be found in refs. \cite{Altarelli:2006,Altarelli:2009a}.

Both types of ordering, normal and inverted are allowed and the above sum rule gives rise to bounds on the lightest neutrino mass.
For normal ordering we have
\bea
m_1&\ge& \sqrt{\frac{\Delta m^2_{sun}}{3}} \left(1-\dd\frac{4\sqrt{3}}{9} r+ ...\right) \approx 0.004~{\rm eV}\nn\\
m_1&\le& \sqrt{\frac{\Delta m^2_{sun}}{3}}\left(1+\dd\frac{4\sqrt{3}}{9} r+...\right) \approx 0.006~{\rm eV}
\label{boundno}
\eea
and for the inverted ordering:
\be
m_3\ge \sqrt{\frac{\Delta m^2_{atm}}{8}} \left(1-\dd\frac{1}{6} r^2+...\right) \approx 0.017~{\rm eV},
\label{boundio}
\ee
where the dots represent terms with higher powers of $r$. Notice that for normal ordering the neutrino mass spectrum is
essentially determined: $m_1\approx 0.005$ eV,  $m_1\approx 0.01$ eV and $m_3\approx 0.05$ eV. Also the possible values of $|m_{ee}|$ are restricted. For normal hierarchy we have
\be
|m_{ee}|\approx \dd\frac{4}{3\sqrt{3}} \Delta m^2_{sun} \approx 0.007~{\rm eV}~~~.
\label{meeno}
\ee
while for inverted hierarchy
\be
|m_{ee}|\ge  \dd\sqrt{\frac{\Delta m^2_{atm}}{8}} \approx 0.017~{\rm eV}~~~.
\label{meeio}
\ee
In a completely general framework, without the restrictions imposed by the flavor symmetry,
$|m_{ee}|$ could vanish in the case of normal hierarchy. In this model $|m_{ee}|$ is always
different from zero, though its value for normal hierarchy is probably too small to be detected
in the next generation of $0\nu\beta\beta$ experiments.

Note that in the charged lepton sector the flavor symmetry $A_4$ is broken by $\langle \varphi_T \rangle$ down to
$G_T$. Actually the above mass terms for charged leptons are the most general allowed by the 
symmetry $G_T$. At leading order in $1/\Lambda$, charged lepton masses are diagonal simply because
there is a low-energy $G_T$ symmetry. In the neutrino sector $A_4$ is broken down to $G_S$,
though neutrino masses in this model are not the most general ones allowed by $G_S$. The additional property which is needed, the invariance under $A_{\mu\tau}$, is obtained by stipulating that there are no $A_4$ breaking flavons transforming like $1'$ and $1''$. In fact, from eq. (\ref{tensorproda4}), we see that the expressions for $(33)'$ and $(33)''$ are not 2-3 symmetric.

At the next level of approximation each term of the superpotential is corrected by operators of higher dimension whose contributions are suppressed by at least one power of VEV's/$\Lambda$.  The corrections to the relevant part of the superpotential determine small deviations from the LO VEV alignment configuration. The NLO corrections to mass and mixing matrices are obtained by inserting the corrected VEV alignment in the LO operators plus the contribution of the new operators evaluated with the unperturbed VEV's. The final result is \cite{Altarelli:2006} that, when the NLO corrections are included, TB mixing is violated by small terms of the same order for all mixing angles:
\bea
\sin^2\theta_{12}&=&\frac{1}{3}+{\cal O}(\varepsilon)\nn \\
\sin^2\theta_{23}&=&\frac{1}{2}+{\cal O}(\varepsilon) \label{corr}\\
\sin\theta_{13}&=&{\cal O}(\varepsilon)\nn 
\eea
where $\varepsilon$ is of order of the typical VEV in units of $\Lambda$.
The fact that TB mixing is well satisfied by the data sets the restriction $\varepsilon < {\cal O}(\lambda_C^2)$.
From the requirement that the Yukawa coupling $y_\tau$ remains in the perturbative regime, we also get
a lower bound on $\varepsilon$ of about $0.01$, the exact value depending on $\tan\beta=v_u/v_d$ and on the 
largest allowed $|y_\tau|$. Thus we approximately have
\be
0.01<\varepsilon<0.05~~~.
\label{range}
\ee
From the see-saw relations in eq. (\ref{nueig}), assuming a coupling $y$ of order one, we 
see that the heavy  RH  neutrino masses are all of order $10^{15}$ GeV,
close to the GUT scale. The cut-off of the theory can be estimated form eq. (\ref{range})
to be close to $10^{17}$ GeV.

The above   
results in eqs. (\ref{mch}-\ref{meeio}) on the lepton mass matrices and the neutrino   
spectrum refer to the LO approximation.
Relations among neutrino masses can be affected by NLO corrections
but, for $\varepsilon$ varying in the range of eq. (\ref{range}), the bounds (\ref{boundno},\ref{boundio}) 
do not appreciably change (see for example ref. \cite{Barry:2010zk} for a numerical study of the deviations induced by vacuum misalignment). Also corrections induced by the renormalization group
evolution of the parameters can modify the above predictions,
but only in the case of sufficiently degenerate mass levels $m_1$ and $m_2$ with equal 
phases, which occurs for inverted mass ordering and far from the lower bound  
(\ref{boundio}) \cite{Lin:2009b}. 
The expansion parameter $\varepsilon$ directly controls also other observables,
such as the CP asymmetries of leptogenesis and 
the rates of lepton flavor violating transitions. 
This provides an interesting link between the physics
in the early universe relevant for leptogenesis and the low energy physics accessible in current experiments.
We will discuss the interplay bewteen discrete flavor symmetries and leptogenesis in Sect. 10.

\section{Possible origin of $A_4$}
There is an interesting relation \cite{Altarelli:2006} between the $A_4$ model considered so far and the modular group. This relation could possibly be relevant to understand the origin of the $A_4$ symmetry from a more fundamental layer of the theory.
The modular group $\Gamma$ is the group of linear fractional transformations acting on a complex variable $z$:
\be
z\to\frac{az+b}{cz+d}~~~,~~~~~~~ad-bc=1~~~,
\label{frac}
\ee
where $a,b,c,d$ are integers. 
There are infinite elements in $\Gamma$, but all of them can be generated by the two
transformations:
\be
s:~~~z\to -\frac{1}{z}~~~,~~~~~~~t:~~~z\to z+1~~~,
\label{st}
\ee
The transformations $s$ and $t$ in (\ref{st}) satisfy the relations
\be
s^2=(st)^3=1
\label{absdef}
\ee
and, conversely, these relations provide an abstract characterization of the modular group.
Since the relations (\ref{pres}) are a particular case of the more general constraint (\ref{absdef}),
it is clear that $A_4$ is a very small subgroup of the modular group and that the $A_4$ representations discussed above are also representations of the modular group.
In string theory the transformations (\ref{st})
operate in many different contexts. For instance the role of the complex 
variable $z$ can be played by a field, whose VEV can be related to a physical
quantity like a compactification radius or a coupling constant. In that case
$s$ in eq. (\ref{st}) represents a duality transformation and $t$ in eq. (\ref{st}) represent the transformation associated to an ''axionic'' symmetry. 

A different way to  
understand the dynamical origin of $A_4$ was recently presented in ref. \cite{Altarelli:2007} where it is shown that the $A_4$ symmetry can be simply  
obtained by orbifolding starting from a model in 6 dimensions (6D).  
In this approach $A_4$ appears as the remnant of the reduction
from 6D to 4D space-time symmetry induced by the 
special orbifolding adopted.  
This approach suggests a deep relation between flavor symmetry 
in 4D and  space-time symmetry in extra dimensions.

The orbifolding is defined as follows.
We consider a quantum field theory in 6 dimensions, with two extra dimensions
compactified on an orbifold $T^2/Z_2$. We denote by $z=x_5+i x_6$ the complex
coordinate describing the extra space. The torus $T^2$ is defined by identifying 
in the complex plane the points related by
\be
\begin{array}{l}
z\to z+1\\
z\to z+\gamma~~~~~~~~~~~~~~~~~\gamma=e^{\dd i\frac{\pi}{3}}~~~,
\label{torus}
\end{array}
\ee
where our length unit, $2\pi R$, has been set to 1 for the time being.
The parity $Z_2$ is defined by
\be
z\to -z
\label{parity}
\ee
and the orbifold $T^2/Z_2$ can be represented by the fundamental region given by the triangle
with vertices $0,1,\gamma$, see Fig. 3. The orbifold has four fixed points, $(z_1,z_2,z_3,z_4)=(1/2,(1+\gamma)/2,\gamma/2,0)$.
The fixed point $z_4$ is also represented by the vertices $1$ and $\gamma$. In the orbifold,
the segments labelled by $a$ in Fig. 1, $(0,1/2)$ and $(1,1/2)$, are 
identified and similarly for those labelled by $b$, $(1,(1+\gamma)/2)$ and 
$(\gamma,(1+\gamma)/2)$, and those labelled by $c$, $(0,\gamma/2)$, $(\gamma,\gamma/2)$. Therefore the orbifold is a regular tetrahedron
with vertices at the four fixed points.

\begin{figure}[]
\centering
$$\hspace{-4mm}
\includegraphics[width=10.0 cm]{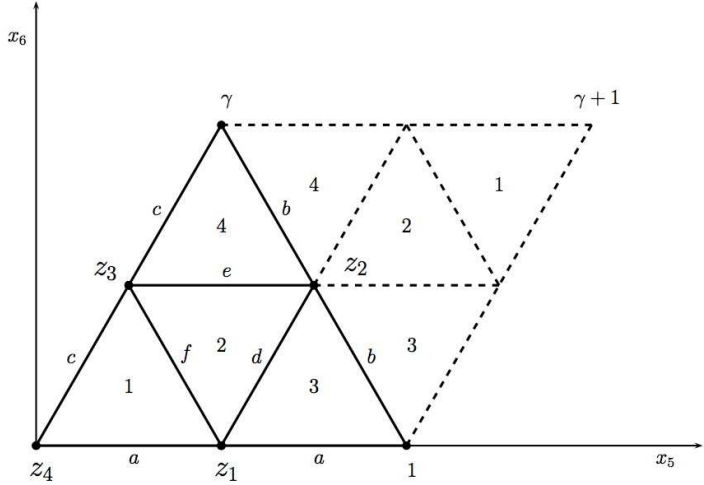}$$
\caption[]{Orbifold $T_2/Z_2$. The regions with the same numbers are 
identified with each other. The four triangles bounded by solid lines form the
fundamental region, where also the edges with the same letters are identified.
The orbifold $T_2/Z_2$ is exactly a regular tetrahedron with 6 edges
$a,b,c,d,e,f$ and four vertices $z_1$, $z_2$, $z_3$, $z_4$, corresponding to 
the four fixed points of the orbifold. }
\end{figure}
The symmetry of the uncompactified 6D space time is broken
by compactification. Here we assume that, before compactification,
the space-time symmetry coincides with the product of 6D translations
and 6D proper Lorentz transformations. The compactification breaks
part of this symmetry.
However, due to the special geometry of our orbifold, 
a discrete subgroup of rotations and translations in the extra space is left
unbroken. This group can be generated by two transformations:
\be
\begin{array}{ll}
{\cal S}:& z\to z+\frac{1}{2}\\
{\cal T}:& z\to \omega z~~~~~~~~~~~~~\omega\equiv\gamma^2~~~~.
\label{rototra}
\end{array}
\ee
Indeed ${\cal S}$ and ${\cal T}$ induce even permutations of the four fixed points:
\be
\begin{array}{cc}
{\cal S}:& (z_1,z_2,z_3,z_4)\to (z_4,z_3,z_2,z_1)\\
{\cal T}:& (z_1,z_2,z_3,z_4)\to (z_2,z_3,z_1,z_4)
\end{array}~~~,
\label{stfix}
\ee
thus generating the group $A_4$.  From 
the previous equations we immediately verify that ${\cal S}$ and ${\cal T}$ satisfy
the characteristic relations obeyed by the generators of $A_4$:
${\cal S}^2={\cal T}^3=({\cal ST})^3=1$.
These relations are actually satisfied not only at the fixed points, but on the whole orbifold,
as can be easily checked from the general definitions of ${\cal S}$ and ${\cal T}$ in eq. (\ref{rototra}),
with the help of the orbifold defining rules in eqs. (\ref{torus}) and (\ref{parity}).

We can exploit this particular geometry of the internal space to build a model 
with $A_4$ flavor symmetry.
There are 4D branes at the four fixed points of the orbifolding and the  
tetrahedral symmetry of $A_4$ connects these branes. The standard  
model fields have components on the fixed point branes while the scalar  
fields necessary for the $A_4$ breaking are in the bulk. Each brane field, either a 
triplet or a singlet, has components on all of the four fixed points (in particular all components are 
equal for a singlet) but the interactions are local, i.e. all vertices involve products of field 
components at the same space-time point. 
In the low-energy limit this model coincides with the one illustrated in the previous section.
Unfortunately in such a limit the 6D construction does not provide additional constraints or predictions.

This construction can be embedded in a SU(5) GUT \cite{Burrows:2009}.
Other discrete groups can arise from the compactification of two extra dimensions on orbifolds and the possibilities have been 
classified in \cite{Adulpravitchai:2009b,Adulpravitchai:2010na} within a field theory approach. In string theory the flavor symmetry can be larger than
the isometry of the compact space. For instance in heterotic orbifold models the orbifold geometry combines with the space group selection 
rules of the string, as shown in \cite{Kobayashi:2007}. Discrete flavor symmetries from magnetized/intersecting D-branes are discussed in \cite{Abe:2009}.
Discrete symmetries can also arise from the spontaneous breaking of continuous ones. Such a possibility has been discussed in ref. \cite{Adulpravitchai:2009c,Berger:2009}.
%
%
\section{Alternative routes to TB mixing}

While $A_4$ is the minimal flavor group leading to TB mixing, alternative flavor groups have been studied in the literature and can lead to interesting variants with some specific features.

Recently, in ref. \cite{Lam:2008}, the claim was made that, in order to obtain the TB mixing "without fine tuning", the finite group must be $S_4$ or a larger group containing $S_4$. For us this claim is not well grounded being based on an abstract mathematical criterium for a natural model (see also \cite{Grimus:2008a}). For us a physical field theory model is natural if the interesting results are obtained from the most general lagrangian compatible with the stated symmetry and the specified representation content for the flavons. For example, we obtain from $A_4$ (which is a subgroup of $S_4$) a natural (in our sense) model for the TB mixing by simply not including symmetry breaking flavons transforming like the $1'$ and the $1''$ representations of $A_4$. This limitation on the transformation properties of the flavons is not allowed by the rules specified in ref. \cite{Lam:2008} which demand that the symmetry breaking is induced by all possible kinds of flavons (note that, according to this criterium, the SM of electroweak interactions would not be natural because only Higgs doublets are introduced!). Rather, for naturalness we also require that additional physical properties like the VEV alignment or the hierarchy of charged lepton masses also follow from the assumed symmetry and are not obtained by fine tuning parameters: for this actually $A_4$ can be more effective than $S_4$ because it possesses three different singlet representations 1, $1'$ and $1''$.  

Models of neutrino mixing based on $S_4$ have in fact been studied \cite{Mohapatra:2004, Hagedorn:2006, Cai:2006, Ma:2007a, Bazzocchi:2008c, Ishimori:2009, Bazzocchi:2009a, Bazzocchi:2009b, Meloni:2009, Dutta:2009, Dutta:2009a, Morisi:2010, Ding:2010,Hagedorn:2010th,Ishimori:2010xk}. The group of the permutations of 4 objects $S_4$ has 24 elements and 5 equivalence classes (the character table is given in Table \ref{tcars}) that correspond to 5 inequivalent irreducible representations, two singlets, one doublet, two triplets: $1_1$, $1_2$, $2$, $3_1$ and $3_2$ (see Table 2). Note that the squares of the dimensions af all these representations add up to 24. 

\begin{table}[h]
\begin{center}
\begin{tabular}{|l|c|c|c|c|c|}
\hline \textbf{Class} & \textbf{$\chi(1_1)$} & \textbf{$\chi(1_2)$} &\textbf{$\chi(2)$}&\textbf{$\chi(3_1)$}&\textbf{$\chi(3_2)$}\\
\hline $C_1$ & 1 & 1& 2& 3& 3\\
\hline $C_2$ & 1 & 1& 2& -1& -1\\
\hline $C_3$ & 1 & -1& 0& 1& -1\\
\hline $C_4$ & 1 & 1& -1& 0& 0\\
\hline $C_5$ & 1 & -1& 0&-1& 1\\
\hline
\end{tabular}
\caption{Characters of $S_4$}
\label{tcars}
\end{center}
\end{table}
For models of TB mixing, one starts from the $S_4$ presentation $A^3=B^4=(BA^2)^2=1$ and identifies, up to a similarity transformation, $B^2=S$ and $A=T$, where $S$ and $T$ are given in eqs. (\ref{trep}, \ref{ta4}). In this presentation one obtains a realisation of the 3-dimensional representation of $S_4$ where the $S$ and $A_{23}$ matrices in eq. (\ref{inv}) that leave invariant the TB form of $m_{\nu}$ in eq. (\ref{gl21}) as well as the matrix $T$ in eq. (\ref{ta4}) of invariance for $m_e^\dagger m_e$, all explicitly appear \cite{Bazzocchi:2009a}. In $S_4$ the $1'$ and $1''$ of $A_4$ are collected in a doublet. When the VEV of the doublet flavon is aligned along the $G_S$ preserving direction the resulting couplings are 2-3 symmetric as needed. In $A_4$ the 2-3 symmetry is only achieved if the $1'$ and $1''$ VEV's are identical (which is the $S_4$ prediction). As discussed in ref. \cite{Bazzocchi:2009a}, in the leptonic sector the main difference between $A_4$ and $S_4$ is that, while in the typical versions of $A_4$  the most general neutrino mass matrix depends on 2 complex parameters (related to the couplings of the singlet and triplet flavons),  in $S_4$ it depends on 3 complex parameters  (because the doublet is present in addition to singlet and triplet flavons). 

Other flavor groups have been considered for models of TB mixing. Some of them include $S_4$ as a subgroup,  like $PSL_2(7)$ (the smallest group with complex triplet representations) \cite{Luhn:2007a, King:2009, King:2009a}, while others, like 
 $\Delta(27)$ (which is a discrete subgroup of $SU(3)$)  \cite{deMedeirosVarzielas:2007, Ma:2007b, Grimus:2007, Luhn:2007, Bazzocchi:2009c} or $Z_7\rtimes Z_3$ \cite{Luhn:2007b},  have no direct relation to $S_4$ \cite{King:2009b}. In Sect. 8 we will consider $S_4$ again in the different context of BM with large corrections from the lepton sector.

A different approach to TB mixing has been proposed and developed in different versions by S. King and collaborators  over the last few years \cite{King:2005, King:2006, deMedeirosVarzielas:2006, deMedeirosVarzielas:2007a, King:2009b}. 
The starting point is the decomposition of the neutrino mass matrix given in eqs. (\ref{1k1},\ref{4k1}) corresponding to exact TB mixing in the diagonal charged
lepton basis:
\begin{equation}
\label{mLL} 
m_\nu= m_1\Phi_1 \Phi_1^T + m_2\Phi_2 \Phi_2^T + m_3\Phi_3 \Phi_3^T
\end{equation}
where $\Phi_1^T=\frac{1}{\sqrt{6}}(2,-1,-1)$,
$\Phi_2^T=\frac{1}{\sqrt{3}}(1,1,1)$, $\Phi_3^T=\frac{1}{\sqrt{2}}(0,-1,1)$,
are the respective columns of $U_{TB}$ and $m_i$ are the neutrino mass eigenvalues.
Such decomposition is purely kinematical and does not possess any dynamical or symmetry
content. In the King models the idea is that the three columns of $U_{TB}$
$\Phi_i$ are promoted to flavon fields whose VEVs break the family symmetry, with the
particular vacuum alignments along the directions
$\Phi_i$. Eq. (\ref{mLL}) directly arises in the see-saw mechanism, $m_\nu=m_D^T M^{-1} m_D$, written in the diagonal RH neutrino mass basis,  $M={\rm diag}(M_1, M_2, M_3)$ when the Dirac
mass matrix is given by $m_D^T=(v_1 \Phi_1,v_2 \Phi_2,v_3 \Phi_3)$, where $v_i$ are mass parameters describing the size of the VEVs.
In this way, to each RH neutrino eigenvalue $M_i$, a particular light neutrino mass $m_i$ is associated. In the case of a strong neutrino hierarchy this idea can be combined with the framework of "Sequential Dominance", where the lightest RH neutrino, with its symmetry properties fixes the heaviest light neutrino and so on. For no pronounced hierarchy the correspondence between $M_i$ and $m_i$ can still hold and one talks of "Form Dominance" \cite{Chen:2009}.
In these models
the underlying family symmetry of the Lagrangian $G_f$ is
completely broken by the the combined action of the $\Phi_i$ VEV's, and the flavor symmetry of the neutrino mass
matrix emerges entirely as an accidental residual
symmetry of the quadratic form of eq. (\ref{mLL}) \cite{King:2009b}. The symmetry $G_f$ plays a less direct role and the name "Indirect Models" is used by the authors.
%
%
\section{Extension to quarks and GUT's}

Much attention  has been devoted to the question whether models with TB mixing in the neutrino sector can be  suitably extended to also successfully describe the observed pattern of quark mixings and masses and whether this more complete framework can be made compatible with (supersymmetric) SU(5) or SO(10) Grand Unification.  For models with approximate TB mixing in the leptonic sector we first consider the extension to quarks without Grand Unification and then the more ambitious task of building grand unified models. In GUT models based on $SU(5)\otimes G_f$ or $SO(10)\otimes G_f$ \footnote {The Pati-Salam group $SU(4)\otimes SU(2)\otimes SU(2)$ has also been considered, for example in \cite{King:2007,Toorop:2010yh}}, where $G_f$ is a flavor group, clearly all fields in a whole representation of $SU(5)$ or $SO(10)$ must have the same transformation properties under $G_f$. This poses a strong constraint on the way quarks and leptons have to transform under $G_f$.

\subsection{Extension to quarks without GUT's}

The simplest attempts of directly extending models based on $A_4$ to quarks have not been  satisfactory.
At first sight the most appealing
possibility is to adopt for quarks the same classification scheme under $A_4$ that one has
used for leptons (see, for example,  \cite{Ma:2001, Ma:2002, Babu:2003, Altarelli:2006}). Thus one tentatively assumes that LH quark doublets $Q$ transform
as a triplet $3$, while the  RH  quarks $(u^c,d^c)$,
$(c^c,s^c)$ and $(t^c,b^c)$ transform as $1$, $1''$ and $1'$, respectively. This leads to $V_u=V_d$ and to the identity matrix for $V_{CKM}=V_u^\dagger V_d$ in the lowest approximation. This at first appears as very promising: a LO approximation where neutrino mixing is TB and $V_{CKM}=1$ is a very good starting point. But there are some problems. First, the corrections 
to $V_{CKM}=1$ turn out to be strongly constrained by the leptonic sector, because lepton mixing angles are very close to the TB values, and, in the simplest models, this constraint leads to a too small $V_{us}$
(i.e. the Cabibbo angle is rather large in comparison to the allowed shifts from the TB mixing angles) \cite{Altarelli:2006}. Also in these models, the quark classification which leads to $V_{CKM}=1$ is not compatible with $A_4$ commuting with SU(5). 
An additional consequence of the above assignment is that the top quark mass would arise from a non-renormalizable dimension five operator. In that case, to reproduce the top mass, we need 
to compensate the cutoff suppression by some extra dynamical mechanism. Alternatively, we have to introduce a separate symmetry breaking parameter for the quark sector, sufficiently close to the cutoff
scale.

Due to this, larger discrete groups have been considered for the description of quarks.
A particularly appealing set of models is based on the discrete group $T'$, the double covering group of $A_4$ \cite{Frampton:1995, Aranda:2000, Aranda:2000a, Carr:2000,  Aranda:2007, Frampton:2007, Frampton:2009, Ding:2009,Feruglio:2007, Chen:2007}. As we see in Table 2 the 
representations of $T'$ are those of $A_4$ plus three independent doublets 2, $2'$ and $2''$. The doublets are interesting for the classification of the first two generations of quarks \cite{Pomarol:1996, Barbieri:1996, Barbieri:1997, Barbieri:1997a}. For example, in ref. \cite{Feruglio:2007} a viable description was obtained, i.e. in the leptonic sector the predictions of the $A_4$ model are reproduced, while the $T'$ symmetry plays an essential role for reproducing the pattern of quark mixing. But, again, the classification adopted in this model is not compatible with Grand Unification.

\subsection{Extension to quarks within GUT's} 

As a result, the group $A_4$ was considered by many authors to be too
limited to also describe quarks and to lead to a grand unified
description. It has been recently shown \cite{Altarelli:2008} that this negative attitude
is not justified and that it is actually possible to construct a
viable model based on $A_4$ which leads to a grand
unified theory (GUT) of quarks and leptons with TB mixing
for leptons and with quark (and charged lepton) masses and mixings compatible with experiment. At the same time this model offers an example of an
extra dimensional SU(5) GUT in which a description of all fermion masses
and mixings is accomplished.  The
formulation of SU(5) in extra dimensions has the usual advantages of
avoiding large Higgs representations to break SU(5) and of solving the
doublet-triplet splitting problem.  The choice of the transformation properties of the two
Higgses $H_5$ and $H_{\overline{5}}$ has a special role in this model. They are chosen to transform 
as two different $A_4$ singlets
$1$ and $1'$. As a consequence, mass terms for the Higgs colour
triplets are  not directly allowed and their masses are
introduced by orbifolding, \`{a} la Kawamura \cite{Witten:1985, Kawamura:2001, Faraggi:2001}.  In this model, proton
decay is dominated by gauge vector boson exchange giving rise to
dimension six operators, while the usual contribution of dimension five operators is forbidden by the selection rules of the model. Given the large $M_{GUT}$ scale of SUSY models and the relatively huge theoretical uncertainties, the decay rate is within the present experimental limits.
A see-saw realization
in terms of an $A_4$ triplet of RH neutrinos $\nu^c$ ensures the
correct ratio of light neutrino masses with respect to the GUT
scale. In this model extra dimensional effects directly
contribute to determine the flavor pattern, in that the two lightest
tenplets $T_1$ and $T_2$ are in the bulk (with a doubling $T_i$ and
$T'_i$, $i=1,2$ to ensure the correct zero mode spectrum), whereas the
pentaplets $F$ and $T_3$ are on the brane. The hierarchy of quark and
charged lepton masses and of quark mixings is determined by a
combination of extra dimensional suppression factors and of $U(1)_{FN}$ charges, both of which only apply to the first two
generations, while the neutrino mixing angles
derive from $A_4$ in the usual way. If the extra dimensional suppression factors and the $U(1)_{FN}$ charges are switched off, only the third generation masses of quarks and charged leptons survive. Thus the charged fermion mass matrices are nearly empty in this limit (not much of $A_4$ effects remain) and the quark mixing angles are determined by the small corrections induced by those effects. The model is natural, since most of the
small parameters in the observed pattern of masses and mixings as well
as the necessary vacuum alignment are  justified by the symmetries of
the model. However, in this case, like in all models based on $U(1)_{FN}$, the number of $\mathcal{O}(1)$ parameters is larger than the number of measurable quantities, so that in the quark sector the model can only account for the orders of magnitude (measured in terms of powers of an expansion parameter) and not for the exact values of mass ratios and mixing angles. A moderate fine tuning is only needed to enhance the Cabibbo mixing angle between the first two generations, which would generically be of $\mathcal{O}(\lambda_C^2)$. 

The problem of constructing GUT models based on  $SU(5)\otimes G_f$ or $SO(10)\otimes G_f$ with approximate TB mixing in the leptonic sector has been considered by many authors (see, for example \cite{Ma:2005b, Ma:2006a, Ma:2006b, Morisi:2007, Grimus:2008, Bazzocchi:2008b, Altarelli:2008, Ciafaloni:2009, Bazzocchi:2009,Antusch:2010es}, based on $A_4$). In our opinion most of the models are incomplete (for example, the crucial issue of VEV alignment is not really treated in depth as it should) and/or involve a number of unjustified, ad hoc fine tuning of parameters. An interesting model based on  $SU(5)\otimes T'$ is discussed in ref. \cite{Chen:2007}. In this model the $SU(5)$ tenplets $T_3$ and $T_a$ ($a=1,2$) of the third and of the first two generations are classified as 1 and 2 of $T'$, respectively, while the $SU(5)$ pentaplets are in a 3 of $T'$. This model provides a good description of fermion masses and mixings and appears simpler than the model in ref. \cite{Altarelli:2008}, which is also based on $SU(5)$. However, the model of ref. \cite{Chen:2007}  is fine tuned. In fact one does not understand how it is possible that, for example, the electron and the muon masses can come out so widely different as observed, given that in this model their left and right components separately transform in an identical way under $T'$.
The reason is that in the second term of eq. 5 of ref. \cite{Chen:2007}, only one of three possible contractions has been taken into account. If the missing ones, which are also allowed by the assumed symmetry properties, are included with generic coefficients, one in fact finds that the $e$ and $\mu$ masses are of the same order in the absence of fine tuning. Given that the expansion parameter in the model is of $\mathcal{O}(\lambda_C)$ the fine tuning which is needed is large. One possible way out would be to invoke some ultraviolet completion of the model where particular heavy field exchanges could justify the presence of only the desired couplings after the heavy fields are integrated out. Also, in the model of ref. \cite{Chen:2007}  there is no discussion of the origin of the required vacuum alignment. Recently some GUT models based on SU(5)$\times S_4$ have
appeared \cite{Ishimori:2010xk,Hagedorn:2010th}. Also in these models the first two generation fermions are in the same $S_4$ representations (either a doublet, for tenplets, or a triplet, for pentaplets). In the absence
of an additional principle the electron and muon mass should naturally be of the same order. In ref. \cite{Ishimori:2010xk} the vanishing of the electron mass at LO is obtained by the ad hoc choice
of one particular minimum of the scalar potential among a continuous family of degenerate solutions (see their eqs. (70-71)). In the case of ref. \cite{Hagedorn:2010th} the problem is solved by introducing new heavy particles with suitable interactions that, once integrated out, produce the desired structure for the mass matrix.  

As for the models based on $SO(10)\otimes G_f$ we select two recent examples with $G_f=S_4$ \cite{Dutta:2009, Dutta:2009a} and $G_f=PSL_2(7)$ \cite{King:2009a}. Clearly the case of $SO(10)$ is even more difficult than that of $SU(5)$ because the neutrino sector is tightly related to that of quarks and charged leptons as all belong to the 16 of $SO(10)$ (for a general analysis of $SO(10)\otimes A_4$ see \cite{Bazzocchi:2008b}). The strategy adopted in refs. \cite{Dutta:2009, Dutta:2009a, King:2009a}  as well as in other $SO(10)$ models, is as follows. One considers renormalisable fermion mass terms with Higgs multiplets of the $SO(10)$ 10 ($h$ terms) and 126 ($f$ terms) representations. The Majorana neutrino mass matrix  arises from the 126. One assumes that the dominant contribution to the Dirac masses of fermions is from the $h$ terms with small corrections from the $f$ terms. In first approximation the $h$ contribution is a matrix of rank 1 with only the third generation mass being non vanishing. The light fermion masses and the quark mixings then arise from the $f$ terms (and from some possible extra terms). The third family dominance is obtained by a term with a double flavon factor in ref. \cite{Dutta:2009, Dutta:2009a}  (based on $S_4$)  which then makes particularly difficult to keep the corrective $f$ terms small (for this fine tuning is needed or a suitable ultraviolet completion). In ref. \cite{King:2009a}  the dominant $h$ terms are induced by a single $PSL_2(7)$ sextet flavon (the existence of complex 3-dim and of 6-dim representations is the peculiarity of $PSL_2(7)$). In both models in the neutrino sector one has a sum of type I and II see-saw contributions of the form:
\beq
m_\nu=fv_L-m_D^T\frac{1}{fv_R}m_D\\
\label{I+II}
\eeq
where the first term is from the exchange of a triplet Higgs with VEV proportional to $v_L$ while the second term is from type I see-saw with RH mass proportional to $v_R$. One must assume that the first term is dominant and the second is negligible. Then the leading approximation for the fermion Dirac masses is from the $h$ terms and for neutrino masses from the $f$ terms. The $f$ terms are diagonalized by the TB mixing unitary matrix. In this way the connection between quarks and neutrinos is relaxed and a completely different pattern of mixing can be realized in the two sectors. 
Clearly for the $f v_L$ dominance in eq. (\ref{I+II}) one needs $v_L>>v^2/v_R$ with $v \sim h/f$ 
. This needs widely different scales  for $v_L$ and $v_R$ in the model and much of the description of the corresponding dynamics, along the lines of refs. \cite{Goh:2004, Mohapatra:2007a}, remains to be studied in detail.
In both of these models the discussion of the alignment is not satisfactory. In particular in ref. \cite{King:2009a}  it is only proven that the arbitrary coefficients appearing in the most general allowed superpotential can be fitted to lead to the required ratios of components in the VEV's (while for a natural model one would require that the alignment automatically follows in a whole region of the parameter space). In conclusion, in our opinion, the problem of constructing a satisfactory natural model based on $SO(10)$ with built-in TB mixing at the LO approximation, remains open.

\section{The $S_4$ group and BM mixing}

If one takes the alternative view that the agreement with TB mixing is accidental and is rather oriented to consider "weak" complementarity as a more attractive guiding principle, then a better starting point could be BM mixing. In the BM scheme $\tan^2{\theta_{12}}= 1$, to be compared with the latest experimental
determination:  $\tan^2{\theta_{12}}= 0.45\pm 0.04$ (at $1\sigma$) \cite{Strumia:2006, Gonzalez-Garcia:2008a, Bandyopadhyay:2008, Fogli:2008, Fogli:2008a, Schwetz:2008, Maltoni:2008} (see fig.2), so that a rather large non leading correction is needed, as already mentioned. A discrete group approach can also work for BM mixing. We now summarise a model \cite{Altarelli:2009b} based on $S_4$ that leads to BM mixing in first approximation while the agreement with the data is restored by large NLO corrections that arise from the charged lepton sector.

The group $S_4$ is particularly suitable for reproducing BM mixing in LO because the unitary matrices
$S_{BM}$, given in eq. (\ref{matS}), and $T_{BM}$,  given in eq. (\ref{ts4}), directly provide a presentation of $S_4$. We recall that $S_{BM}$ leaves invariant the most general mass matrix for BM mixing in the charged lepton diagonal basis, eq. (\ref{gl2}), while $T_{BM}$ leaves invariant the most general diagonal matrix $m_e^\dagger m_e$ for charged leptons (see eqs. (\ref{invS}, \ref{Tdiag})). In fact, from Table 2, we see that a possible presentation of $S_4$ is given by:
\beq
A^4=B^2=(AB)^3=1\\
\label{press4}
\eeq
In terms of 3x3 matrices, we can make the identifications $A=T_{BM}$ and $B=S_{BM}$ and eqs. (\ref{press4}) are satisfied. As was the case for the $A_4$ models, again in this model the invariance under $A_{23}$, which is also necessary to specify BM mixing according to eq. (\ref{invS}), arises accidentally as a consequence of the specific field content and is limited to the contribution of the dominant terms to the neutrino mass matrix.

In the model the 3 generations of LH lepton doublets $l$ and of RH neutrinos $\nu^c$ to two triplets $3$, while the RH charged leptons $e^c$, $\mu^c$ and $\tau^c$ transform as $1$, $1'$ and $1$, respectively. The $S_4$ symmetry is then broken by suitable triplet flavons. All the flavon fields are singlets under the Standard Model gauge group.  Additional symmetries are needed, as usual, to prevent unwanted couplings and to obtain a natural hierarchy among $m_e$, $m_\mu$ and $m_\tau$. The complete flavor symmetry of the model is $S_4\times Z_4\times U(1)_{FN}$.  A flavon $\theta$, carrying a negative unit of the $U(1)_{FN}$ charge F, acquires a vacuum expectation value (VEV) and breaks $U(1)_{FN}$. A supersymmetric context is adopted, so that two Higgs doublets $h_{u,d}$, invariant under $S_4$, are present in the model as well as the $U(1)_R$ symmetry related to R-parity and the driving fields in the flavon superpotential. Supersymmetry also helps producing and maintaining the hierarchy $\langle h_{u,d}\rangle=v_{u,d}\ll \Lambda$ where $\Lambda$ is the cut-off scale of the theory.

The fields in the model and their classification under the symmetry are summarized in Table \ref{table:TransformationsS}. The fields  $\psi_l^0$, $\chi_l^0$ , $\xi_\nu^0$ and $\phi_\nu^0$ are the driving fields. \begin{table}[h]
\begin{center}
\begin{tabular}{|c||c|c|c|c|c|c||c||c|c|c|c||c|c|c|c|}
  \hline
  &&&&&&&&&&&&&&&\\[-0,3cm]
  & $l$ & $e^c$ & $\mu^c$ & $\tau^c$ & $\nu^c$ & $h_{u,d}$ & $\theta$ & $\phi_l$ & $\chi_l$ & $\psi_l^0$ & $\chi_l^0$ & $\xi_\nu$ &$\phi_\nu$ & $\xi_\nu^0$ & $\phi_\nu^0$ \\
  &&&&&&&&&&&&&&&\\[-0,3cm]
  \hline
  &&&&&&&&&&&&&&&\\[-0,3cm]
  $S_4$ & 3 & 1 & $1^\prime$ & 1 & 3 & 1 & 1 & 3 & $3^\prime$ & 2 & $3'$ & 1 & 3 & 1 & 3  \\
  &&&&&&&&&&&&&&&\\[-0,3cm]
  $Z_4$ & 1 & -1 & -i & -i & 1 & 1 & 1 & i & i & -1 & -1 & 1 & 1 & 1 & 1 \\
  &&&&&&&&&&&&&&&\\[-0,3cm]
  $U(1)_{FN}$ & 0 & 2 & 1 & 0 & 0 & 0 & -1 & 0 & 0 & 0 & 0 & 0 & 0 & 0 & 0  \\
  &&&&&&&&&&&&&&&\\[-0,3cm]
  $U(1)_R$ & 1 & 1 & 1 & 1 & 1 & 1 & 0 & 0 & 0 & 2 & 2 & 0 & 0 & 2 & 2  \\
  \hline
  \end{tabular}
\end{center}
\caption{\label{table:TransformationsS}Transformation properties of all the fields.}
\end{table}
The complete superpotential can be written as $w=w_l+w_\nu+w_d$.
The $w_d$ term is responsible for the alignment. It was discussed in ref. \cite{Altarelli:2009b} and this discussion will not be repeated here. The terms $w_l$ and $w_\nu$
determine the lepton mass matrices (we indicate with $(\ldots)$ the singlet 1, with $(\ldots)^\prime$ the singlet $1^\prime$ and with $(\ldots)_V$ ($V=2,\,3,\,3'$) the representation V)
\bea
w_l\;=&&\frac{y_e^{(1)}}{\La^2}\frac{\theta^2}{\La^2}e^c(l\phi_l\phi_l)+ \frac{y_e^{(2)}}{\La^2}\frac{\theta^2}{\La^2}e^c(l\chi_l\chi_l)+ \frac{y_e^{(3)}}{\La^2}\frac{\theta^2}{\La^2}e^c(l\phi_l\chi_l)+\nn\\
&+&\frac{y_\mu}{\La}\frac{\theta}{\La}\mu^c(l\chi_l)^\prime+\frac{y_\tau}{\La}\tau^c(l\phi_l)+\dots
\label{wl}\\
\nn\\
w_\nu\;=&&y(\nu^cl)+M \Lambda (\nu^c\nu^c)+a(\nu^c\nu^c\xi_\nu)+b(\nu^c\nu^c\phi_\nu)+\dots\\
\label{wd}\nn
\eea
where $a$ and $b$ are complex coefficients. Again, to keep our formulae compact, we omit to write the Higgs fields
$h_{u,d}$.  For instance
$y_\tau \tau^c(l\phi_l)/\La$ stands for $y_\tau \tau^c(l\phi_l)h_d/\La$,
$y(\nu^cl)$ stands for $y(\nu^cl) h_u$. The powers of the cutoff $\La$ also take  into account the presence of the omitted Higgs fields. Note that the parameters $M$, $M_\phi$, $M_\xi$ and $M'_\xi$ defined above are dimensionless.
In the above expression for the superpotential $w$, only the lowest order operators
in an expansion in powers of $1/\Lambda$ are explicitly shown. Dots stand for higher
dimensional operators that will be discussed later on. The stated symmetries ensure that, for the leading terms, the flavons that appear in $w_l$ cannot contribute to $w_\nu$ and viceversa.

The potential corresponding to $w_d$ possesses an isolated minimum for the following VEV configuration:
\beq
\dd\frac{\mean{\phi_l}}{\La}=\left(
                     \begin{array}{c}
                       0 \\
                       1 \\
                       0 \\
                     \end{array}
                   \right)A\qquad\qquad
                   \dd\frac{\mean{\chi_l}}{\La}=\left(
                     \begin{array}{c}
                       0 \\
                       0 \\
                       1 \\
                     \end{array}
                   \right)B
\label{vev:charged:best}
\eeq
\beq
\hspace{-1.5cm}
\dd\frac{\mean{\phi_\nu}}{\La}=\left(
                     \begin{array}{c}
                       0 \\
                       1 \\
                       -1 \\
                     \end{array}
                   \right)C\qquad\quad
\dd\frac{\mean{\xi_\nu}}{\La}=D
\label{vev:neutrinos}
\eeq
where the factors  $A$, $B$, $C$, $D$ should obey to the relations:
\bea
&\sqrt{3}f_1A^2+\sqrt{3}f_2B^2+f_3AB=0
\label{AB}\\
\nn\\
&D=-\dd\frac{M_\phi}{g_2}\qquad\qquad C^2=\dd\frac{g_2^2M_\xi^2+g_3M_\phi^2-g_2M_\phi M'_\xi}{2 g_2^2g_4}
\label{CD}\;.
\eea
Similarly, the Froggatt-Nielsen flavon $\theta$ gets a VEV, determined by the D-term associated to the local $U(1)_{FN}$ symmetry, and it is denoted by
\beq
\frac{\mean{\theta}}{\La}= t\;.
\label{deft}
\eeq

With this VEV's configuration, the charged lepton mass matrix is diagonal
\beq
m_l=\left(
         \begin{array}{ccc}
           (y_e^{(1)}B^2-y_e^{(2)}A^2+y_e^{(3)}AB)t^2 & 0 & 0 \\
           0 & y_\mu Bt & 0 \\
           0 & 0 & y_\tau A \\
         \end{array}
       \right) v_d
\eeq
so that at LO there is no contribution to the $U_{PMNS}$ mixing matrix from the diagonalization of charged lepton masses.
In the neutrino sector for the Dirac and RH Majorana matrices we have
\beq
m_\nu^D=\left(
          \begin{array}{ccc}
            1 & 0 & 0 \\
            0 & 0 & 1 \\
            0 & 1 & 0 \\
          \end{array}
        \right)yv_u\qquad\qquad
M_N=\left(
              \begin{array}{ccc}
                2M+2aD & -2bC & -2bC \\
                -2bC & 0 & 2M+2aD \\
                -2bC & 2M+2aD & 0 \\
              \end{array}
            \right)\Lambda\;.
\label{Feq:RHnu:masses}
\eeq
The matrix $M_N$ can be diagonalized by the BM mixing matrix $U_{BM}$, which represents the full lepton mixing at the LO, and the eigenvalues are
\beq
M_1=2|M+aD-\sqrt{2}bC|\Lambda\qquad M_2=2|M+aD+\sqrt{2}bC|\Lambda\qquad M_3=2|M+aD|\Lambda\;.
\eeq
After see-saw, since the Dirac neutrino mass matrix commutes with $M_N$ and its square is a matrix proportional to unity,
the light neutrino Majorana mass matrix, given by the see-saw relation \mbox{$m_\nu=(m_\nu^D)^TM_N^{-1}m_\nu^D$}, is also diagonalized by the BM mixing matrix and the eigenvalues are
\beq
|m_1|=\frac{|y^2|v_u^2}{2|M+aD-\sqrt{2}bC|}\dd\frac{1}{\Lambda}\qquad
|m_2|=\frac{|y^2|v_u^2}{2|M+aD+\sqrt{2}bC|}\dd\frac{1}{\Lambda}\qquad
|m_3|=\frac{|y^2|v_u^2}{2|M+aD|}\dd\frac{1}{\Lambda}\;.
\label{spec}
\eeq
The light neutrino mass matrix depends on only 2 effective parameters, at LO, indeed the terms $M$ and $aD$ enter the mass matrix in the combination $F\equiv M+a D$. The coefficients $y_e^{(i)}$, $y_\mu$, $y_\tau$, $y$, $a$ and $b$ are all expected to be of $\mathcal{O}(1)$. A priori $M$ could be of $\mathcal{O}(1)$, corresponding to a RH neutrino Majorana mass of $\mathcal{O}(\Lambda)$, but, actually,  it must be of the same order as $C$ and $D$. In the context of a grand unified theory this would correspond to
the requirement that $M$ is of $\mathcal{O}(M_{GUT})$ rather than of $\mathcal{O}(M_{Planck})$.

We expect a common order of magnitude for the VEV's (scaled by the cutoff $\Lambda$):
\beq
A \sim B \sim v\;,~~~~~~~~~~C \sim D \sim v'\;.
\eeq
However, due to the different minimization conditions that determine $(A,B)$ and $(C,D)$, we may tolerate a moderate hierarchy
between $v$ and $v'$. Similarly the order of magnitude of $t$ is in principle unrelated to those of $v$ and $v'$.
It is possible to estimate the values of $v$ and $t$ by looking at the mass ratios of charged leptons (while $v'$ only enters in the neutrino sector):
and the result is that $t \sim 0.06$ and $v \sim 0.08$ (modulo coefficients of $\mathcal{O}(1)$).

So far we have shown that, at LO, we have diagonal and hierarchical charged leptons together with the exact BM mixing for neutrinos. It is clear that substantial NLO corrections are needed to bring the model to agree with the data on $\theta_{12}$. A crucial feature of the model is that the neutrino sector flavons  $\phi_\nu$ and $\xi_\nu$ are invariant under $Z_4$ which is not the case for the charged lepton sector flavons $\phi_l$ and $\chi_l$. The consequence is that $\phi_\nu$ and $\xi_\nu$  can contribute at NLO to the corrections in the charged lepton sector, while at NLO $\phi_l$ and $\chi_l$ cannot modify the neutrino sector couplings. As a results the dominant corrections to the BM mixing matrix only occur  at NLO through the diagonalization of the charged leptons. 
In fact, at NLO the neutrino mass matrix is still diagonalized by $U_{BM}$ but the mass matrix of charged leptons is no more diagonal. Including these additional terms from the diagonalization of charged leptons the $U_{PMNS}$ matrix can be written as
\beq
U_{PMNS}=U_l^\dag U_{BM}\;,
\eeq
and therefore the corrections from $U_l$ affect the neutrino mixing angles at NLO according to
\bac{l}
\sin^2\theta_{12}=\dd\frac{1}{2}-\frac{1}{\sqrt{2}}(V_{12}+V_{13})v'\\[0.2cm]
\sin^2\theta_{23}=\dd\frac{1}{2}\\[0.2cm]
\sin\theta_{13}=\dd\frac{1}{\sqrt{2}}(V_{12}-V_{13})v'\;.
\label{sinNLO}
\eac
where the coefficients $V_{ij}$ arise from $U_l$.
By comparing these expressions with the current experimental values of the mixing angles in Table 1, we see that, to correctly reproduce $\theta_{12}$ we need a parameter $v'$ of the order of the Cabibbo angle $\lambda_C$. Moreover, barring cancellations of/among some the $V_{ij}$ coefficients, also $\theta_{13}$ is corrected by a similar amount, while $\theta_{23}$ is unaffected at the NLO. A salient feature of this model is that, at NLO accuracy, the large corrections of $\mathcal{O}(\lambda_C)$ only apply to $\theta_{12}$ and $\theta_{13}$ while $\theta_{23}$ is unchanged at this order. As a correction of $\mathcal{O}(\lambda_C)$ to $\theta_{23}$ is hardly compatible with the present data (see Table 1) this feature is very crucial for the phenomenological success of this model. It is easy to see that this essential property depends on the selection in the neutrino sector of flavons $\xi_\nu$ and $\phi_\nu$ that transform as 1 and 3 of $S_4$, respectively. If, for example, the singlet $\xi_\nu$ is replaced by a doublet $\psi_\nu$ (and correspondingly the singlet driving field  $\xi_\nu^0$ is replaced by a doublet $\psi_\nu^0$), all other quantum numbers being the same, one can construct a variant of the model along similar lines, but, in this case, all the 3 mixing angles are corrected by terms of the same order. This confirms that a  particular set of $S_4$ breaking flavons is needed in order to preserve $\theta_{23}$ from taking as large corrections as the other two mixing angles.

All this discussion applies at the NLO  and we expect that at the NNLO the value of $\theta_{23}$ will eventually be modified with deviations of about $\mathcal{O}(\lambda^2_C)$. The next generation of experiments, in particular those exploiting a high intensity neutrino beam, will probably reduce the experimental error on $\theta_{23}$ and the sensitivity on $\theta_{13}$ to few degrees. All quantitative estimates are clearly affected by large uncertainties due to the presence of unknown parameters of order one, but in this model a value of $\theta_{13}$ much smaller than the present upper bound would be unnatural. If in the forthcoming generation of experiments no significant deviations from zero of $\theta_{13}$ will be detected, this construction will be strongly disfavoured.
\section{Lepton flavor violation}
Neutrino oscillations provide evidence of flavor conversion in the lepton sector.
This indicates that lepton flavor violation (LFV) might take place, at least at some level, 
also in other processes such as those involving charged leptons. Flavor violating decays of charged leptons, 
strictly forbidden in the SM, are indeed allowed as soon as neutrino mass terms are considered. 
If neutrino masses are the only source of LFV, the effects are too small to be detected, 
but in most extensions of the SM where new particles and new interactions
with a characteristic scale $\Lambda_{NP}$ are included, the presence of new sources of flavor violation,
in both quark and lepton sectors, is a generic feature.  
The scale $\Lambda_{NP}$ can be much smaller than the cut-off scale $\Lambda$ introduced before.
Indeed there are several indications suggesting new physics at the TeV scale, such as 
a successful gauge coupling unification, viable solutions to the hierarchy problem and realistic dark matter candidates.
In a low-energy description, the associated effects can be parametrized
by higher-dimensional operators. 
The dominant terms are represented by dimension six operators, suppressed by two powers of $\Lambda_{NP}$:
\be
{\cal L}_{eff}=
i\frac{e}{\Lambda_{NP}^2} {e^c}_i H^\dagger \sigma^{\mu\nu} F_{\mu\nu} {\cal Z}_{ij}  l_j +\dd\frac{1}{\Lambda_{NP}^2}[\tt 4-fermion~~ operators]+h.c.
\label{leff}
\ee
where $e$ is the electric charge and
${\cal Z}_{ij}$ denotes an adimensional complex matrix with indices in flavor space.
If the underlying theory is weakly interacting with a typical coupling constant $g_{NP}$ and predicts new particles of mass $m_{NP}$
we expect $\Lambda_{NP}\approx 4\pi m_{NP}/g_{NP}$.
The present bounds on the branching ratios \cite{Raidal:2008} of the rare charged lepton decays set stringent limits on combinations
of the scale $\Lambda_{NP}$ and the coefficients of the involved operators. 
For instance, from $BR(\mu\to e \gamma)<1.2\times 10^{-11}$ \cite{Brooks:1999,Adam:2009} we get $|{\cal Z}_{\mu e}|<10^{-8}\times [\Lambda_{NP}({\rm TeV})/1~{\rm TeV}]^2$.
Typically, for coefficients of order one, the existing bounds
require a large scale $\Lambda_{NP}$, several orders of magnitude larger than the TeV scale.
Conversely, to allow for new physics close to the TeV scale, coefficients much smaller than one
are required, which may indicate the effect of a flavor symmetry. 

In theories with a flavor symmetry group $G_f$ spontaneously broken by a set
of small parameters $\varepsilon$, the coefficients of the effective lagrangian in eq. (\ref{leff}) become functions of $\varepsilon$. 
The low-energy Lagrangian of eq. (\ref{leff}) is derived from the theory defined close to the cut-off scale $\Lambda$, 
where all operators are invariant under $G_f$ thanks to their dependence on the flavon multiplets.
Below the flavor symmetry breaking scale the flavons are replaced by their VEVs, which enter the coefficients of ${\cal L}_{eff}$ through the dimensionless combination $\varepsilon\approx VEV/\Lambda$.
Exploiting the smallness of the parameters $\varepsilon$ we can keep in ${\cal L}_{eff}$ the first few terms of a power series expansion.
For instance:
\be
{\cal Z}_{ij}\equiv{\cal Z}_{ij}\left(\varepsilon\right)={\cal Z}_{ij}^{(0)}+{\cal Z}_{ij}^{(1)}~\varepsilon+{\cal Z}_{ij}^{(2)}~\varepsilon^2...
\ee
Notice that the same symmetry breaking parameters that control lepton masses and mixing angles also control
the flavor pattern of the operators in ${\cal L}_{eff}$. This result is interesting in several respects. First of all the presence of the factors $\varepsilon^n$ can help in
suppressing the rates of rare charged lepton decays while allowing for a relatively small and accessible scale $\Lambda_{NP}$.
Second, once the above expansion has been determined in a given model,
it could be possible to establish characteristic relations among LFV processes as a consequence
of flavor symmetries and of their pattern of symmetry breaking. 
Finally, if $\Lambda_{NP}$ is sufficiently small, this opens the possibility that new particles might be produced and detected at the LHC, 
with features that could additionally confirm or reject the assumed symmetry pattern.
All this allows, at least  in principle, to realize an independent test of the flavor symmetry in the charged lepton sector. While the size of the scale $\Lambda_{NP}$ could be relatively small, in our presentation we assume that
the flavour scale or cutoff $\Lambda$ is extremely large, possibly as large as the GUT scale.
Then all low-energy effects due to the flavon dynamics are essentially those associated to their VEVs, which enter the effective higher dimensional operators through the dimensionless
combination $\epsilon$.  Virtual flavon exchanges give rise to other higher
dimensional operators which are depleted by inverse power of $\Lambda$ and can be safely neglected. A much richer variety of effects due to the flavour dynamics
would be possible if the scale $\Lambda$ were much smaller, close to the 100 TeV energy
range, but we do not consider this possibility here.

The effects described by ${\cal L}_{eff}$ are well-known. In a field basis where the kinetic terms are canonical and the charged lepton
mass matrix is diagonal the real and imaginary parts of the diagonal matrix elements
${\cal Z}_{ii}$ are proportional to the anomalous magnetic moments (MDM) $a_i$
and to the electric dipole moments (EDM) $d_i$ of charged leptons, respectively:
\begin{equation}
a_i=2 m_i^{ch} \frac{v}{\sqrt{2} \Lambda_{NP}^2}Re {\cal Z}_{ii}~~~,~~~~~~~d_i=e \frac{v}{\sqrt{2} \Lambda_{NP}^2}Im {\cal Z}_{ii}~~~.
\label{dm}
\end{equation}
The off-diagonal elements 
${\cal Z}_{ij}$ describe the amplitudes for the radiative decays of the charged leptons:
\begin{equation}
R_{ij}=\frac{BR(l_i\to l_j\gamma)}{BR(l_i\to l_j\nu_i{\bar \nu_j})}=\frac{12\sqrt{2}\pi^3 \alpha}{G_F^3 {m_i^{\tt ch}}^2 \Lambda_{NP}^4}\left(\vert{\cal Z}_{ij}\vert^2+\vert{\cal Z}_{ji}\vert^2\right)
\label{dt}
\end{equation}
where $\alpha$ is the fine structure constant, $G_F$ is the Fermi constant and $m_i^{ch}$ is the mass of the lepton $l_i$.
Finally the four-fermion operators, together with the dipole
operators controlled by ${\cal Z}$, describe other flavor violating processes
like $\mu\to eee$, $\tau\to\mu\mu\mu$, $\tau\to eee$. 

An interesting example of flavor symmetry is that of minimal flavor violation (MFV) \cite{Chivukula:1987,Hall:1990,Ciuchini:1998,Buras:2001,D'Ambrosio:2002,Cirigliano:2005,Cirigliano:2006,Davidson:2006,Cirigliano:2007} whose (minimal) flavor symmetry group in the lepton sector is $G_f=SU(3)_{e^c}\times SU(3)_l$.
Electroweak singlets $e^c$ and doublets $l$ transform as $(3,1)$ and $(1,\bar{3})$, respectively. The flavon fields or, better, their VEVs are the Yukawa couplings of the charged leptons, $Y_l=m_l/v$,
and the adimensional coupling constants $\eta$ of the five-dimensional operator $O_5$ in eq. (\ref{O5}). They
transform as $(\bar{3},3)$ and $(1,6)$, respectively. In a basis where the charged leptons are diagonal, we have
\begin{equation}
Y_l=\frac{\sqrt{2}}{v} m_l^{\rm diag}~~~,~~~~~~~~~~\eta=\frac{M}{v^2} U^* m_\nu^{\rm diag} U^\dagger~~~,
\end{equation}
where $M$ here denotes the mass scale suppressing the operator $O_5$. In MFV models the leading off-diagonal elements of ${\cal Z}_{ij}$ are given by:
\begin{eqnarray}
{\cal Z}_{ij}&=&c~ (Y_l~ \eta^\dagger \eta)_{ij}\nonumber\\
&=&\sqrt{2}c~\frac{m_i^{\tt ch}}{v}\frac{M^2}{v^4}\left[\Delta m^2_{sol} U_{i2} U^*_{j2}\pm\Delta m^2_{atm} U_{i3} U^*_{j3}\right]
\label{zij}
\end{eqnarray}
where $c$ is an overall coefficient of order one and the plus (minus) sign refers to the case of normal (inverted) hierarchy.
We see that, due to the presence of the ratio $M^2/v^2$ the overall scale of these matrix elements is poorly constrained.
This is due to the fact that MFV does not restrict the overall strength of the coupling constants $\eta$, apart from the requirement
that they remain in the perturbative regime. 
Very small or relatively large (but smaller than one) $\eta$ can be accommodated by adjusting the scale $M$.
Thus, even after fixing $\Lambda_{NP}$ close to the TeV scale, in MFV the non-observation of $l_i\to l_j \gamma$ could be justified by choosing a small $M$, while a positive signal
in $\mu\to e \gamma$ with a branching ratio in the range $1.2\times 10^{-11}\div 10^{-13}$ could also be fitted by an appropriate $M$,
apart from a small region of the $\theta_{13}$ angle, around $\theta_{13}\approx0.02$ where a cancellation can take place in the left-hand side of eq. (\ref{zij}).
The dependence on the scales $M$ and $\Lambda_{NP}$ can be eliminated by considering ratios of branching ratios. For instance:
\begin{equation}
\dd\frac{R_{\mu e}}{R_{\tau \mu}}=
\left\vert\frac{2\Delta m^2_{sol}}{3\Delta m^2_{atm}}\pm \sqrt{2}\sin\theta_{13} e^{i\delta}\right\vert^2<1~~~,
\label{mfv}
\end{equation}
where we took the TB ansatz to fix $\theta_{12}$ and $\theta_{23}$.
We see that $BR(\mu\to e\gamma)<BR(\tau\to \mu\gamma)$ always in MFV. Moreover, for $\theta_{13}$ above approximately $0.07$, $BR(\mu\to e\gamma)<1.2\times 10^{-11}$ implies $BR(\tau\to \mu\gamma)<10^{-9}$. For $\theta_{13}$ below $0.07$, apart possibly from a small region around $\theta_{13}\approx0.02$, both the transitions $\mu\to e \gamma$ and $\tau\to\mu\gamma$
might be above the sensitivity of the future experiments. 
The present limits are $BR(\tau\to\mu\gamma)<1.6\times 10^{-8}$ and $BR(\tau\to e\gamma)<9.4\times 10^{-8}$. A future super B factory
might improve them by about one order of magnitude.
In the SUSY case there are two doublets
in the low-energy Lagrangian and we should take into account the $\tan\beta$ dependence.

A different result for the matrix ${\cal Z}$ is obtained in
the model described in Sect. 4 where $G_f=A_4\times Z_3\times U(1)_{FN}$. Starting from the relevant set of invariant operators, after the breaking of the flavor and electroweak symmetries, 
and after moving to a basis with canonical kinetic terms and diagonal mass matrix for charged leptons, we find \cite{Feruglio:2009,Feruglio:2008}:
\begin{equation}
\mathcal{Z} = \left( \begin{array}{ccc}
        \mathcal{O}(t^2 \varepsilon) & \mathcal{O}(t^2 \varepsilon^2) & \mathcal{O}(t^2 \varepsilon^2)\\
        \mathcal{O}(t \varepsilon^2) & \mathcal{O}(t \varepsilon) &  \mathcal{O}(t \varepsilon^2)\\
        \mathcal{O}(\varepsilon^2) & \mathcal{O}(\varepsilon^2) & \mathcal{O}(\varepsilon)
\end{array}
\right)
\label{hatM}
\end{equation}
where each matrix element is known only up to an unknown order-one dimensionless coefficient. There are two independent symmetry breaking parameters. The parameter
$t=\langle\theta\rangle/\Lambda$ controls the charged lepton mass hierarchy and $\varepsilon= v_T/\Lambda$ describes the breaking of $A_4$. 
Notice that the uncertainty in the overall scale of the matrix elements ${\cal Z}_{ij}$ is related to the parameter $\varepsilon$ and 
is much smaller than the corresponding uncertainty in MFV.
We can see that MDMs and EDMs arise at the first order in the parameter $\varepsilon$. By assuming that the unknown coefficients have absolute values and phases of order one, from eqs. (\ref{dm}) and (\ref{hatM}) we have:
\begin{equation}
a_i=\mathcal{O}\left(2\dd\frac{{m_i^{\tt ch}}^2}{\Lambda_{NP}^2}\right)~~~,~~~~~~~~~~d_i=\mathcal{O}\left(e\dd\frac{m_i^{\tt ch}}{\Lambda_{NP}^2}\right)~~~.
\label{oom}
\end{equation}
From the
existing limits on MDMs and EDMs and by using eqs. (\ref{oom}) as exact equalities we find the results shown in table 7.
\begin{table}[!ht] 
\centering
                \begin{tabular}{|c|c|}
                    \hline
                    & \\[-9pt]
                    $d_e<1.6\times 10^{-27}~~e~cm$&$\Lambda_{NP}>80~~{\rm TeV}$\\[3pt]
                    \hline
                    &\\[-9pt]
                   $d_\mu<2.8\times 10^{-19}~~e~cm$&$\Lambda_{NP}>80~~{\rm GeV}$\\[3pt]
                    \hline
                    &\\[-9pt]
                    $\delta a_e<3.8\times 10^{-12}$&$\Lambda_{NP}>350~~{\rm GeV}$\\[3pt]
                    \hline
                    &\\[-9pt]
                    $\delta a_\mu= 302\pm88\times 10^{-11}$&$\Lambda_{NP}\approx 2.7~~{\rm TeV}$\\[3pt]
                    \hline
                \end{tabular}
            \caption{Experimental limits on lepton MDMs and EDMs and corresponding bounds on the scale $\Lambda_{NP}$, derived from eq. (\ref{oom}).
The data on the $\tau$ lepton have not been reported since they are much less constraining. For the anomalous magnetic moment of the muon,
$\delta a_\mu$ stands for the deviation of the experimental central value from the SM expectation \cite{Bennett:2004,Passera:2009}.}
            \end{table}
\vskip 0.2cm
\noindent
Concerning the flavor violating dipole transitions, from eq. ({\ref{hatM}) we see that the dominant contribution to the rate for $l_i\to l_j\gamma$ is 
given by:
\begin{equation}
\frac{BR(l_i\to l_j\gamma)}{BR(l_i\to l_j\nu_i{\bar \nu_j})}=\frac{48\pi^3 \alpha}{G_F^2 \Lambda_{NP}^4}\vert w_{ij} ~\varepsilon\vert^2
\label{LFV}
\end{equation}
where $w_{ij}$ are numbers of order one.
As a consequence, the branching ratios of the three
transitions $\mu\to e\gamma $, $\tau\to\mu\gamma$ and $\tau\to e\gamma$ are all expected be of the same order:
\begin{equation}
BR(\mu\to e \gamma)\approx BR(\tau\to\mu\gamma)\approx BR(\tau\to e \gamma)~~~.
\label{equalbr}
\end{equation}
This is a distinctive feature of this class of models. 
 Given the present experimental bound on $BR(\mu\to e \gamma)$,
eq. (\ref{equalbr}) implies that $\tau\to\mu\gamma$ and $\tau\to e \gamma$ have rates much below the present and expected future sensitivity.
Moreover, from the current (future) experimental limit on $BR(\mu\to e \gamma)$ \cite{Brooks:1999,Adam:2009,Meg} and assuming $\vert w_{\mu e}\vert=1$, we derive the following bound
on $\vert \varepsilon/\Lambda_{NP}^2\vert$:
\begin{equation}
BR(\mu\to e \gamma)<1.2\times 10^{-11}~(10^{-13})~~~~~~~
\left\vert\dd\frac{\varepsilon}{\Lambda_{NP}^2}\right\vert<1.2\times 10^{-11}~(1.1\times 10^{-12})~~{\rm GeV}^{-2}~~~.
\end{equation}
\noindent
Taking two extreme values for the parameter $\vert \varepsilon\vert$ we find
\begin{eqnarray}
\Lambda_{NP}>20~(67)~~{\rm TeV}~&[\vert \varepsilon\vert = 0.005]\nn\\
\Lambda_{NP}>65~(210)~~{\rm TeV}&~~~[\vert \varepsilon\vert = 0.05]~~~.
\end{eqnarray}
This model also allows for four-fermion operators that are not suppressed by any power of the small parameter $t$ or $\varepsilon$ and that violate the individual lepton numbers $L_i$ \cite{Feruglio:2010qu}.
They are all characterized by the selection rule
$\Delta L_e \Delta L_\mu \Delta L_\tau=2$. For instance, one such operator is
\be
({\bar l} l)'({\bar l} l)''=
\left[{\bar l}_e l_\tau {\bar l}_\mu l_\tau+
{\bar l}_\mu l_e {\bar l}_\tau l_e+
{\bar l}_\tau l_\mu {\bar l}_e l_\mu+h.c.\right]+...~~~.
\ee
where dots stand for additional flavor conserving contributions.
These operators can contribute to LFV decays such as $\tau^-\to \mu^+ e^- e^-$,
$\tau^-\to e^+ \mu^- \mu^-$ and their conjugate, whose branching ratios have upper bounds of the order of $10^{-7}$\cite{Amsler:2008}.
Through a rough dimensional estimate we find a lower bound on the scale $\Lambda_{NP}$ of the order of $15$ TeV.
From the previous considerations we see that, even invoking a cancellation in the imaginary part
of ${\cal{Z}}_{ee}$ to suppress the contribution to the electron EDM, it is difficult to avoid the conclusion that the scale
$\Lambda_{NP}$ should lie considerably above the TeV range. We recall that if the operator in eq. (\ref{leff})
originates from one-loop diagrams via the exchange of weakly interacting particles of masses $m_{NP}$,
then in our normalization a lower bound on $\Lambda_{NP}$ of 20 TeV corresponds to a lower bound on $m_{NP}$
of about $g_{NP}\Lambda_{NP}/(4\pi)\approx1~$ TeV, assuming $g_{NP}$ similar to the SU(2) gauge coupling.

All the previous estimates are based on an effective Lagrangian approach, with no explicit reference to the 
dynamics at the scale $\Lambda_{NP}$. If the degrees of freedom associated to the new physics at the scale $\Lambda_{NP}$ 
and their interactions are known, it is possible to directly compute the amplitudes of interest.
For instance, the SUSY model of Sect. 4 can be completed by adding a set of soft SUSY breaking terms,
which are constrained by the invariance under $G_f=A_4\times Z_3\times U(1)_{FN}$ and its pattern of symmetry breaking \cite{Ishimori:2008,Hayakawa:2009,feruglio:2009a,Ding:2009b}.
LFV amplitudes arise at one-loop level, via exchange of sleptons, charginos and neutralinos with masses of order $m_{SUSY}$.
An explicit computation of $BR(l_i\to l_j\gamma)$ confirms both the predictions of eq. (\ref{equalbr}) and the behaviour of eq. (\ref{LFV}), with $\Lambda_{NP}=(4\pi/g) m_{SUSY}$
.
The coefficients $w_{ij}$ are typically of ${\cal O}(0.1)$. When $\varepsilon$ is small, which also entails small $\tan\beta$ in our model,
relatively light SUSY particles are allowed, while for $\varepsilon$ close to its upper limit, 0.05, SUSY particle masses of several hundred
GeV or close to the TeV are needed to satisfy the present bound on $BR(\mu\to e \gamma)$, particularly if $\tan\beta$ is larger than 10. 
In either case there is only a very limited region of the parameter space where it is possible to explain the observed discrepancy in the muon MDM and to satisfy at the same time the current limit on
$BR(\mu\to e \gamma)$.
An interesting special case is that of universal SUSY breaking terms, giving rise to a cancellation in the elements of ${\cal Z}_{ij}$
below the diagonal \cite{feruglio:2009b,feruglio:2009a}. Under these circumstances $BR(l_i\to l_j\gamma)$ scale as $\varepsilon^4$ rather than as $\varepsilon^2$, with the possibility of much lighter SUSY particles.
In SUSY $A_4$ models also LFV 4-fermion operators are depleted by powers of $\varepsilon$ and the corresponding bounds on $m_{SUSY}$ are relaxed.

In the model discussed in Sect. 8, with $G_f=S_4\times Z_4\times U(1)_{FN}$, the matrix ${\cal Z}$ is given by \cite{masiero:2009}:
\begin{equation}
\mathcal{Z} = \left( \begin{array}{ccc}
        \mathcal{O}(t^2 v^2) & \mathcal{O}(t^2 v^2 v') & \mathcal{O}(t^2 v^2 v')\\
        \mathcal{O}(t v v') & \mathcal{O}(t v) &  \mathcal{O}(t v v'^2)\\
        \mathcal{O}(v v') & \mathcal{O}(v v'^2) & \mathcal{O}(v)
\end{array}
\right)
\label{hatM2}
\end{equation}
Predictions for EDMs and MDMs and corresponding bounds are similar to those discussed above in the case of the $A_4$ model and
summarized in Table 7. Concerning the radiative decays of the charged leptons we find that $R_{\mu e}$ and $R_{\tau e}$
scale as $v'^2/\Lambda_{NP}^4$, whereas $R_{\tau\mu}$ scales as $v'^4/\Lambda_{NP}^4$. In this case the symmetry breaking parameter 
$v'$ is considerably larger than the parameter $\varepsilon$ of the $A_4$ model and this
gives rise to more restrictive bounds on the scale of new physics $\Lambda_{NP}$.
From $BR(\mu\to e \gamma)<1.2\times 10^{-11}~(10^{-13})$ we get:
\begin{eqnarray}
\Lambda_{NP}>90~(300)~~{\rm TeV}~&[v' = 0.1]\nn\\
\Lambda_{NP}>130~(430)~~{\rm TeV}&~~~[v' = 0.2]~~~.
\end{eqnarray}
The model also predicts:
\be
BR(\mu\to e \gamma)\approx BR(\tau\to e\gamma)\gg BR(\tau\to \mu \gamma)~~~.
\ee

Summarizing, in models with discrete flavor symmetries LFV processes are generically suppressed by the presence of small symmetry breaking
parameters. However such a suppression is not completely efficient, at least in the explored models, to guarantee 
a scale of new physics close to the TeV. The best case is the one of the $A_4$ model, thanks to the very small expansion parameter
$\varepsilon$. In specific SUSY realizations of the $A_4$ symmetry the present limits on the branching ratios of LFV processes
still allow for a relatively light spectrum of superparticles, in a region of masses of interest to LHC.

%
%
\section{Leptogenesis}
The violation of $B-L$ implied by the see-saw mechanism suggests an interesting link between neutrino physics and the mechanism that produced
the observed baryon asymmetry in the early universe. If the baryon asymmetry originates well above the electroweak scale, $B-L$ violation represents a necessary condition, 
since any initial $B+L$ asymmetry would be erased in the subsequent evolution of the universe.
According to leptogenesis the asymmetry is determined by the CP violating, out-of-equilibrium decays of the heavy
RH neutrinos \cite{Fukugita:1986}. Through $B-L$ non-conservation of neutrino interactions, the asymmetry is first generated in the leptonic number and then partly converted into the observed baryonic one
via sphaleron interactions. Depending on whether the relevant decays occur at a sufficiently high temperature or not, we have an unflavored regime, where the leptons
in the final state are indistinguishable, or a flavored regime, where the specific interactions of the different leptons in the decay products cannot be neglected \cite{Abada:2006,Nardi:2006}.
It is also quite remarkable that, at least in its simplest implementation, leptogenesis requires light neutrino masses below the eV scale \cite{Buchmuller:2003,Buchmuller:2004,Giudice:2004,Buchmuller:2005},
in a range which is fully compatible with  other experimental constraints.
Unfortunately, without any additional assumptions, it is difficult to promote this elegant picture into a testable theory, due to the large number of independent parameters of the see-saw model.

Models of lepton masses based on flavor symmetries typically depend on a restricted number of parameters,
thus opening the interesting possibility of relating the baryon asymmetry
to other low-energy observables.
As a general rule, to provide a realistic description of lepton masses and mixing angles
the flavor symmetry should always be broken. The breaking is described by a set of small dimensionless 
quantities $\varepsilon$, which provide efficient expansion parameters. 
As we have seen in the previous sections, small observable quantities such as charged lepton mass
ratios, $\theta_{13}$, $\theta_{23}-\pi/4$ can be expanded in power series of $\varepsilon$, and the predictions are
dominated by the lowest (positive) power. 

In the context of leptogenesis, given the extreme smallness of the baryon asymmetry \cite{Komatsu:2009}
\be\label{etaBobs}
\eta_B^{\rm CMB} = (6.2 \pm 0.15)\times 10^{-10} \, ,
\ee
it can be convenient, at least in a certain regime, that the $C\!P$ asymmetries in the RH
neutrino decays are also suppressed by powers of $\varepsilon$.
If the baryon asymmetry is dominated by the decay of a single RH neutrino, we can write
\footnote{We will denote the $C\!P$ asymmetries with $\xi$ and we keep the letter $\varepsilon$ to indicate the generic expansion parameter of a spontaneously broken flavor symmetry.}:
\be
\eta_B= d~ \xi~ k
\ee 
where $d$ describes the combined effect of sphaleron conversion and dilution from photon production,
$\xi$ is the relevant $C\!P$ asymmetry and $k$ takes into account the wash-out effects.
Typically we expect a dilution factor $d$ of order
$10^{-2}$ and, barring fine-tuning of the parameters, a wash-out factor $k$ in the range $10^{-3}\div 10^{-2}$,
which favors $\xi$ around $10^{-6}\div 10^{-5}$. 
Such $C\!P$ asymmetry arises from the interference of the tree-level and the one-loop decay amplitudes and depends
quadratically on the neutrino Yukawa couplings. In models like the ones discussed in Sects. 4 and 8,
where the RH neutrino masses are very large, close to $10^{14}$ GeV, and the corresponding neutrino Yukawa couplings are 
of $\mathcal{O}(1)$, a rough estimate of the total $C\!P$ asymmetry would give $\xi=\mathcal{O}(1/(8\pi))$, by far too large
compared to $10^{-6}\div 10^{-5}$. It is therefore interesting to analyze under which conditions the $C\!P$ asymmetries
vanish in the limit of exact symmetry, so that the first non-vanishing contribution is given by some power of the symmetry breaking parameters 
$\varepsilon$. If the $C\!P$ asymmetry relevant for leptogenesis is suppressed by powers of $\varepsilon$,
this opens the very interesting possibility of relating the observed baryon asymmetry $\eta_B$ to other low-energy observable quantities  \cite{Lin:2009a, Mohapatra:2005a, Mohapatra:2005b}
such as $\theta_{13}$, $\theta_{23}-\pi/4$, $BR(l_i\to l_j \gamma)$. 

The total $C\!P$ asymmetries 
in the decay of a RH neutrino $\nu^c_i$ are
\be
\xi_i=\dd\frac{\Gamma_i-\ov{\Gamma}_i}{\Gamma_i+\ov{\Gamma}_i}
\ee
where $\Gamma_i$ $(\ov{\Gamma}_i)$ is the decay rate of $\nu^c_i$ into leptons (antileptons).
In the flavored regime the relevant asymmetries $\xi_{if}$ involve final states
with a specific lepton flavor $f$. The flavored regime takes place for $M_i\le c~10^{12}$ GeV
where $c=1$ $(1+\tan^2\beta)$ in the ordinary (SUSY) case. The unflavored regime occurs for RH neutrino masses
above that threshold.
At one-loop we have~\cite{Covi:1996}:
\begin{eqnarray}
  \label{eq:flCPasymm}
 \xi_{if} &=&
\frac{1}{8\pi \hcY_{ii}}\sum_{j\neq i}
\left\{ {\rm Im}\left[ \hcY_{ij} \hat Y_{if} \hat Y^*_{jf}\right] f_{ij} +
{\rm Im}\left[ \hcY_{ji} \hat Y_{if}\hat Y^*_{jf}\right] g_{ij}\right\} \\
  \label{eq:CPasymm}
 \xi_{i} &=& \sum_f
 \xi_{if} =
\frac{1}{8\pi \hcY_{ii}}\sum_{j\neq i}
 {\rm Im} \left[\hcY_{ij}^2\right] f_{ij} \, ,
  \end{eqnarray}
where $\cY$ is a combination of the neutrino Yukawa couplings $Y=m_\nu^D/v_u$
\begin{equation}
  \label{eq:YY}
  \cY_{ij} = \left(Y Y^\dagger\right)_{ij}~~~.
\end{equation}
and the hat in eqs. (\ref{eq:flCPasymm},\ref{eq:CPasymm}) denotes a basis where the mass matrix $M$ of heavy Majorana neutrinos and that of charged leptons, $m_l$, are diagonal.
The functions $f_{ij}$ and $g_{ij}$ depend on the mass ratios of the RH neutrino masses $M_i$. 
From eqs. (\ref{eq:flCPasymm},\ref{eq:CPasymm}) we see that both $\xi_{if}$ and $\xi_i$ vanish if $\hcY$ is diagonal. The total
asymmetries $\xi_i$ vanish also if $\hcY$ has real non-diagonal entries.
A necessary and sufficient condition for a diagonal $\hcY$ is:
\be
{\cY} M-M {\cY}^T =0~~~,
\label{com1}
\ee
where the matrices $\cY$ and $M$ are evaluated in any basis.

If the model is invariant under the action of a flavor symmetry group $G_f$ we have an interesting sufficient condition for the vanishing of the $C\!P$ asymmetries.
If the heavy RH neutrinos transform in a (three-dimensional) irreducible representation of $G_f$, then in the limit of exact symmetry, 
where the symmetry breaking parameters $\varepsilon$ go to zero, all $C\!P$ asymmetries vanish \cite{Bertuzzo:2009}. In this limit it is possible to show that $\cY$ becomes proportional to the unit matrix
as a consequence of a completely general group theoretical property. Thus, from eqs. (\ref{eq:flCPasymm},\ref{eq:CPasymm}) we conclude that
the asymmetries $\xi_i$ and $\xi_{if}$ vanish. Notice that irreducible representations
of dimension larger than one are only possible if $G_f$ is non-abelian. Beyond the symmetry limit, in general $\hcY$ gets corrections and develops complex off-diagonal entries 
at some order $\varepsilon^p$. If the spectrum of RH neutrinos is non-degenerate in the symmetry limit, we expect $\xi_i={\cal O}(\varepsilon^{2p})$ and 
$\xi_{if}={\cal O}(\varepsilon^{p})$. Degeneracy of RH neutrinos can modify this behavior through the dependence on $\varepsilon$ of the functions $f_{ij}$ and $g_{ij}$.
This result applies to both the models described in Sects. 4 and 8, where the RH neutrinos transform in the three-dimensional representions
of $A_4$ and $S_4$, respectively. 
In the limit of exact flavor symmetry we find in both cases $\cY=|y|^2~{\bf 1}$ where ${\bf 1}$ denotes the identity matrix. This equality
holds in any basis, in particular in the mass eigenstate basis of RH neutrinos and we have $\xi_i=0$ in the
symmetry limit. In both models all RH neutrino are very heavy, with masses well above $10^{12}$ GeV, and the unflavored regime applies.

In the $A_4$ model of Sect. 4, $\hcY$ acquires complex off-diagonal entries of order $\varepsilon\approx v_T/\Lambda$. 
The $C\!P$ asymmetries $\xi_i$ depend only on three real parameters: two independent real symmetry breaking parameters $\varepsilon_i$ and the lightest neutrino mass. 
In particular there is only one independent phase which is determined by the lightest neutrino mass up to an overall sign. 
We have approximately \cite{Jenkins:2008}
\be
\xi_i\approx \dd\frac{\varepsilon^2}{8\pi}
\label{cpa}
\ee
More precisely \cite{Bertuzzo:2009,Hagedorn:2009,Riva:2010}, for normal ordering of the neutrino mass spectrum all asymmetries $\xi_i$ are of the same order of magnitude.
For inverted ordering the two asymmetries $\xi_{1,2}$ get enhanced 
compared to the approximate estimate of eq. (\ref{cpa}) by a factor $~10^{3}$ coming from the functions $f_{12}$ and $f_{21}$, as a result of the near degeneracy 
of two heavy RH neutrinos. To reproduce the observed baryon asymmetry, eq. (\ref{etaBobs}), different wash-out effects are required in
the two cases. In the case of normal ordering the experimental value in eq. (\ref{etaBobs}) is obtained when the parameter $\varepsilon$ is in its natural window,
$5\times 10^{-3}\div 5\times 10^{-2}$, for a wide range of neutrino Yukawa couplings $y$. 
For inverted ordering a much larger wash-out suppression is needed. 
When $\varepsilon$ falls in the optimal range $5\times 10^{-3}\div 5\times 10^{-2}$
this can be accommodated by restricting both $y\times\sin\beta$ and $m_3$ in a limited range.
It is quite remarkable that in both cases the range of the symmetry breaking
parameter $\varepsilon$ suggested by the constraints on lepton masses and mixing angles corresponds to that required to get the observed
baryon asymmetry through leptogenesis. 

In the $S_4$ model discussed in Sect. 8,  $\hcY$ acquires complex off-diagonal entries at the order $v^4/v'$ and the 
$C\!P$ asymmetries are expected to be of order $v^8/(v'^2 8\pi)$. Assuming a typical wash-out suppression of order $10^{-2}$, 
the observed baryon asymmetry can be obtained for values of $(v,v')$ close to the range selected to fit charged lepton masses
and mixing angles.

Another class of models where the $C\!P$ asymmetries vanish is the one of type I see-saw models where the Dirac
and Majorana neutrino mass matrix $m_\nu^D$ and $M$ as well as their see-saw combination $m_\nu$ are form-diagonalizable.
A matrix $A$ depending on a set of parameters $\alpha_i$ is said to be form-diagonalizable \cite{Low:2003} if it is diagonalized by unitary transformations 
that do not depend on $\alpha_i$:
\be
U_L^\dagger A(\alpha) U_R = A^d(\alpha)
\ee
where $A^d(\alpha)$ is diagonal and the unitary matrices $U_{L,R}$ are independent from $\alpha$. Examples of form-diagonalizable matrix are $m_\nu$
in eqs. (\ref{1k1},\ref{4k1}) and  (\ref{1k},\ref{4k}). The parameters are the eigenvalues $m_{1,2,3}$ and the diagonalizing matrices are $U_R=U_L^*=U_{TB}$ and
$U_R=U_L^*=U_{BM}$, respectively.
As we have seen in section 2, form-diagonalizable matrices naturally arise in the context of models with discrete flavor symmetries.
It is possible to show that if in a type I see-saw $m_\nu^D$, $M$ and $m_\nu$ are all form-diagonalizable, then the matrix $\hcY$ is diagonal 
and the $C\!P$ asymmetries vanish \cite{Aristizabal:2009,Felipe:2009}. In realistic models $m_\nu^D$, $M$ and $m_\nu$ are typically form-diagonalizable only in some symmetry
limit. Symmetry breaking terms usually spoil this property and allow for small non-vanishing $C\!P$ asymmetries.

So far we have discussed the regime of large RH masses and large neutrino Yukawa couplings.
When the smallest RH neutrino mass is below the so-called Davidson-Ibarra bound \cite{Davidson:2002} $(4\times 10^8\div 2\times 10^{9})$ GeV, and we are
in the regime of strong hierarchy among RH neutrino masses, the $C\!P$ asymmetry 
associated to the lightest RH neutrino decay is too small to allow for a successful leptogenesis.
To evade the Davidson-Ibarra bound we should depart from the strong hierarchical regime. Under certain conditions
a significant enhancement of the $C\!P$ asymmetry can be achieved even for RH neutrino mass ratios as small as 0.1 \cite{Hambye:2004,Raidal:2006}. 
Alternatively, we can exploit the regime of resonant leptogenesis \cite{Pilaftsis:1997}, occurring when the decaying RH neutrino is quasi-degenerate in mass with some other RH neutrino, the mass differences being comparable with
the RH neutrino decay width. A quasi-degeneracy of the RH neutrino spectrum is better understood and dynamically controlled in the presence of an underlying flavor symmetry.
Several symmetries have been proposed in the literature such as $G_f=SU(3)_{e^c}\times SU(3)_l\times O(3)_{\nu^c}$ in minimal lepton flavor violation \cite{Cirigliano:2005,Cirigliano:2006,Davidson:2006,Cirigliano:2007,Cirigliano:2007a,Branco:2007}
or $G_f=SO(3)$ in ref. \cite{Pilaftsis:2005}. In these two examples the light neutrino masses and their mixing angles
are not explained but just accommodated. An interesting model based on a flavor symmetry group $G_f=A_4\times Z_3\times Z_4$ 
is that of ref. \cite{Branco:2009}. Like the model discussed in Sect. 4 it predicts a lepton mixing close to TB.
Due to the presence of an additional discrete factor in the symmetry group, the RH neutrino spectrum is degenerate at LO,
and the degeneracy is lifted by radiative corrections or small soft breaking terms, allowing for successful resonant leptogenesis, for a wide range of RH neutrino masses.

\section{Summary and conclusion}

We have reviewed the motivation, the formalism and the implications of applying non abelian discrete flavor groups to the theory of neutrino mixing. The data on neutrino mixing are by now quite precise. It is a fact that, to a precision comparable with the measurement accuracy, the TB mixing pattern is well approximated by the data  (see Fig. (2)). If this experimental result is not a mere accident but a real indication that a dynamical mechanism is at work to guarantee the validity of TB mixing in the leading approximation, corrected by small non leading terms, then non abelian discrete flavor groups emerge as the main road to an understanding of this mixing pattern. Indeed the entries of the TB mixing matrix are clearly suggestive of "rotations" by simple, very specific angles. In fact the group $A_4$, the simplest group used to explain TB mixing, is specified by the set of those rotations that leave a regular tetrahedron invariant. We have started by recalling some basic notions about finite groups and then we have concentrated on those symmetries, like $A_4$ and $S_4$, that are found to be the main candidates for obtaining TB mixing. 
We have discussed the general mechanism that realizes TB mixing within the framework of discrete flavor symmetries. 
The symmetry is broken down to two different subgroups in the charged lepton sector and in the neutrino sector, and 
the mixing matrix arises from the mismatch between the two different residual simmetries. TB mixing requires
a flavor symmetry group possessing appropriate residual subgroups.
The breaking can be realized in a natural way through the specific vacuum alignments of a set of scalar  flavons. 
We have described a set of models where TB mixing is indeed derived at leading order within this mechanism.
There are many variants of such models (in particular with or without see-saw) with different detailed predictions for the spectrum of neutrino masses and for deviations from the TB values of the mixing angles. 
In general at NLO the different mixing angles receive corrections of the same order of magnitude, which are constrained to be small due to the experimental results which are very close to the TB values. 
Indeed the small experimental error on $\theta_{12}$, which nicely agrees with the value predicted by TB mixing, suggests that the NLO corrections should be of order of few percent, at most.
Additional symmetries are needed, typically of the $U(1)_{FN}$ or $Z_N$ type, in order to reproduce the mass hierarchy of charged leptons. In the neutrino sector there is no reason for the mass eigenvalues not to be of the same order in absolute value. Thus the smallness of the ratio $\sqrt{r} \sim 0.2$, where $r$ is defined in eq. (\ref{r}), is accidental in most of these models. Both normal and inverse hierarchy spectra can be realized. The phenomenology of the models was summarized. We have also discussed the implications of models based on discrete flavor groups for lepton flavor violation and for leptogenesis. Lepton flavor violating processes, the muon g-2 and the EDM's of leptons impose strong constraints on every new physics model. This is also true for the models considered here. But the specific suppression factors and selection rules induced by the finite flavor symmetry group, in particular by $A_4$, may help to improve the consistency of the model even in the presence of new physics at the TeV scale. The observed baryon asymmetry in the Universe, explained in terms of leptogenesis from the decay of heavy Majorana neutrinos, is found to be compatible with models based on discrete groups. Neutrino Yukawa couplings of order one and RH neutrino masses of order $10^{14}\div 10^{15}$ GeV
would typically lead to CP asymmetries too large to reproduce the observed baryon asymmetry. However, as a consequence of a general group theoretical property, in all models where the three RH neutrinos transform
in a single irreducible representation of the flavor group, the unflavored CP asymmetries vanish in the limit of exact symmetry and small values can be generated through NLO corrections.

An obvious question is whether some additional indication for discrete flavor groups can be obtained by considering the extension of the models to the quark sector, perhaps in a Grand Unified context. The answer appears to be that, while the quark masses and mixings can indeed be reproduced in models where TB mixing is realized  in the leptonic sector through the action of discrete groups, there are no specific additional hints in favour of discrete groups that come from the quark sector. Examples of Grand Unified descriptions of all fermion masses and mixings with TB mixing for neutrinos have been produced and have been discussed in this review. For quarks, only the third generation masses are present at leading order in these models. The other entries of the mass matrices are small due to additional symmetries or other dynamical reasons (for example, suppression factors from extra dimensions), and the small mass ratios and the small mixing angles are generated by these corrective effects and are not due to the discrete group. As a consequence, the action of the discrete flavor group is only clearly manifest among the comparable neutrino sector masses, in the basis where charged leptons are diagonal.

Different forms of neutrino mixing other than TB mixing are also amenable to a description in terms of discrete groups. In alternative to TB mixing, in sect. 8 we have discussed the possibility that actually a more appropriate starting point, could be BM mixing, corrected by large terms of $O(\lambda_C)$, with $\lambda_C$ being the Cabibbo angle ("weak complementarity"), arising from the diagonalization of charged leptons. By suitably modifying the construction in terms of discrete groups adopted in the case of TB mixing, we have identified the group $S_4$ as a good candidate to also provide, in a different presentation, the basis for naturally obtaining BM mixing in first approximation.  In the model described the NLO terms are such that the dominant corrections only affect $\theta_{12}$ and $\theta_{13}$ (which receive $O(\lambda_C)$ shifts), while $\theta_{23}$ receives smaller corrections. A value of $\theta_{13}$ near the present bound would support this possibility.

In the near future the improved experimental precision on neutrino mixing angles, in particular on $\theta_{13}$, could make the case for TB mixing stronger and then, as a consequence, also the case for discrete flavor groups would be strenghtened. Further important input could come from the LHC. In fact, new physics at the weak scale could have important feedback on the physics of neutrino masses and mixing.

\section*{Acknowledgements}
We recognize that this work has been partly supported by the Italian Ministero 
dell'Universit\`a e della Ricerca Scientifica, under the COFIN program (PRIN 2008) 
and by the European Commission
under the networks "Heptools" and "Quest for Unification" and contracts, MRTN-CT-2006-035505 and  PITN-GA-2009-237920 (UNILHC).
\vfill
\newpage



\begin{thebibliography}{100}
\expandafter\ifx\csname url\endcsname\relax
  \def\url#1{{\tt #1}}\fi
\expandafter\ifx\csname urlprefix\endcsname\relax\def\urlprefix{URL }\fi
\providecommand{\eprint}[2][]{\url{#2}}

\bibitem{Altarelli:2004}
Altarelli G and Feruglio F 2004 {\em New J. Phys.\/} {\bf 6} 106
  (\textit{Preprint} \eprint{hep-ph/0405048})

\bibitem{Altarelli:2009}
Altarelli G 2009 {\em Nuovo Cim.\/} {\bf 032C} 91--102 (\textit{Preprint}
  \eprint{0905.3265})

\bibitem{Mohapatra:2006}
Mohapatra R~N and Smirnov A~Y 2006 {\em Ann. Rev. Nucl. Part. Sci.\/} {\bf 56}
  569--628 (\textit{Preprint} \eprint{hep-ph/0603118})

\bibitem{Mohapatra:2007}
Mohapatra R~N {\em et~al.\/} 2007 {\em Rept. Prog. Phys.\/} {\bf 70} 1757--1867
  (\textit{Preprint} \eprint{hep-ph/0510213})

\bibitem{Grimus:2006}
Grimus W 2006 {\em PoS\/} {\bf P2GC} 001 (\textit{Preprint}
  \eprint{hep-ph/0612311})

\bibitem{Gonzalez-Garcia:2008}
Gonzalez-Garcia M~C and Maltoni M 2008 {\em Phys. Rept.\/} {\bf 460} 1--129
  (\textit{Preprint} \eprint{0704.1800})

\bibitem{Super-Kamiokande:2006}
Hosaka J {\em et~al.\/} (Super-Kamkiokande) 2006 {\em Phys. Rev.\/} {\bf D73}
  112001 (\textit{Preprint} \eprint{hep-ex/0508053})

\bibitem{Super-Kamiokande:2008}
Cravens J~P {\em et~al.\/} (Super-Kamiokande) 2008 {\em Phys. Rev.\/} {\bf D78}
  032002 (\textit{Preprint} \eprint{0803.4312})

\bibitem{SNO:2009}
Aharmim B {\em et~al.\/} (SNO) 2010 {\em Phys. Rev.\/} {\bf C81} 055504
  (\textit{Preprint} \eprint{0910.2984})

\bibitem{Super-Kamiokande:2006a}
Abe K {\em et~al.\/} (Super-Kamiokande) 2006 {\em Phys. Rev. Lett.\/} {\bf 97}
  171801 (\textit{Preprint} \eprint{hep-ex/0607059})

\bibitem{Super-Kamiokande:2006b}
Hosaka J {\em et~al.\/} (Super-Kamiokande) 2006 {\em Phys. Rev.\/} {\bf D74}
  032002 (\textit{Preprint} \eprint{hep-ex/0604011})

\bibitem{KamLAND:2008}
Abe S {\em et~al.\/} (KamLAND) 2008 {\em Phys. Rev. Lett.\/} {\bf 100} 221803
  (\textit{Preprint} \eprint{0801.4589})

\bibitem{K2K:2006}
Ahn M~H {\em et~al.\/} (K2K) 2006 {\em Phys. Rev.\/} {\bf D74} 072003
  (\textit{Preprint} \eprint{hep-ex/0606032})

\bibitem{Kafka:2010zz}
Kafka T (MINOS) 2010 {\em Prog. Part. Nucl. Phys.\/} {\bf 64} 184--186

\bibitem{Agafonova:2010dc}
Agafonova N {\em et~al.\/} (OPERA) 2010  (\textit{Preprint} \eprint{1006.1623})

\bibitem{LSND:1996}
Athanassopoulos C {\em et~al.\/} (LSND) 1996 {\em Phys. Rev. Lett.\/} {\bf 77}
  3082--3085 (\textit{Preprint} \eprint{nucl-ex/9605003})

\bibitem{LSND:1998}
Athanassopoulos C {\em et~al.\/} (LSND) 1998 {\em Phys. Rev.\/} {\bf C58}
  2489--2511 (\textit{Preprint} \eprint{nucl-ex/9706006})

\bibitem{LSND:1998a}
Athanassopoulos C {\em et~al.\/} (LSND) 1998 {\em Phys. Rev. Lett.\/} {\bf 81}
  1774--1777 (\textit{Preprint} \eprint{nucl-ex/9709006})

\bibitem{KARMEN:2002}
Armbruster B {\em et~al.\/} (KARMEN) 2002 {\em Phys. Rev.\/} {\bf D65} 112001
  (\textit{Preprint} \eprint{hep-ex/0203021})

\bibitem{MiniBooNE:2009}
Aguilar-Arevalo A~A {\em et~al.\/} (MiniBooNE) 2009 {\em Phys. Rev. Lett.\/}
  {\bf 102} 101802 (\textit{Preprint} \eprint{0812.2243})

\bibitem{MiniBooNE:2009a}
Aguilar-Arevalo A~A {\em et~al.\/} (MiniBooNE) 2009 {\em Phys. Rev. Lett.\/}
  {\bf 103} 111801 (\textit{Preprint} \eprint{0904.1958})

\bibitem{Strumia:2006}
Strumia A and Vissani F 2006  (\textit{Preprint} \eprint{hep-ph/0606054})

\bibitem{Gonzalez-Garcia:2008a}
Gonzalez-Garcia M~C and Maltoni M 2008 {\em Phys. Lett.\/} {\bf B663} 405--409
  (\textit{Preprint} \eprint{0802.3699})

\bibitem{Bandyopadhyay:2008}
Bandyopadhyay A, Choubey S, Goswami S, Petcov S~T and Roy D~P 2008
  (\textit{Preprint} \eprint{0804.4857})

\bibitem{Fogli:2008}
Fogli G~L, Lisi E, Marrone A, Palazzo A and Rotunno A~M 2008 {\em Phys. Rev.
  Lett.\/} {\bf 101} 141801 (\textit{Preprint} \eprint{0806.2649})

\bibitem{Fogli:2008a}
Fogli G~L, Lisi E, Marrone A, Palazzo A and Rotunno A~M 2008
  (\textit{Preprint} \eprint{0809.2936})

\bibitem{Schwetz:2008}
Schwetz T, Tortola M~A and Valle J~W~F 2008 {\em New J. Phys.\/} {\bf 10}
  113011 (\textit{Preprint} \eprint{0808.2016})

\bibitem{Maltoni:2008}
Maltoni M and Schwetz T 2008 {\em PoS\/} {\bf IDM2008} 072 (\textit{Preprint}
  \eprint{0812.3161})

\bibitem{Lesgourgues:2006}
Lesgourgues J and Pastor S 2006 {\em Phys. Rept.\/} {\bf 429} 307--379
  (\textit{Preprint} \eprint{astro-ph/0603494})

\bibitem{Avignone:2008}
Avignone III F~T, Elliott S~R and Engel J 2008 {\em Rev. Mod. Phys.\/} {\bf 80}
  481--516 (\textit{Preprint} \eprint{0708.1033})

\bibitem{Kraus:2005}
Kraus C {\em et~al.\/} 2005 {\em Eur. Phys. J.\/} {\bf C40} 447--468
  (\textit{Preprint} \eprint{hep-ex/0412056})

\bibitem{Fogli:2008b}
Fogli G~L {\em et~al.\/} 2008 {\em Phys. Rev.\/} {\bf D78} 033010
  (\textit{Preprint} \eprint{0805.2517})

\bibitem{Weinberg:1979}
Weinberg S 1979 {\em Phys. Rev. Lett.\/} {\bf 43} 1566--1570

\bibitem{Minkowski:1977}
Minkowski P 1977 {\em Phys. Lett.\/} {\bf B67} 421

\bibitem{Yanagida:1979}
Yanagida T 1979  In Proceedings of the Workshop on the Baryon Number of the
  Universe and Unified Theories, Tsukuba, Japan, 13-14 Feb 1979

\bibitem{Gell-Mann:1979}
Gell-Mann M, Ramond P and Slansky R 1979  Print-80-0576 (CERN)

\bibitem{Glashow:1980}
Glashow S~L 1980  In Quarks and Leptons, Cargese, ed. M. Levy et al., Plenum,
  1980 New York, p. 707

\bibitem{Mohapatra:1980}
Mohapatra R~N and Senjanovic G 1980 {\em Phys. Rev. Lett.\/} {\bf 44} 912

\bibitem{Feruglio:2002}
Feruglio F, Strumia A and Vissani F 2002 {\em Nucl. Phys.\/} {\bf B637}
  345--377 (\textit{Preprint} \eprint{hep-ph/0201291})

\bibitem{Harrison:2002}
Harrison P~F, Perkins D~H and Scott W~G 2002 {\em Phys. Lett.\/} {\bf B530} 167
  (\textit{Preprint} \eprint{hep-ph/0202074})

\bibitem{Harrison:2002a}
Harrison P~F and Scott W~G 2002 {\em Phys. Lett.\/} {\bf B535} 163--169
  (\textit{Preprint} \eprint{hep-ph/0203209})

\bibitem{Harrison:2003}
Harrison P~F and Scott W~G 2003 {\em Phys. Lett.\/} {\bf B557} 76
  (\textit{Preprint} \eprint{hep-ph/0302025})

\bibitem{Harrison:2004}
Harrison P~F and Scott W~G 2004  (\textit{Preprint} \eprint{hep-ph/0402006})

\bibitem{Albright:2008}
Albright C~H and Rodejohann W 2008 {\em Phys. Lett.\/} {\bf B665} 378--383
  (\textit{Preprint} \eprint{0804.4581})

\bibitem{Bertolini:2006}
Bertolini S, Schwetz T and Malinsky M 2006 {\em Phys. Rev.\/} {\bf D73} 115012
  (\textit{Preprint} \eprint{hep-ph/0605006})

\bibitem{Plentinger:2008}
Plentinger F and Seidl G 2008 {\em Phys. Rev.\/} {\bf D78} 045004
  (\textit{Preprint} \eprint{0803.2889})

\bibitem{Frampton:1995}
Frampton P~H and Kephart T~W 1995 {\em Int. J. Mod. Phys.\/} {\bf A10}
  4689--4704 (\textit{Preprint} \eprint{hep-ph/9409330})

\bibitem{Ishimori:2010au}
Ishimori H {\em et~al.\/} 2010  (\textit{Preprint} \eprint{1003.3552})

\bibitem{Ma:2001}
Ma E and Rajasekaran G 2001 {\em Phys. Rev.\/} {\bf D64} 113012
  (\textit{Preprint} \eprint{hep-ph/0106291})

\bibitem{Ma:2002}
Ma E and Rajasekaran G 2001 {\em Phys. Rev.\/} {\bf D64} 113012
  (\textit{Preprint} \eprint{hep-ph/0106291})

\bibitem{Babu:2003}
Babu K~S, Ma E and Valle J~W~F 2003 {\em Phys. Lett.\/} {\bf B552} 207--213
  (\textit{Preprint} \eprint{hep-ph/0206292})

\bibitem{Hirsch:2004}
Hirsch M, Romao J~C, Skadhauge S, Valle J~W~F and Villanova~del Moral A 2004
  {\em Phys. Rev.\/} {\bf D69} 093006 (\textit{Preprint}
  \eprint{hep-ph/0312265})

\bibitem{Ma:2004}
Ma E 2004 {\em Phys. Rev.\/} {\bf D70} 031901 (\textit{Preprint}
  \eprint{hep-ph/0404199})

\bibitem{Ma:2004a}
Ma E 2004 {\em New J. Phys.\/} {\bf 6} 104 (\textit{Preprint}
  \eprint{hep-ph/0405152})

\bibitem{Chen:2005}
Chen S~L, Frigerio M and Ma E 2005 {\em Nucl. Phys.\/} {\bf B724} 423--431
  (\textit{Preprint} \eprint{hep-ph/0504181})

\bibitem{Altarelli:2005}
Altarelli G and Feruglio F 2005 {\em Nucl. Phys.\/} {\bf B720} 64--88
  (\textit{Preprint} \eprint{hep-ph/0504165})

\bibitem{Ma:2005}
Ma E 2005 {\em Phys. Rev.\/} {\bf D72} 037301 (\textit{Preprint}
  \eprint{hep-ph/0505209})

\bibitem{Hirsch:2005}
Hirsch M, Villanova~del Moral A, Valle J~W~F and Ma E 2005 {\em Phys. Rev.\/}
  {\bf D72} 091301 (\textit{Preprint} \eprint{hep-ph/0507148})

\bibitem{Babu:2005}
Babu K~S and He X~G 2005  (\textit{Preprint} \eprint{hep-ph/0507217})

\bibitem{Ma:2005a}
Ma E 2005 {\em Mod. Phys. Lett.\/} {\bf A20} 2601--2606 (\textit{Preprint}
  \eprint{hep-ph/0508099})

\bibitem{Zee:2005}
Zee A 2005 {\em Phys. Lett.\/} {\bf B630} 58--67 (\textit{Preprint}
  \eprint{hep-ph/0508278})

\bibitem{Ma:2006}
Ma E 2006 {\em Phys. Rev.\/} {\bf D73} 057304 (\textit{Preprint}
  \eprint{hep-ph/0511133})

\bibitem{He:2006}
He X~G, Keum Y~Y and Volkas R~R 2006 {\em JHEP\/} {\bf 04} 039
  (\textit{Preprint} \eprint{hep-ph/0601001})

\bibitem{Adhikary:2006}
Adhikary B, Brahmachari B, Ghosal A, Ma E and Parida M~K 2006 {\em Phys.
  Lett.\/} {\bf B638} 345--349 (\textit{Preprint} \eprint{hep-ph/0603059})

\bibitem{Altarelli:2006}
Altarelli G and Feruglio F 2006 {\em Nucl. Phys.\/} {\bf B741} 215--235
  (\textit{Preprint} \eprint{hep-ph/0512103})

\bibitem{Lavoura:2006}
Lavoura L and Kuhbock H 2007 {\em Mod. Phys. Lett.\/} {\bf A22} 181
  (\textit{Preprint} \eprint{hep-ph/0610050})

\bibitem{Ma:2007}
Ma E 2007 {\em Mod. Phys. Lett.\/} {\bf A22} 101--106 (\textit{Preprint}
  \eprint{hep-ph/0610342})

\bibitem{Hirsch:2007}
Hirsch M, Joshipura A~S, Kaneko S and Valle J~W~F 2007 {\em Phys. Rev. Lett.\/}
  {\bf 99} 151802 (\textit{Preprint} \eprint{hep-ph/0703046})

\bibitem{Altarelli:2007}
Altarelli G, Feruglio F and Lin Y 2007 {\em Nucl. Phys.\/} {\bf B775} 31--44
  (\textit{Preprint} \eprint{hep-ph/0610165})

\bibitem{Yin:2007}
Yin F 2007 {\em Phys. Rev.\/} {\bf D75} 073010 (\textit{Preprint}
  \eprint{0704.3827})

\bibitem{Bazzocchi:2008}
Bazzocchi F, Kaneko S and Morisi S 2008 {\em JHEP\/} {\bf 03} 063
  (\textit{Preprint} \eprint{0707.3032})

\bibitem{Bazzocchi:2008a}
Bazzocchi F, Morisi S and Picariello M 2008 {\em Phys. Lett.\/} {\bf B659}
  628--633 (\textit{Preprint} \eprint{0710.2928})

\bibitem{Honda:2008}
Honda M and Tanimoto M 2008 {\em Prog. Theor. Phys.\/} {\bf 119} 583--598
  (\textit{Preprint} \eprint{0801.0181})

\bibitem{Brahmachari:2008}
Brahmachari B, Choubey S and Mitra M 2008 {\em Phys. Rev.\/} {\bf D77} 073008
  (\textit{Preprint} \eprint{0801.3554})

\bibitem{Adhikary:2008}
Adhikary B and Ghosal A 2008 {\em Phys. Rev.\/} {\bf D78} 073007
  (\textit{Preprint} \eprint{0803.3582})

\bibitem{Hirsch:2008}
Hirsch M, Morisi S and Valle J~W~F 2008 {\em Phys. Rev.\/} {\bf D78} 093007
  (\textit{Preprint} \eprint{0804.1521})

\bibitem{Frampton:2008}
Frampton P~H and Matsuzaki S 2008  (\textit{Preprint} \eprint{0806.4592})

\bibitem{Csaki:2008}
Csaki C, Delaunay C, Grojean C and Grossman Y 2008 {\em JHEP\/} {\bf 10} 055
  (\textit{Preprint} \eprint{0806.0356})

\bibitem{Altarelli:2008}
Altarelli G, Feruglio F and Hagedorn C 2008 {\em JHEP\/} {\bf 03} 052--052
  (\textit{Preprint} \eprint{0802.0090})

\bibitem{Morisi:2009}
Morisi S 2009 {\em Phys. Rev.\/} {\bf D79} 033008 (\textit{Preprint}
  \eprint{0901.1080})

\bibitem{Lin:2009}
Lin Y 2009 {\em Nucl. Phys.\/} {\bf B813} 91--105 (\textit{Preprint}
  \eprint{0804.2867})

\bibitem{Lin:2009a}
Lin Y 2009 {\em Phys. Rev.\/} {\bf D80} 076011 (\textit{Preprint}
  \eprint{0903.0831})

\bibitem{Altarelli:2009a}
Altarelli G and Meloni D 2009 {\em J. Phys.\/} {\bf G36} 085005
  (\textit{Preprint} \eprint{0905.0620})

\bibitem{Ma:2005b}
Ma E 2005 {\em Mod. Phys. Lett.\/} {\bf A20} 2767--2774 (\textit{Preprint}
  \eprint{hep-ph/0506036})

\bibitem{Ma:2006a}
Ma E, Sawanaka H and Tanimoto M 2006 {\em Phys. Lett.\/} {\bf B641} 301--304
  (\textit{Preprint} \eprint{hep-ph/0606103})

\bibitem{Ma:2006b}
Ma E 2006 {\em Mod. Phys. Lett.\/} {\bf A21} 2931--2936 (\textit{Preprint}
  \eprint{hep-ph/0607190})

\bibitem{Morisi:2007}
Morisi S, Picariello M and Torrente-Lujan E 2007 {\em Phys. Rev.\/} {\bf D75}
  075015 (\textit{Preprint} \eprint{hep-ph/0702034})

\bibitem{Grimus:2008}
Grimus W and Kuhbock H 2008 {\em Phys. Rev.\/} {\bf D77} 055008
  (\textit{Preprint} \eprint{0710.1585})

\bibitem{Ciafaloni:2009}
Ciafaloni P, Picariello M, Torrente-Lujan E and Urbano A 2009 {\em Phys.
  Rev.\/} {\bf D79} 116010 (\textit{Preprint} \eprint{0901.2236})

\bibitem{Bazzocchi:2009}
Bazzocchi F, Morisi S, Picariello M and Torrente-Lujan E 2009 {\em J. Phys.\/}
  {\bf G36} 015002 (\textit{Preprint} \eprint{0802.1693})

\bibitem{Bazzocchi:2008b}
Bazzocchi F, Frigerio M and Morisi S 2008 {\em Phys. Rev.\/} {\bf D78} 116018
  (\textit{Preprint} \eprint{0809.3573})

\bibitem{delAguila:2010}
del Aguila F, Carmona A and Santiago J 2010  (\textit{Preprint}
  \eprint{1001.5151})

\bibitem{Kadosh:2010rm}
Kadosh A and Pallante E 2010  (\textit{Preprint} \eprint{1004.0321})

\bibitem{Antusch:2010es}
Antusch S, King S~F and Spinrath M 2010  (\textit{Preprint} \eprint{1005.0708})

\bibitem{Aranda:2000}
Aranda A, Carone C~D and Lebed R~F 2000 {\em Phys. Lett.\/} {\bf B474} 170--176
  (\textit{Preprint} \eprint{hep-ph/9910392})

\bibitem{Aranda:2000a}
Aranda A, Carone C~D and Lebed R~F 2000 {\em Phys. Rev.\/} {\bf D62} 016009
  (\textit{Preprint} \eprint{hep-ph/0002044})

\bibitem{Carr:2000}
Carr P~D and Frampton P~H 2007  (\textit{Preprint} \eprint{hep-ph/0701034})

\bibitem{Aranda:2007}
Aranda A 2007 {\em Phys. Rev.\/} {\bf D76} 111301 (\textit{Preprint}
  \eprint{0707.3661})

\bibitem{Frampton:2007}
Frampton P~H and Kephart T~W 2007 {\em JHEP\/} {\bf 09} 110 (\textit{Preprint}
  \eprint{0706.1186})

\bibitem{Frampton:2009}
Frampton P~H and Matsuzaki S 2009 {\em Phys. Lett.\/} {\bf B679} 347--349
  (\textit{Preprint} \eprint{0902.1140})

\bibitem{Ding:2009}
Ding G~J 2008 {\em Phys. Rev.\/} {\bf D78} 036011 (\textit{Preprint}
  \eprint{0803.2278})

\bibitem{Feruglio:2007}
Feruglio F, Hagedorn C, Lin Y and Merlo L 2007 {\em Nucl. Phys.\/} {\bf B775}
  120--142 (\textit{Preprint} \eprint{hep-ph/0702194})

\bibitem{Chen:2007}
Chen M~C and Mahanthappa K~T 2007 {\em Phys. Lett.\/} {\bf B652} 34--39
  (\textit{Preprint} \eprint{0705.0714})

\bibitem{Mohapatra:2004}
Mohapatra R~N, Parida M~K and Rajasekaran G 2004 {\em Phys. Rev.\/} {\bf D69}
  053007 (\textit{Preprint} \eprint{hep-ph/0301234})

\bibitem{Hagedorn:2006}
Hagedorn C, Lindner M and Mohapatra R~N 2006 {\em JHEP\/} {\bf 06} 042
  (\textit{Preprint} \eprint{hep-ph/0602244})

\bibitem{Cai:2006}
Cai Y and Yu H~B 2006 {\em Phys. Rev.\/} {\bf D74} 115005 (\textit{Preprint}
  \eprint{hep-ph/0608022})

\bibitem{Ma:2007a}
Ma E 2006 {\em Phys. Lett.\/} {\bf B632} 352--356 (\textit{Preprint}
  \eprint{hep-ph/0508231})

\bibitem{Bazzocchi:2008c}
Bazzocchi F and Morisi S 2009 {\em Phys. Rev.\/} {\bf D80} 096005
  (\textit{Preprint} \eprint{0811.0345})

\bibitem{Ishimori:2009}
Ishimori H, Shimizu Y and Tanimoto M 2009 {\em Prog. Theor. Phys.\/} {\bf 121}
  769--787 (\textit{Preprint} \eprint{0812.5031})

\bibitem{Bazzocchi:2009a}
Bazzocchi F, Merlo L and Morisi S 2009 {\em Nucl. Phys.\/} {\bf B816} 204--226
  (\textit{Preprint} \eprint{0901.2086})

\bibitem{Bazzocchi:2009b}
Bazzocchi F, Merlo L and Morisi S 2009 {\em Phys. Rev.\/} {\bf D80} 053003
  (\textit{Preprint} \eprint{0902.2849})

\bibitem{Meloni:2009}
Meloni D 2010 {\em J. Phys.\/} {\bf G37} 055201 (\textit{Preprint}
  \eprint{0911.3591})

\bibitem{Dutta:2009}
Dutta B, Mimura Y and Mohapatra R~N 2009 {\em Phys. Rev.\/} {\bf D80} 095021
  (\textit{Preprint} \eprint{0910.1043})

\bibitem{Dutta:2009a}
Dutta B, Mimura Y and Mohapatra R~N 2010 {\em JHEP\/} {\bf 05} 034
  (\textit{Preprint} \eprint{0911.2242})

\bibitem{Ding:2010}
Ding G~J 2010 {\em Nucl. Phys.\/} {\bf B827} 82--111 (\textit{Preprint}
  \eprint{0909.2210})

\bibitem{Morisi:2010}
Morisi S and Peinado E 2010 {\em Phys. Rev.\/} {\bf D81} 085015
  (\textit{Preprint} \eprint{1001.2265})

\bibitem{Hagedorn:2010th}
Hagedorn C, King S~F and Luhn C 2010  (\textit{Preprint} \eprint{1003.4249})

\bibitem{Ishimori:2010xk}
Ishimori H, Saga K, Shimizu Y and Tanimoto M 2010  (\textit{Preprint}
  \eprint{1004.5004})

\bibitem{deMedeirosVarzielas:2007}
de~Medeiros~Varzielas I, King S~F and Ross G~G 2007 {\em Phys. Lett.\/} {\bf
  B648} 201--206 (\textit{Preprint} \eprint{hep-ph/0607045})

\bibitem{Ma:2007b}
Ma E 2008 {\em Phys. Lett.\/} {\bf B660} 505--507 (\textit{Preprint}
  \eprint{0709.0507})

\bibitem{Grimus:2007}
Grimus W and Lavoura L 2008 {\em JHEP\/} {\bf 09} 106 (\textit{Preprint}
  \eprint{0809.0226})

\bibitem{Luhn:2007}
Luhn C, Nasri S and Ramond P 2007 {\em J. Math. Phys.\/} {\bf 48} 073501
  (\textit{Preprint} \eprint{hep-th/0701188})

\bibitem{Bazzocchi:2009c}
Bazzocchi F and de~Medeiros~Varzielas I 2009 {\em Phys. Rev.\/} {\bf D79}
  093001 (\textit{Preprint} \eprint{0902.3250})

\bibitem{Everett:2009}
Everett L~L and Stuart A~J 2009 {\em Phys. Rev.\/} {\bf D79} 085005
  (\textit{Preprint} \eprint{0812.1057})

\bibitem{Luhn:2007a}
Luhn C, Nasri S and Ramond P 2007 {\em J. Math. Phys.\/} {\bf 48} 123519
  (\textit{Preprint} \eprint{0709.1447})

\bibitem{King:2009}
King S~F and Luhn C 2009 {\em Nucl. Phys.\/} {\bf B820} 269--289
  (\textit{Preprint} \eprint{0905.1686})

\bibitem{King:2009a}
King S~F and Luhn C 2009  (\textit{Preprint} \eprint{0912.1344})

\bibitem{Luhn:2007b}
Luhn C, Nasri S and Ramond P 2007 {\em Phys. Lett.\/} {\bf B652} 27--33
  (\textit{Preprint} \eprint{0706.2341})

\bibitem{King:2005}
King S~F 2005 {\em JHEP\/} {\bf 08} 105 (\textit{Preprint}
  \eprint{hep-ph/0506297})

\bibitem{King:2006}
King S~F and Malinsky M 2006 {\em JHEP\/} {\bf 11} 071 (\textit{Preprint}
  \eprint{hep-ph/0608021})

\bibitem{deMedeirosVarzielas:2006}
de~Medeiros~Varzielas I and Ross G~G 2006 {\em Nucl. Phys.\/} {\bf B733} 31--47
  (\textit{Preprint} \eprint{hep-ph/0507176})

\bibitem{deMedeirosVarzielas:2007a}
de~Medeiros~Varzielas I, King S~F and Ross G~G 2007 {\em Phys. Lett.\/} {\bf
  B644} 153--157 (\textit{Preprint} \eprint{hep-ph/0512313})

\bibitem{Adulpravitchai:2009c}
Adulpravitchai A, Blum A and Lindner M 2009 {\em JHEP\/} {\bf 09} 018
  (\textit{Preprint} \eprint{0907.2332})

\bibitem{Berger:2009}
Berger J and Grossman Y 2010 {\em JHEP\/} {\bf 02} 071 (\textit{Preprint}
  \eprint{0910.4392})

\bibitem{Xing:2002}
Xing Z~z 2002 {\em Phys. Lett.\/} {\bf B533} 85--93 (\textit{Preprint}
  \eprint{hep-ph/0204049})

\bibitem{Matias:2005}
Matias J and Burgess C~P 2005 {\em JHEP\/} {\bf 09} 052 (\textit{Preprint}
  \eprint{hep-ph/0508156})

\bibitem{Luo:2005}
Luo S and Xing Z~z 2006 {\em Phys. Lett.\/} {\bf B632} 341--348
  (\textit{Preprint} \eprint{hep-ph/0509065})

\bibitem{Grimus:2005}
Grimus W and Lavoura L 2006 {\em JHEP\/} {\bf 01} 018 (\textit{Preprint}
  \eprint{hep-ph/0509239})

\bibitem{Koide:2007}
Koide Y 2007  (\textit{Preprint} \eprint{0707.0899})

\bibitem{Grimus:2009}
Grimus W and Lavoura L 2009 {\em JHEP\/} {\bf 04} 013 (\textit{Preprint}
  \eprint{0811.4766})

\bibitem{Babu:2010bx}
Babu K~S and Gabriel S 2010  (\textit{Preprint} \eprint{1006.0203})

\bibitem{GonzalezGarcia:2010}
Gonzalez-Garcia M~C, Maltoni M and Salvado J 2010 {\em JHEP\/} {\bf 04} 056
  (\textit{Preprint} \eprint{1001.4524})

\bibitem{Raidal:2004}
Raidal M 2004 {\em Phys. Rev. Lett.\/} {\bf 93} 161801 (\textit{Preprint}
  \eprint{hep-ph/0404046})

\bibitem{Minakata:2004}
Minakata H and Smirnov A~Y 2004 {\em Phys. Rev.\/} {\bf D70} 073009
  (\textit{Preprint} \eprint{hep-ph/0405088})

\bibitem{Altarelli:2004a}
Altarelli G, Feruglio F and Masina I 2004 {\em Nucl. Phys.\/} {\bf B689}
  157--171 (\textit{Preprint} \eprint{hep-ph/0402155})

\bibitem{Frampton:2005}
Frampton P~H and Mohapatra R~N 2005 {\em JHEP\/} {\bf 01} 025
  (\textit{Preprint} \eprint{hep-ph/0407139})

\bibitem{Ferrandis:2005}
Ferrandis J and Pakvasa S 2005 {\em Phys. Rev.\/} {\bf D71} 033004
  (\textit{Preprint} \eprint{hep-ph/0412038})

\bibitem{Kang:2005}
Kang S~K, Kim C~S and Lee J 2005 {\em Phys. Lett.\/} {\bf B619} 129--135
  (\textit{Preprint} \eprint{hep-ph/0501029})

\bibitem{Minakata:2005}
Minakata H 2005  (\textit{Preprint} \eprint{hep-ph/0505262})

\bibitem{Li:2005}
Li N and Ma B~Q 2005 {\em Phys. Rev.\/} {\bf D71} 097301 (\textit{Preprint}
  \eprint{hep-ph/0501226})

\bibitem{Cheung:2005}
Cheung K, Kang S~K, Kim C~S and Lee J 2005 {\em Phys. Rev.\/} {\bf D72} 036003
  (\textit{Preprint} \eprint{hep-ph/0503122})

\bibitem{Xing:2005}
Xing Z~z 2005 {\em Phys. Lett.\/} {\bf B618} 141--149 (\textit{Preprint}
  \eprint{hep-ph/0503200})

\bibitem{Datta:2005}
Datta A, Everett L and Ramond P 2005 {\em Phys. Lett.\/} {\bf B620} 42--51
  (\textit{Preprint} \eprint{hep-ph/0503222})

\bibitem{Ohlsson:2005}
Ohlsson T 2005 {\em Phys. Lett.\/} {\bf B622} 159--164 (\textit{Preprint}
  \eprint{hep-ph/0506094})

\bibitem{Antusch:2005}
Antusch S, King S~F and Mohapatra R~N 2005 {\em Phys. Lett.\/} {\bf B618}
  150--161 (\textit{Preprint} \eprint{hep-ph/0504007})

\bibitem{Lindner:2005}
Lindner M, Schmidt M~A and Smirnov A~Y 2005 {\em JHEP\/} {\bf 07} 048
  (\textit{Preprint} \eprint{hep-ph/0505067})

\bibitem{King:2005a}
King S~F 2005 {\em JHEP\/} {\bf 08} 105 (\textit{Preprint}
  \eprint{hep-ph/0506297})

\bibitem{Dighe:2006}
Dighe A, Goswami S and Roy P 2006 {\em Phys. Rev.\/} {\bf D73} 071301
  (\textit{Preprint} \eprint{hep-ph/0602062})

\bibitem{Schmidt:2006}
Schmidt M~A and Smirnov A~Y 2006 {\em Phys. Rev.\/} {\bf D74} 113003
  (\textit{Preprint} \eprint{hep-ph/0607232})

\bibitem{Chauhan:2007}
Chauhan B~C, Picariello M, Pulido J and Torrente-Lujan E 2007 {\em Eur. Phys.
  J.\/} {\bf C50} 573--578 (\textit{Preprint} \eprint{hep-ph/0605032})

\bibitem{Hochmuth:2007}
Hochmuth K~A and Rodejohann W 2007 {\em Phys. Rev.\/} {\bf D75} 073001
  (\textit{Preprint} \eprint{hep-ph/0607103})

\bibitem{Plentinger:2007}
Plentinger F, Seidl G and Winter W 2007 {\em Phys. Rev.\/} {\bf D76} 113003
  (\textit{Preprint} \eprint{0707.2379})

\bibitem{Plentinger:2008a}
Plentinger F, Seidl G and Winter W 2008 {\em Nucl. Phys.\/} {\bf B791} 60--92
  (\textit{Preprint} \eprint{hep-ph/0612169})

\bibitem{Altarelli:2009b}
Altarelli G, Feruglio F and Merlo L 2009 {\em JHEP\/} {\bf 05} 020
  (\textit{Preprint} \eprint{0903.1940})

\bibitem{Kajiyama:2007}
Kajiyama Y, Raidal M and Strumia A 2007 {\em Phys. Rev.\/} {\bf D76} 117301
  (\textit{Preprint} \eprint{0705.4559})

\bibitem{Rodejohann:2009}
Rodejohann W 2009 {\em Phys. Lett.\/} {\bf B671} 267--271 (\textit{Preprint}
  \eprint{0810.5239})

\bibitem{Adulpravitchai:2009}
Adulpravitchai A, Blum A and Rodejohann W 2009 {\em New J. Phys.\/} {\bf 11}
  063026 (\textit{Preprint} \eprint{0903.0531})

\bibitem{Kubo:2003}
Kubo J, Mondragon A, Mondragon M and Rodriguez-Jauregui E 2003 {\em Prog.
  Theor. Phys.\/} {\bf 109} 795--807 (\textit{Preprint}
  \eprint{hep-ph/0302196})

\bibitem{Kubo:2004}
Kubo J 2004 {\em Phys. Lett.\/} {\bf B578} 156--164 (\textit{Preprint}
  \eprint{hep-ph/0309167})

\bibitem{Kubo:2006}
Morisi S and Picariello M 2006 {\em Int. J. Theor. Phys.\/} {\bf 45} 1267--1277
  (\textit{Preprint} \eprint{hep-ph/0505113})

\bibitem{Chen:2004}
Chen S~L, Frigerio M and Ma E 2004 {\em Phys. Rev.\/} {\bf D70} 073008
  (\textit{Preprint} \eprint{hep-ph/0404084})

\bibitem{Lavoura:2005}
Lavoura L and Ma E 2005 {\em Mod. Phys. Lett.\/} {\bf A20} 1217--1226
  (\textit{Preprint} \eprint{hep-ph/0502181})

\bibitem{Dermisek:2005}
Dermisek R and Raby S 2005 {\em Phys. Lett.\/} {\bf B622} 327--338
  (\textit{Preprint} \eprint{hep-ph/0507045})

\bibitem{Caravaglios:2005}
Caravaglios F and Morisi S 2005  (\textit{Preprint} \eprint{hep-ph/0503234})

\bibitem{Caravaglios:2005a}
Caravaglios F and Morisi S 2005  (\textit{Preprint} \eprint{hep-ph/0510321})

\bibitem{Grimus:2006a}
Grimus W and Lavoura L 2006 {\em JHEP\/} {\bf 01} 018 (\textit{Preprint}
  \eprint{hep-ph/0509239})

\bibitem{Koide:2006}
Koide Y 2006 {\em Phys. Rev.\/} {\bf D73} 057901 (\textit{Preprint}
  \eprint{hep-ph/0509214})

\bibitem{Teshima:2006}
Teshima T 2006 {\em Phys. Rev.\/} {\bf D73} 045019 (\textit{Preprint}
  \eprint{hep-ph/0509094})

\bibitem{Haba:2006}
Haba N and Yoshioka K 2006 {\em Nucl. Phys.\/} {\bf B739} 254--284
  (\textit{Preprint} \eprint{hep-ph/0511108})

\bibitem{Tanimoto:2006}
Tanimoto M and Yanagida T 2006 {\em Phys. Lett.\/} {\bf B633} 567--572
  (\textit{Preprint} \eprint{hep-ph/0511336})

\bibitem{Koide:2006a}
Koide Y 2006 {\em Eur. Phys. J.\/} {\bf C48} 223--228 (\textit{Preprint}
  \eprint{hep-ph/0508301})

\bibitem{Morisi:2006}
Morisi S 2006  (\textit{Preprint} \eprint{hep-ph/0604106})

\bibitem{Picariello:2006}
Picariello M 2008 {\em Int. J. Mod. Phys.\/} {\bf A23} 4435--4448
  (\textit{Preprint} \eprint{hep-ph/0611189})

\bibitem{Mohapatra:2006a}
Mohapatra R~N, Nasri S and Yu H~B 2006 {\em Phys. Lett.\/} {\bf B636} 114--118
  (\textit{Preprint} \eprint{hep-ph/0603020})

\bibitem{Mohapatra:2006b}
Mohapatra R~N, Nasri S and Yu H~B 2006 {\em Phys. Lett.\/} {\bf B639} 318--321
  (\textit{Preprint} \eprint{hep-ph/0605020})

\bibitem{Kaneko:2007}
Kaneko S, Sawanaka H, Shingai T, Tanimoto M and Yoshioka K 2007 {\em Prog.
  Theor. Phys.\/} {\bf 117} 161--181 (\textit{Preprint}
  \eprint{hep-ph/0609220})

\bibitem{Koide:2007a}
Koide Y 2007 {\em Eur. Phys. J.\/} {\bf C50} 809--816 (\textit{Preprint}
  \eprint{hep-ph/0612058})

\bibitem{Chen:2008}
Chen C~Y and Wolfenstein L 2008 {\em Phys. Rev.\/} {\bf D77} 093009
  (\textit{Preprint} \eprint{0709.3767})

\bibitem{Feruglio:2007a}
Feruglio F and Lin Y 2008 {\em Nucl. Phys.\/} {\bf B800} 77--93
  (\textit{Preprint} \eprint{0712.1528})

\bibitem{Grimus:2004}
Grimus W, Joshipura A~S, Kaneko S, Lavoura L and Tanimoto M 2004 {\em JHEP\/}
  {\bf 07} 078 (\textit{Preprint} \eprint{hep-ph/0407112})

\bibitem{Adulpravitchai:2009a}
Adulpravitchai A, Blum A and Hagedorn C 2009 {\em JHEP\/} {\bf 03} 046
  (\textit{Preprint} \eprint{0812.3799})

\bibitem{Blum:2004}
Blum A, Hagedorn C and Lindner M 2008 {\em Phys. Rev.\/} {\bf D77} 076004
  (\textit{Preprint} \eprint{0709.3450})

\bibitem{Blum:2008}
Blum A, Hagedorn C and Hohenegger A 2008 {\em JHEP\/} {\bf 03} 070
  (\textit{Preprint} \eprint{0710.5061})

\bibitem{Zhang:2007}
Zhang H 2007 {\em Phys. Lett.\/} {\bf B655} 132--140 (\textit{Preprint}
  \eprint{hep-ph/0612214})

\bibitem{Fukuyama:1997}
Fukuyama T and Nishiura H 1997  (\textit{Preprint} \eprint{hep-ph/9702253})

\bibitem{Mohapatra:1999}
Mohapatra R~N and Nussinov S 1999 {\em Phys. Rev.\/} {\bf D60} 013002
  (\textit{Preprint} \eprint{hep-ph/9809415})

\bibitem{Ma:2001a}
Ma E and Raidal M 2001 {\em Phys. Rev. Lett.\/} {\bf 87} 011802
  (\textit{Preprint} \eprint{hep-ph/0102255})

\bibitem{Lam:2001}
Lam C~S 2001 {\em Phys. Lett.\/} {\bf B507} 214--218 (\textit{Preprint}
  \eprint{hep-ph/0104116})

\bibitem{Kitabayashi:2003}
Kitabayashi T and Yasue M 2003 {\em Phys. Rev.\/} {\bf D67} 015006
  (\textit{Preprint} \eprint{hep-ph/0209294})

\bibitem{Grimus:2003}
Grimus W and Lavoura L 2004 {\em J. Phys.\/} {\bf G30} 73--82
  (\textit{Preprint} \eprint{hep-ph/0309050})

\bibitem{Koide:2004}
Koide Y 2004 {\em Phys. Rev.\/} {\bf D69} 093001 (\textit{Preprint}
  \eprint{hep-ph/0312207})

\bibitem{Ghosal:2003}
Ghosal A 2003  (\textit{Preprint} \eprint{hep-ph/0304090})

\bibitem{Grimus:2005a}
Grimus W {\em et~al.\/} 2005 {\em Nucl. Phys.\/} {\bf B713} 151--172
  (\textit{Preprint} \eprint{hep-ph/0408123})

\bibitem{deGouvea:2004}
de~Gouvea A 2004 {\em Phys. Rev.\/} {\bf D69} 093007 (\textit{Preprint}
  \eprint{hep-ph/0401220})

\bibitem{Mohapatra:2005}
Mohapatra R~N and Rodejohann W 2005 {\em Phys. Rev.\/} {\bf D72} 053001
  (\textit{Preprint} \eprint{hep-ph/0507312})

\bibitem{Kitabayashi:2005}
Kitabayashi T and Yasue M 2005 {\em Phys. Lett.\/} {\bf B621} 133--138
  (\textit{Preprint} \eprint{hep-ph/0504212})

\bibitem{Mohapatra:2005a}
Mohapatra R~N and Nasri S 2005 {\em Phys. Rev.\/} {\bf D71} 033001
  (\textit{Preprint} \eprint{hep-ph/0410369})

\bibitem{Mohapatra:2005b}
Mohapatra R~N, Nasri S and Yu H~B 2005 {\em Phys. Lett.\/} {\bf B615} 231--239
  (\textit{Preprint} \eprint{hep-ph/0502026})

\bibitem{Mohapatra:2005c}
Mohapatra R~N, Nasri S and Yu H~B 2005 {\em Phys. Rev.\/} {\bf D72} 033007
  (\textit{Preprint} \eprint{hep-ph/0505021})

\bibitem{Ahn:2006}
Ahn Y~H, Kang S~K, Kim C~S and Lee J 2006 {\em Phys. Rev.\/} {\bf D73} 093005
  (\textit{Preprint} \eprint{hep-ph/0602160})

\bibitem{Ge:2010}
Ge S~F, He H~J and Yin F~R 2010 {\em JCAP\/} {\bf 1005} 017 (\textit{Preprint}
  \eprint{1001.0940})

\bibitem{Chankowski:2001mx}
Chankowski P~H and Pokorski S 2002 {\em Int. J. Mod. Phys.\/} {\bf A17}
  575--614 (\textit{Preprint} \eprint{hep-ph/0110249})

\bibitem{Cabibbo:1978}
Cabibbo N 1978 {\em Phys. Lett.\/} {\bf B72} 333--335

\bibitem{Wolfenstein:1978}
Wolfenstein L 1978 {\em Phys. Rev.\/} {\bf D18} 958--960

\bibitem{Froggatt:1979}
Froggatt C~D and Nielsen H~B 1979 {\em Nucl. Phys.\/} {\bf B147} 277

\bibitem{Barry:2010zk}
Barry J and Rodejohann W 2010 {\em Phys. Rev.\/} {\bf D81} 093002
  (\textit{Preprint} \eprint{1003.2385})

\bibitem{Lin:2009b}
Lin Y, Merlo L and Paris A 2010 {\em Nucl. Phys.\/} {\bf B835} 238--261
  (\textit{Preprint} \eprint{0911.3037})

\bibitem{Burrows:2009}
Burrows T~J and King S~F 2010 {\em Nucl. Phys.\/} {\bf B835} 174--196
  (\textit{Preprint} \eprint{0909.1433})

\bibitem{Adulpravitchai:2009b}
Adulpravitchai A, Blum A and Lindner M 2009 {\em JHEP\/} {\bf 07} 053
  (\textit{Preprint} \eprint{0906.0468})

\bibitem{Adulpravitchai:2010na}
Adulpravitchai A and Schmidt M A 2010  
 (\textit{Preprint} \eprint{1001.3172})
  
\bibitem{Kobayashi:2007}
Kobayashi T, Nilles H~P, Ploger F, Raby S and Ratz M 2007 {\em Nucl. Phys.\/}
  {\bf B768} 135--156 (\textit{Preprint} \eprint{hep-ph/0611020})

\bibitem{Abe:2009}
Abe H, Choi K~S, Kobayashi T and Ohki H 2009 {\em Nucl. Phys.\/} {\bf B820}
  317--333 (\textit{Preprint} \eprint{0904.2631})

\bibitem{Lam:2008}
Lam C~S 2008 {\em Phys. Rev. Lett.\/} {\bf 101} 121602 (\textit{Preprint}
  \eprint{0804.2622})

\bibitem{Grimus:2008a}
Grimus W, Lavoura L and Ludl P~O 2009 {\em J. Phys.\/} {\bf G36} 115007
  (\textit{Preprint} \eprint{0906.2689})

\bibitem{King:2009b}
King S~F 2010 {\em AIP Conf. Proc.\/} {\bf 1200} 103--111 (\textit{Preprint}
  \eprint{0909.2969})

\bibitem{Chen:2009}
Chen M~C and King S~F 2009 {\em JHEP\/} {\bf 06} 072 (\textit{Preprint}
  \eprint{0903.0125})

\bibitem{King:2007}
King S~F and Malinsky M 2007 {\em Phys. Lett.\/} {\bf B645} 351--357
  (\textit{Preprint} \eprint{hep-ph/0610250})

\bibitem{Toorop:2010yh}
Toorop R~d~A, Bazzocchi F and Merlo L 2010  (\textit{Preprint}
  \eprint{1003.4502})

\bibitem{Pomarol:1996}
Pomarol A and Tommasini D 1996 {\em Nucl. Phys.\/} {\bf B466} 3--24
  (\textit{Preprint} \eprint{hep-ph/9507462})

\bibitem{Barbieri:1996}
Barbieri R, Dvali G~R and Hall L~J 1996 {\em Phys. Lett.\/} {\bf B377} 76--82
  (\textit{Preprint} \eprint{hep-ph/9512388})

\bibitem{Barbieri:1997}
Barbieri R, Hall L~J, Raby S and Romanino A 1997 {\em Nucl. Phys.\/} {\bf B493}
  3--26 (\textit{Preprint} \eprint{hep-ph/9610449})

\bibitem{Barbieri:1997a}
Barbieri R, Hall L~J and Romanino A 1997 {\em Phys. Lett.\/} {\bf B401} 47--53
  (\textit{Preprint} \eprint{hep-ph/9702315})

\bibitem{Witten:1985}
Witten E 1985 {\em Nucl. Phys.\/} {\bf B258} 75

\bibitem{Kawamura:2001}
Kawamura Y 2001 {\em Prog. Theor. Phys.\/} {\bf 105} 999--1006
  (\textit{Preprint} \eprint{hep-ph/0012125})

\bibitem{Faraggi:2001}
Faraggi A~E 2001 {\em Phys. Lett.\/} {\bf B520} 337--344 (\textit{Preprint}
  \eprint{hep-ph/0107094})

\bibitem{Goh:2004}
Goh H~S, Mohapatra R~N and Nasri S 2004 {\em Phys. Rev.\/} {\bf D70} 075022
  (\textit{Preprint} \eprint{hep-ph/0408139})

\bibitem{Mohapatra:2007a}
Mohapatra R~N, Okada N and Yu H~B 2007 {\em Phys. Rev.\/} {\bf D76} 015013
  (\textit{Preprint} \eprint{0704.3258})

\bibitem{Raidal:2008}
Raidal M {\em et~al.\/} 2008 {\em Eur. Phys. J.\/} {\bf C57} 13--182
  (\textit{Preprint} \eprint{0801.1826})

\bibitem{Brooks:1999}
Brooks M~L {\em et~al.\/} (MEGA) 1999 {\em Phys. Rev. Lett.\/} {\bf 83}
  1521--1524 (\textit{Preprint} \eprint{hep-ex/9905013})

\bibitem{Adam:2009}
Adam J {\em et~al.\/} (MEG) 2010 {\em Nucl. Phys.\/} {\bf B834} 1--12
  (\textit{Preprint} \eprint{0908.2594})

\bibitem{Chivukula:1987}
Chivukula R~S and Georgi H 1987 {\em Phys. Lett.\/} {\bf B188} 99

\bibitem{Hall:1990}
Hall L~J and Randall L 1990 {\em Phys. Rev. Lett.\/} {\bf 65} 2939--2942

\bibitem{Ciuchini:1998}
Ciuchini M, Degrassi G, Gambino P and Giudice G~F 1998 {\em Nucl. Phys.\/} {\bf
  B534} 3--20 (\textit{Preprint} \eprint{hep-ph/9806308})

\bibitem{Buras:2001}
Buras A~J, Gambino P, Gorbahn M, Jager S and Silvestrini L 2001 {\em Phys.
  Lett.\/} {\bf B500} 161--167 (\textit{Preprint} \eprint{hep-ph/0007085})

\bibitem{D'Ambrosio:2002}
D'Ambrosio G, Giudice G~F, Isidori G and Strumia A 2002 {\em Nucl. Phys.\/}
  {\bf B645} 155--187 (\textit{Preprint} \eprint{hep-ph/0207036})

\bibitem{Cirigliano:2005}
Cirigliano V, Grinstein B, Isidori G and Wise M~B 2005 {\em Nucl. Phys.\/} {\bf
  B728} 121--134 (\textit{Preprint} \eprint{hep-ph/0507001})

\bibitem{Cirigliano:2006}
Cirigliano V and Grinstein B 2006 {\em Nucl. Phys.\/} {\bf B752} 18--39
  (\textit{Preprint} \eprint{hep-ph/0601111})

\bibitem{Davidson:2006}
Davidson S and Palorini F 2006 {\em Phys. Lett.\/} {\bf B642} 72--80
  (\textit{Preprint} \eprint{hep-ph/0607329})

\bibitem{Cirigliano:2007}
Grinstein B, Cirigliano V, Isidori G and Wise M~B 2007 {\em Nucl. Phys.\/} {\bf
  B763} 35--48 (\textit{Preprint} \eprint{hep-ph/0608123})

\bibitem{Feruglio:2009}
Feruglio F, Hagedorn C, Lin Y and Merlo L 2009 {\em Nucl. Phys.\/} {\bf B809}
  218--243 (\textit{Preprint} \eprint{0807.3160})

\bibitem{Feruglio:2008}
Feruglio F, Hagedorn C, Lin Y and Merlo L 2008  (\textit{Preprint}
  \eprint{0808.0812})

\bibitem{Bennett:2004}
Bennett G~W {\em et~al.\/} (Muon g-2) 2004 {\em Phys. Rev. Lett.\/} {\bf 92}
  161802 (\textit{Preprint} \eprint{hep-ex/0401008})

\bibitem{Passera:2009}
Passera M, Marciano W~J and Sirlin A 2009 {\em AIP Conf. Proc.\/} {\bf 1078}
  378--381 (\textit{Preprint} \eprint{0809.4062})

\bibitem{Meg}
MEG 1999  Proposal of the MEG experiment, at http://meg.web.psi.ch.

\bibitem{Feruglio:2010qu}
Feruglio F and Paris A 2010  (\textit{Preprint} \eprint{1005.5526})

\bibitem{Amsler:2008}
Amsler C {\em et~al.\/} (Particle Data Group) 2008 {\em Phys. Lett.\/} {\bf
  B667} 1

\bibitem{Ishimori:2008}
Ishimori H, Kobayashi T, Omura Y and Tanimoto M 2008 {\em JHEP\/} {\bf 12} 082
  (\textit{Preprint} \eprint{0807.4625})

\bibitem{Hayakawa:2009}
Hayakawa A, Ishimori H, Shimizu Y and Tanimoto M 2009 {\em Phys. Lett.\/} {\bf
  B680} 334--342 (\textit{Preprint} \eprint{0904.3820})

\bibitem{feruglio:2009a}
Feruglio F, Hagedorn C, Lin Y and Merlo L 2009  (\textit{Preprint}
  \eprint{0911.3874})

\bibitem{Ding:2009b}
Ding G~J and Liu J~F 2010 {\em JHEP\/} {\bf 05} 029 (\textit{Preprint}
  \eprint{0911.4799})

\bibitem{feruglio:2009b}
Feruglio F, Hagedorn C and Merlo L 2010 {\em JHEP\/} {\bf 03} 084
  (\textit{Preprint} \eprint{0910.4058})

\bibitem{masiero:2009}
Masiero L 2009  Graduation thesis, University of Padova, 2009

\bibitem{Fukugita:1986}
Fukugita M and Yanagida T 1986 {\em Phys. Lett.\/} {\bf B174} 45

\bibitem{Abada:2006}
Abada A, Davidson S, Josse-Michaux F~X, Losada M and Riotto A 2006 {\em JCAP\/}
  {\bf 0604} 004 (\textit{Preprint} \eprint{hep-ph/0601083})

\bibitem{Nardi:2006}
Nardi E, Nir Y, Roulet E and Racker J 2006 {\em JHEP\/} {\bf 01} 164
  (\textit{Preprint} \eprint{hep-ph/0601084})

\bibitem{Buchmuller:2003}
Buchmuller W, Di~Bari P and Plumacher M 2003 {\em Nucl. Phys.\/} {\bf B665}
  445--468 (\textit{Preprint} \eprint{hep-ph/0302092})

\bibitem{Buchmuller:2004}
Buchmuller W, Di~Bari P and Plumacher M 2004 {\em New J. Phys.\/} {\bf 6} 105
  (\textit{Preprint} \eprint{hep-ph/0406014})

\bibitem{Giudice:2004}
Giudice G~F, Notari A, Raidal M, Riotto A and Strumia A 2004 {\em Nucl.
  Phys.\/} {\bf B685} 89--149 (\textit{Preprint} \eprint{hep-ph/0310123})

\bibitem{Buchmuller:2005}
Buchmuller W, Di~Bari P and Plumacher M 2005 {\em Ann. Phys.\/} {\bf 315}
  305--351 (\textit{Preprint} \eprint{hep-ph/0401240})

\bibitem{Komatsu:2009}
Komatsu E {\em et~al.\/} (WMAP) 2009 {\em Astrophys. J. Suppl.\/} {\bf 180}
  330--376 (\textit{Preprint} \eprint{0803.0547})

\bibitem{Covi:1996}
Covi L, Roulet E and Vissani F 1996 {\em Phys. Lett.\/} {\bf B384} 169--174
  (\textit{Preprint} \eprint{hep-ph/9605319})

\bibitem{Bertuzzo:2009}
Bertuzzo E, Di~Bari P, Feruglio F and Nardi E 2009 {\em JHEP\/} {\bf 11} 036
  (\textit{Preprint} \eprint{0908.0161})

\bibitem{Jenkins:2008}
Jenkins E~E and Manohar A~V 2008 {\em Phys. Lett.\/} {\bf B668} 210--215
  (\textit{Preprint} \eprint{0807.4176})

\bibitem{Hagedorn:2009}
Hagedorn C, Molinaro E and Petcov S~T 2009 {\em JHEP\/} {\bf 09} 115
  (\textit{Preprint} \eprint{0908.0240})

\bibitem{Riva:2010}
Riva F 2010  (\textit{Preprint} \eprint{1004.1177})

\bibitem{Low:2003}
Low C~I and Volkas R~R 2003 {\em Phys. Rev.\/} {\bf D68} 033007
  (\textit{Preprint} \eprint{hep-ph/0305243})

\bibitem{Aristizabal:2009}
Aristizabal~Sierra D, Bazzocchi F, de~Medeiros~Varzielas I, Merlo L and Morisi
  S 2010 {\em Nucl. Phys.\/} {\bf B827} 34--58 (\textit{Preprint}
  \eprint{0908.0907})

\bibitem{Felipe:2009}
Felipe R~G and Serodio H 2010 {\em Phys. Rev.\/} {\bf D81} 053008
  (\textit{Preprint} \eprint{0908.2947})

\bibitem{Davidson:2002}
Davidson S and Ibarra A 2002 {\em Phys. Lett.\/} {\bf B535} 25--32
  (\textit{Preprint} \eprint{hep-ph/0202239})

\bibitem{Hambye:2004}
Hambye T, Lin Y, Notari A, Papucci M and Strumia A 2004 {\em Nucl. Phys.\/}
  {\bf B695} 169--191 (\textit{Preprint} \eprint{hep-ph/0312203})

\bibitem{Raidal:2006}
Raidal M, Strumia A and Turzynski K 2005 {\em Phys. Lett.\/} {\bf B609}
  351--359 (\textit{Preprint} \eprint{hep-ph/0408015})

\bibitem{Pilaftsis:1997}
Pilaftsis A 1997 {\em Phys. Rev.\/} {\bf D56} 5431--5451 (\textit{Preprint}
  \eprint{hep-ph/9707235})

\bibitem{Cirigliano:2007a}
Cirigliano V, Isidori G and Porretti V 2007 {\em Nucl. Phys.\/} {\bf B763}
  228--246 (\textit{Preprint} \eprint{hep-ph/0607068})

\bibitem{Branco:2007}
Branco G~C, Buras A~J, Jager S, Uhlig S and Weiler A 2007 {\em JHEP\/} {\bf 09}
  004 (\textit{Preprint} \eprint{hep-ph/0609067})

\bibitem{Pilaftsis:2005}
Pilaftsis A and Underwood T~E~J 2005 {\em Phys. Rev.\/} {\bf D72} 113001
  (\textit{Preprint} \eprint{hep-ph/0506107})

\bibitem{Branco:2009}
Branco G~C, Gonzalez~Felipe R, Rebelo M~N and Serodio H 2009 {\em Phys. Rev.\/}
  {\bf D79} 093008 (\textit{Preprint} \eprint{0904.3076})

\end{thebibliography}
\providecommand{\newblock}{}

\end{document}